\documentclass[usegraphicx,usenatbib]{mn2e}
\usepackage[usenames,dvipsnames]{color}
\usepackage{epsf,url}

\def\lsim{ \lower .75ex \hbox{$\sim$} \llap{\raise .27ex \hbox{$<$}} }
\def\gsim{ \lower .75ex \hbox{$\sim$} \llap{\raise .27ex \hbox{$>$}} }

\newcommand{\thinhline}{\noalign{\global\arrayrulewidth0.005cm}\hline
                      \noalign{\global\arrayrulewidth1pt}}

\begin{document}
	
\title[How sensitive are predicted galaxy luminosities to the choice of SPS model?]{How sensitive are predicted galaxy luminosities to the choice of stellar population synthesis model?}
 
\author[]{
\parbox[t]{\textwidth}{
\vspace{-1.0cm}
V.\,Gonzalez-Perez$^{1,2}$,
C.\,G.\,Lacey$^{2}$,
C.\,M.\,Baugh$^{2}$,
C.\,D.\,P.\,Lagos$^{3}$,
J.\,Helly$^{2}$,
D.\,J.\,R.\,Campbell$^{2}$,
P.\,D.\,Mitchell$^{2}$.
}
\\
$^{1}$Centre de Physique des Particules de Marseille, Aix-Marseille
Universit\'e,  CNRS/IN2P3, Marseille, France.
\\
$^{2}$Institute for Computational Cosmology, Department of Physics, 
University of Durham, South Road, Durham, DH1 3LE, U.K.
\\
$^{3}$European Southern Observatory, Karl-Schwarzschild-Strasse 2, 85748, Garching, Germany.
}
 
\maketitle

\begin{abstract}
We present a new release of the {\sc galform} semi-analytical
model of galaxy formation and evolution, which exploits a Millennium
Simulation-class N-body run performed with the WMAP7 cosmology. We use this new model to study the impact of the choice of stellar
population synthesis (SPS) model on the predicted evolution of the
galaxy luminosity function. The semi-analytical model is run
using seven different SPS models. In each case we obtain the rest-frame
luminosity function in the far-ultra-violet, optical and
near-infrared (NIR) wavelength ranges. We find that both the predicted rest-frame ultra-violet and optical luminosity function  are
insensitive to the choice of SPS model. However, we find that the
predicted evolution of the rest-frame NIR luminosity function
depends strongly on the treatment of the thermally pulsating asymptotic
giant branch (TP-AGB) stellar phase in
the SPS models, with differences larger than a factor of 2 for
model galaxies brighter than $M_{\rm AB}(K)-5$log$h<-22$ ($\sim$L$_*$ for
$0\leq  z\leq 1.5$). We have also explored the predicted
number counts of galaxies, finding remarkable agreement between the results with
different choices of SPS model, except when selecting galaxies with
very red optical-NIR colours. The predicted number
counts of these extremely red galaxies appear to be more affected by
the treatment of star
formation in disks than by the treatment of TP-AGB stars in the SPS models. 

\end{abstract}


\section{Introduction}
The luminosity function is a basic observable of the
galaxy population and is of fundamental importance since it encodes
information about the physics of galaxy formation and evolution. If galaxies populate their host dark matter haloes in a
simple way such that the mass-to-light ratio of all the galaxies
within a halo is constant, we would expect to observe more galaxies at both bright
and faint luminosities than are seen \citep{baugh06}. The formation of galaxies
at extreme luminosities is prevented by different mechanisms related to the
processes affecting the cooling of gas \citep{whiterees78,kauffmann93,benson03,delucia10,lacey11}. At high
redshift, photoionization of the intergalactic medium (IGM) prevents the formation of faint
galaxies, through the suppression of both the accretion of baryons and gas
cooling in low circular velocity haloes \citep{benson02,okamoto08}. However,
this is not enough by itself to explain the observed faint end of the
galaxy luminosity function at the present day. Stellar feedback, the
reheating of the gas by supernovae (SNe) in the disks
of galaxies forming stars, also has an impact on
the faint end of the luminosity function
\citep{cole94,benson03}. Models of galaxy formation and evolution
suggest that the bright end of the
luminosity function is mostly affected by the feedback from Active
Galactic Nuclei (AGN) and the duration of starbursts \citep[][ Lacey et al., in preparation]{croton06,bower06}.

Galaxy formation models are useful tools to understand which physical
processes shape the evolution of observed galaxies
properties, such as their luminosities. Stellar population synthesis (SPS) models are used to calculate the
luminosity of galaxies from the predicted star formation and metal enrichment histories directly predicted by the galaxy formation models. Thus, the comparison between model
predictions and observations is subject to the accuracy of the SPS models \citep{conroy1,chen10}. The performance of SPS models is mainly limited by the uncertainties in poorly
constrained stellar evolutionary phases, in particular, the asymptotic
giant branch and later phases \citep[e.g.][]{charlot96}. \citet{mancone12} studied the colours modelled using
  different SPS models, and found that they agree best in the optical
  for old ages and solar metallicities. In their study they pointed
  out that the dependency of SPS uncertainties on both age and wavelength
  will translate into uncertainties which change with redshift in a non-trivial way.

Over the last decade SPS models have incorporated improved
descriptions of stars in a TP-AGB phase
\citep{bc03,mn05,conroy3}, and also of horizontal branch stars which are particularly relevant for the
ultra-violet (UV) emission from old stellar populations
\citep{conroy3,smith12}. The contribution from TP-AGB stars is notably controversial, since some
observations, in particular those of the characteristics of
globular clusters, favour models with stronger emission from
TP-AGB stars \citep{goud06,eminian08,macarthur10,riffel11,lyu12,andreon13}, while other observations, such as those
focusing on post-starburst galaxies, favour models with a smaller contribution from TP-AGB stars
\citep{kriek10,melbourne12,zibetti13}. Stars in the TP-AGB phase can
dominate the near-infrared (NIR) luminosity \citep{mn05}. Rest-frame NIR luminosities are less
affected by dust than shorter wavelengths and thus, have been used to derive stellar
masses. However, the
uncertainties regarding the TP-AGB phase could hinder these determinations
and better constraints are needed \citep{kelson10,mcgaugh13}. 

Semi-analytical galaxy formation models combined with a SPS model with a strong TP-AGB
phase are able to reproduce the observed numbers of extremely red
galaxies in place at $z>1$
\citep{tonini09,tonini10,fontanot10,henriques11,henriques12}. However, the need for a strong TP-AGB phase to match this galaxy
population is not found in all models \citep{eros1}. Here we
assess the robustness of the predicted
luminosities at different wavelengths by combining a new version of the {\sc galform} semi-analytical model with
seven different SPS models. Among these,
there are two which include a strong TP-AGB phase \citep{mn05,cb07} and
one in which the intensity of the TP-AGB phase can be modified \citep{conroy1}. All of the SPS models studied cover a large
range in wavelength, from the UV to the NIR and beyond in some cases, and most are publicly
available.

The model used in this study is a new development of {\sc galform},
the semi-analytical model for the formation and evolution of galaxies
developed mainly at Durham University \citep{cole00}.  Much of the
previous work with  {\sc galform} has used the Millennium
Simulation. The cosmology assumed for the Millennium Simulation, which is close to that derived from the WMAP1 observations \citep{wmap1}, is inconsistent with current observations. Here, we have updated the \citet{lagos12} model to the
WMAP7 cosmology \citep{wmap7}, which is also close to Planck cosmology
\citep{planck} on the scales relevant to galaxy formation. In order to
do this update, we have implemented the {\sc galform} semi-analytical model
using the dark matter halo trees derived from the MS-W7 N-body
simulation (Lacey et al. in preparation), which is based on a WMAP7 cosmology. A
similar MS-W7 simulation has previously been used in combination with
a semi-analytical model of galaxy formation to explore how to rescale
the model predictions from those for an underlying cosmology similar
to WMAP1 to the WMAP7 one \citep{qi12}. In order to update the {\sc galform} semi-analytical model to the
WMAP7 cosmology, we have found the best-fitting model parameters to
reproduce a set of observations as discussed in \S\ref{sec:model}. 

\citet{lagos11} introduced a new
implementation of the star formation law which follows the atomic and
molecular hydrogen in the interstellar medium based on the empirical
relation between the star formation rate and the surface density of
molecular hydrogen inferred by \citet{blitz06}. This new implementation
is more realistic than the simplified one used before in the majority
of semi-analytical models, in which the star
formation was assumed to be proportional to the total mass of cold gas in
a galaxy \citep[e.g.][]{cole00}. 

The new model presented here uses a single initial mass function (IMF)
in all modes of star formation and can be considered as an alternative
to the Lacey et al. (in
preparation) model, which is also implemented in a WMAP7 cosmology but which
does not assume a universal IMF. The universality of the IMF has been
challenged by different observations
\citep[e.g.][]{sw08,cappellari12,geha13,labarbera13}. In {\sc galform}, a top-heavy
IMF has been found to be needed in order to reproduce the observed
numbers and redshift distribution of submillimeter galaxies, i.e. luminous dusty galaxies at redshifts around
$1<z<3$ (see Fig.\ref{fig:submm} in the Appendix for further details). However, it is very uncertain how and in which
environments the IMF changes. The comparison of the model predictions
using different SPS models is simpler for a model with a single IMF,
as is the case in this paper.

In Section \S\ref{sec:model} we describe the new model of galaxy formation and evolution
used in this study. Section \S\ref{intro:sps} summarises the
differences between the SPS models considered. In Sections
\S\ref{sec:lf} and  \S\ref{sec:nc} we compare the luminosity functions
and number counts predicted using different SPS models with
observations. Our conclusions are presented in \S\ref{sec:conclusions}. The Appendix contains further
comparisons between observations and predictions of the new model
presented in this work. This model is available in the Millennium
Archive Data Base\footnote{\url{http://galaxy-catalogue.dur.ac.uk:8080/Millennium/}, \url{http://gavo.mpa-garching.mpg.de/Millennium/}}. There, it is also possible to get the post-procesing calculation introduced by \citet{lagos12}
for estimating the emission of the most widely used tracer of molecular hydrogen: the carbon monoxide, $^{12}$CO.

\section{Galaxy formation model}\label{sec:model}
We present a new development of {\sc galform} \citep{cole00}, a semi-analytical galaxy formation 
model set in a $\Lambda$ cold dark matter universe. Semi-analytical models 
use simple, physically motivated equations to follow the 
fate of baryons in a universe in which structure grows hierarchically 
through gravitational instability \citep[see Baugh 2006 and][for an overview of 
hierarchical galaxy formation models]{benson10}. 

{\sc galform} models the main processes which shape the formation and
evolution of galaxies. These include: (i) the collapse and merging of
dark matter haloes; (ii) the shock-heating and radiative cooling of
gas inside dark matter haloes, leading to the formation of galactic
discs; (iii) quiescent star formation in galactic discs; (iv) feedback
from SNe, from AGN  and from
photoionization of the IGM; (v) chemical enrichment of the stars and gas; (vi) galaxy
mergers driven by dynamical friction within common dark matter haloes,
leading to the formation of stellar spheroids, which also may trigger
bursts of star formation. The model also computes the scale size of
the disk and bulge component of the galaxy. Galaxy luminosities are
obtained by combining a SPS model with the star formation and metal
enrichment histories predicted for each model galaxy. The attenuation
of starlight by dust is modelled in a physically self-consistent way,
based on the results of a radiative transfer calculation \citep{ferrara99} for a
realistic geometry in which stars are distributed in a disk plus bulge
system \citep{cole00,lacey11,drops}. The end product of the calculation is a prediction of the number and properties of galaxies that reside
within dark matter haloes of different masses. The free parameters in the semi-analytical model are chosen in order to reproduce the set of observations described in section \S\ref{sec:w7}.

The model presented here is an updated version of the \citet{lagos12}
model. The \citeauthor{lagos11} model adopts the cosmological parameters of the
Millennium Simulation \citep{springel05}, which correspond
approximately to the results of the first year of the WMAP satellite
\citep[WMAP1][]{wmap1}. For the work presented here we have updated
this model to the best fitting cosmological parameters of the WMAP 7 year dataset \citep[WMAP7][]{wmap7}. In \S\ref{sec:w7} we give details of how this updated model differs from the original one.

\citeauthor{lagos12} introduced a new calculation of the star
formation rate in galaxies compared with that used in the
\citet{bower06} model. \citeauthor{lagos12} also increased the starburst
duration, without further changes to other parameters. The
\citeauthor{lagos12} model was not actually retuned to obtain a good
match to observational data, but nevertheless, gave reasonable
agreement with the observed  b$_j$ and K-band luminosity functions at $z=0$, the HI and H$_2$ mass functions at $z=$0, the observed cold gas content of
galaxies up to $z=2$, the bimodality in the star formation rate versus stellar
mass plane for local galaxies and the cosmic density evolution of HI \citep{lagos10,lagos11,lagos12}.

In \citeauthor{bower06} the star formation rate is assumed to be simply
proportional to the mass of cold gas present in the galaxy  and inversely
proportional to the dynamical time. Recent
observations have motivated a more sophisticated calculation in which the quiescent
surface density of the star formation rate is assumed to be proportional to the surface density of molecular
hydrogen mass in the interstellar medium (ISM) \citep{blitz06,leroy08,bigiel08}. In this calculation, the ratio between the molecular and total gas  is determined by a pressure law \citep{blitz06}.  \citet{lagos11} implemented a self consistent
calculation into {\sc galform} in
which HI and H$_2$ are tracked explicitly and the star formation in
disks is assumed to
depend on the amount of molecular gas, H$_2$, rather than on the total
mass of cold gas. The introduction of this observationally motivated
model of the star formation rate leads to a reduction in the size of the
available parameter space, as the star formation parameters are
determined empirically from the observations. Furthermore, the
constraints on the model are extended through the ability to make new
predictions such as the HI and H$_2$ mass functions.

\subsection{Dark matter halo merger trees}\label{sec:halos}

\begin{table}
  \begin{tabular}{ c | c | c }
  \hline                       
  & Lagos12 & Gonzalez-Perez14 \\ 
  Parameter & (WMAP1) & (WMAP7) \\
  \hline                        
  $\Omega_{\rm m0}$ & 0.25 & 0.272 \\
  $\Omega_{\Lambda 0}$ & 0.75 & 0.728 \\
  $\Omega_{\rm b0}$ & 0.045 & 0.0455 \\
  $\sigma_{8}$ & 0.9 & 0.810 \\
  $h$          & 0.73 & 0.704 \\
  \hline  
  $y$ & 0.020 & 0.021 \\
  $R$ & 0.39 & 0.44 \\
  \hline  
  $V_{\rm hot}$ & 485 & 425 \\
  $\alpha_{\rm cool}$ & 0.58 & 0.6 \\
  $\tau_{\rm min}$ &  0.1  & 0.05 \\
  $f_{\rm dyn}$ &  50  & 10 \\
  \hline                        
\end{tabular}
  \caption{The parameters varied from the \citet{lagos12} model
    (second column), which is
    implemented in the WMAP1 cosmology and the updated model
    presented here (third column). The new model is
    implemented in the WMAP7 cosmology, with an improved algorithm for constructing N-body merger trees \citep{jiang13},
    and for which we fixed the metal yield, $y$, and recycled fraction,
    $R$, to be consistent with the chosen Kennicutt IMF. The first column provides the names of
    the parameters (see the text for their definition). Top rows:
    cosmological parameters; Center rows: $y$, and $R$, which are not free
    parameters in our new model but are determined by the choice of IMF; Bottom rows: modified galaxy formation parameters
    (defined in \S\ref{sec:w7}).
}
\label{tab:param}
\end{table}

The model that we present here is implemented in a new Millennium Simulation run with
the WMAP7 cosmology (MS-W7; Lacey et al. in preparation). This cosmology is very similar to the best fit for the recently released Planck data \citep{planck}; the power spectra corresponding to the best fitting models to WMAP7 and Planck data are very close to one another on the scales relevant to galaxy formation. The MS-W7 N-body simulation uses $2160^3$ particles, each with a
mass of $9.35\times10^8h^{-1}M_{\odot}$, in a box of side $500
h^{-1}$Mpc and with a starting redshift of $z=127$ \citep[see][for details of the original Millennium Simulation]{springel05}. The MS-W7 simulation
uses the WMAP7 cosmological parameters \citep{wmap7}, that are also
summarized in Table \ref{tab:param}: matter density, $\Omega_{\rm m0}=0.272$, cosmological constant, 
$\Omega_{\Lambda 0} = 0.728$, baryon density, $\Omega_{\rm b0}=0.045$, a normalisation of density fluctuations given by $\sigma_{8}=0.810$ and a Hubble constant
today of $H_0=100\,h$ km$\,{\rm s}^{-1}$Mpc$^{-1}$, with
$h=0.704$. The 61 outputs of the simulation are approximately spread
evenly in the logarithm of the expansion factor and so do not correspond to round numbers in redshift.

{\sc galform} calculates the evolution of galaxies in halo merger
trees which describe the assembly and merger histories of cold dark
matter halos. For the model presented here, the halo merger trees have
been extracted from the MS-W7 N-body simulation.

Briefly, the construction of the merger trees starts with two
consecutive steps: 1) groups of dark matter particles are identified
in each simulation snapshot using the {\it Friends-Of-Friends}
algorithm \citep{fof}; 2) self-bound, locally over-dense sub-groups
are identified using the algorithm {\sc subfind} \citep{springel01}. In some cases, this procedure identifies structures as groups that might better to be considered as distinct haloes for the purpose of implementing the semi-analytical galaxy formation model. The merger tree algorithm used in this work deals with such cases and ensures that the resulting trees are strictly hierarchical, i.e. once two haloes are considered to have merged they remain merged at all later times. A full description of the construction method of the halo merger trees is presented in \citet{jiang13}. This new method is similar to that described in
\citet{merson13} but differs in two
points: 

\begin{enumerate}
      \item When determining the main progenitor for each {\sc subfind} sub-group, the new scheme tries to identify the most bound ``core'' of dark matter particles, rather than the most massive one.
       \item When looking for descendants, the new scheme  looks for the same most bound ``core'' of particles. As with (i), before this was done by looking for the most massive progenitor.
\end{enumerate}

These changes were mainly driven by issues that arise in the
construction of trees in very high resolution N-body simulations and do not have a noticeable impact here.

\begin{figure}
\hspace{-0.5cm}\includegraphics[width=9.5cm]{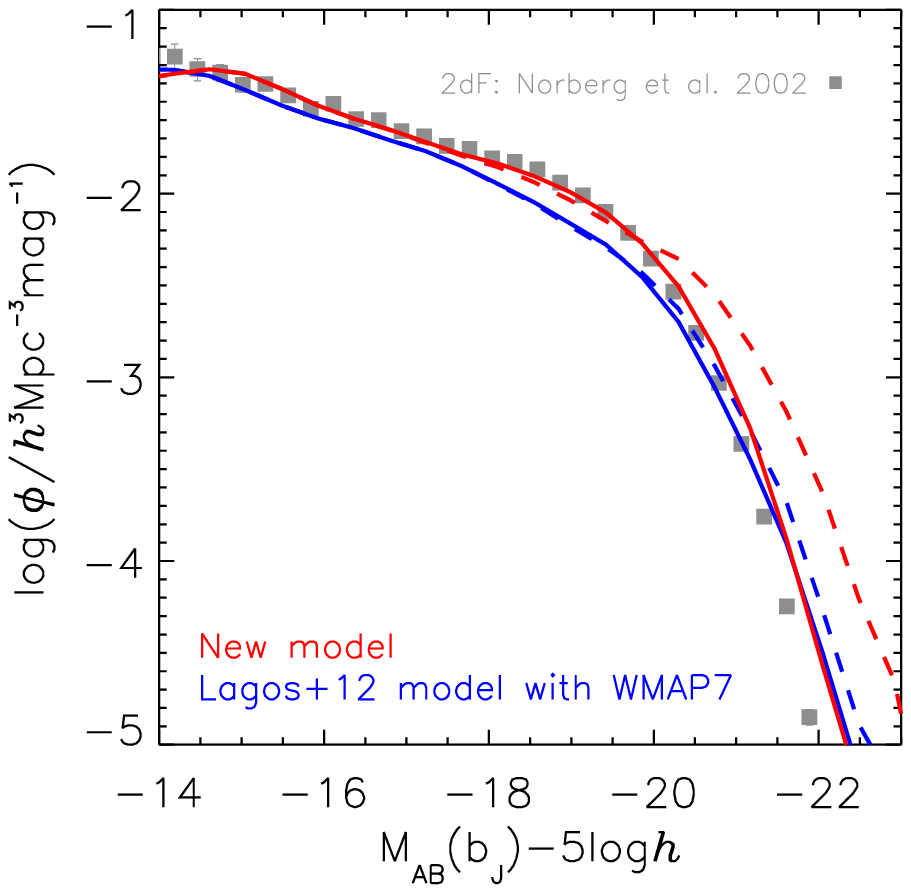}

\hspace{-0.5cm}\includegraphics[width=9.5cm]{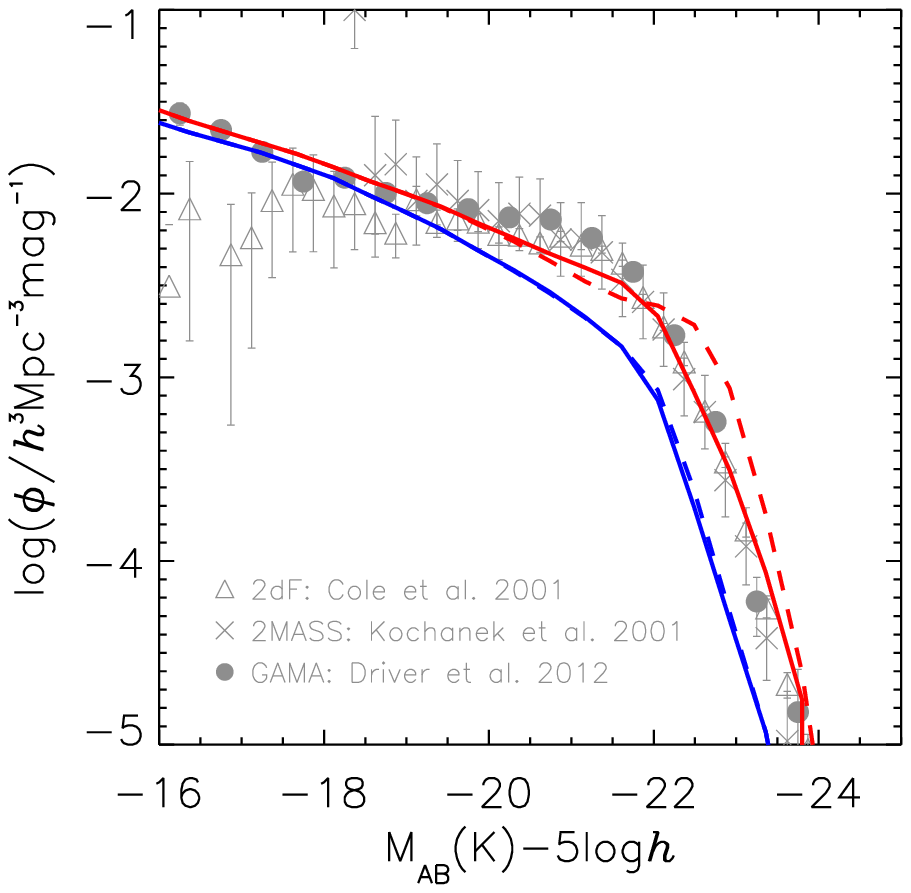}

\caption{The predicted luminosity functions at $z=$0 (solid lines), in
  the b$_{\rm J}$-band ($\lambda _{\rm eff}= 4500$\AA, top)
  and in the K-band ($\lambda _{\rm eff}=2.2\mu$m, bottom), compared with observations from
  \citet{cole01} (triangles), \citet{kochanek01} (crosses), \citet{norberg02} (squares)
  and \citet{driver12} (circles). The blue lines show the predictions from the
  \citeauthor{lagos12} model run in the WMAP7 cosmology, while the
  red lines show the predictions from the new model presented here. The dashed
  lines show the predicted luminosity function without dust
  attenuation. 
}
\label{fig:model}
\end{figure}
\subsection{{\sc galform} in the WMAP7 cosmology}\label{sec:w7}

The new model presented here is essentially an update of the
\citeauthor{lagos12} model to the WMAP7 cosmology. Table
\ref{tab:param} summarises the cosmological parameters used in the
\citeauthor{lagos12} model, which are close to those obtained from
WMAP1 \citep{wmap1}, and the WMAP7 parameters used here
\citep{wmap7}. The growth of structure in the WMAP1 and WMAP7
cosmologies can be compared by looking at the parameter
combination $ \Omega_{m0}^{0.55} \sigma_{8}$ \citep{Eke96,Linder05}.
The differences between the best-fitting values of $\Omega_{m0}$ and
$\sigma_{8}$ in these two cosmologies almost compensate one another in this
parameter combination, so that $ \Omega_{m0}^{0.55} \sigma_{8}$
differs by only 6 per cent. Although this change result in only a little difference in the
mass function of dark matter haloes in these cosmologies, it does have
a visible impact on the predicted luminosity functions at $z=0$. Fig.\ref {fig:model} shows the luminosity
functions at $z=0$ in $b_J$ and K-bands predicted using the original
set of galaxy formation parameters from the \citeauthor{lagos11} model but
implemented in the WMAP7 cosmology. This figure shows the effect that changing the cosmological parameters has on the model of galaxy evolution which, when implemented in the WMAP1 cosmology, satisfactorily reproduced the observations.

In this section we describe the parameters that control the physical
processes in the model which were changed in order to reproduce the b$_J$ and
K-band luminosity functions at $z=$0 and to give reasonable evolution of the predicted rest-frame
UV and K-band luminosity functions.

Before starting to tune the model parameters, for this work we decided
not to consider the metal yield\footnote{Here we
  assume the instantaneous recycling approximation and, thus, the
  metal yield is defined as the mass of metals recycled to the ISM per mass of stars formed.
}, $y$, and recycled fraction\footnote{Given a particular IMF and the
  instantaneous recycling approximation, the recycled fraction is
  defined as the fraction of mass recycled to the ISM per mass of
  stars formed.}, $R$, as free parameters. These two parameters are set to
those values calculated using the Padova stellar evolution  models for the
\citet{kennicutt_imf} IMF: yield = 0.021 and recycled fraction =
0.44 \citep[for a discussion of the uncertainties on these values we
refer the reader to][]{cole00}. Note that these values are different from those
used in \citet{bower06} and \citet{lagos12}. 


Due to the slightly slower growth of structure in the WMAP7 cosmology compared
with that with the WMAP1 parameters, the
feedback efficiencies regulating star formation
should be reduced in order to reproduce the observed luminosity function at $z=0$. This
was previously noted by \citet{qi12} in their comparison of galaxy
formation in the WMAP1 and WMAP7 cosmologies. The SNe feedback
efficiency is quantified in the model in terms of the rate at which
cold gas is reheated and thus ejected into the halo, $\dot{M}_{\rm
  reheated}$, per unit mass of stars formed, $\psi$. In {\sc
  galform}, this rate is assumed to scale as:
\begin{equation}
\dot{M}_{\rm reheated}=\psi\Big(\frac{v_{\rm circ}}{v_{\rm hot}}\Big)^{-\alpha _{\rm hot}},
\end{equation}
where $v_{\rm circ}$ is the circular velocity of either the disk or the bulge (i.e. that characterising
the potential well of the galaxy) and $\alpha _{\rm hot}$ and $v_{\rm
  hot}$ are adjustable parameters. By default, in the
\citeauthor{lagos11} model these parameters are set to $\alpha
_{\rm hot}=3.2$ and $v_{\rm hot}=485\, {\rm km\, s}^{-1}$, as used in \citeauthor{bower06} The SNe
feedback efficiency can be turned down by reducing the value of $v_{\rm
  hot}$. Making such a change alone is enough to produce a luminosity
function at $z=$0 that is in good
agreement with observations in both the $b_J$ and K-bands. However, in
order to update the  \citeauthor{lagos11} model to the WMAP7 cosmology
we have also paid attention to predicting simultaneously reasonable
evolution of both the rest-frame UV and K-band luminosity functions. We find
that, in general, the UV luminosity function agrees better with observations of star
forming galaxies at high redshifts if longer duration bursts are
assumed. However, extending the duration of the bursts worsens the agreement between the model and observations in
rest-frame K-band at the bright end. \citet{henriques13} found that without changing the physical dependencies for gas ejecta,
no combination of parameters could simultaneously reproduce both low and high redshift observations.

In {\sc galform} it is assumed that the available reservoir of cold
gas is consumed during a starburst event with a finite duration. In
our model, as well as in \citeauthor{bower06} and
\citeauthor{lagos11}, starbursts can be triggered by both mergers and
disks becoming dynamically unstable. The
timescale of a starburst event is considered to be proportional to
the bulge dynamical time, $\tau_{\rm bulge}$, through a parameter
$f_{\rm dyn}$, except when this timescale is too small, then a
minimimal duration time is considered, $\tau_{\rm min}$ \citep[for
further details see][]{lacey11}. Thus, due to the scaling with the
dynamical time of the bulge, the duration of starbursts will
change with redshift. Starting from the values advocated by \citeauthor{lagos11}, we find a compromise by decreasing $\tau_{\rm min}$ from 0.1 Gyr to 0.05 Gyr, while increasing $f_{\rm dyn}$ from 2 to 10. A model including these further changes slightly overpredicts the bright end of the K-band luminosity function at $z=$0. Therefore, we increased very slightly the strength of the AGN feedback.

The onset of the AGN
suppression of the cooling flow in the model is governed by a comparison of the
cooling time of the gas, $t_{\rm cool}$ with the free-fall time for the gas to reach
the centre of the halo, $t_{\rm ff}$. Galaxies with  $t_{\rm
  cool}>t_{\rm ff}/ \alpha_{\rm cool}$, where $\alpha_{\rm cool}$ is a
model parameter, are considered to be hosted by
haloes undergoing quasi-hydrostatic cooling. In such haloes, cooling is suppressed if the luminosity released by gas accreted on to a central
supermassive black hole balances or exceeds the cooling luminosity
\citep[see][for further details]{bower06,fanidakis11}. Increasing $\alpha_{\rm cool}$ slightly from 0.58 to 0.6 was found to be enough to compensate the changes in the burst duration which affected the K-band luminosity function. This change implies that slightly more haloes will be quasi-static, leading to the possible quenching of gas cooling and hence star formation.

Table \ref{tab:param} lists the parameters changed from the values
used in the
\citet{lagos12} model in order to update it to the WMAP7
cosmology, once the value of the yield and recycled fractions were
fixed to be consistent with the predictions from stellar evolution
models for the Kennicutt IMF used here. The predicted b$_J$ and K-band
luminosity function at $z=$0 can be seen in Fig.\ref{fig:model} compared with
observations. Further predictions of the model introduced here are presented in the Appendix.

\section{Stellar population synthesis models}\label{intro:sps}
\begin{table*}
  \begin{tabular}{ c | c | c | c | c | c }
    \hline
  Model & Age range (Gyr) & Z range & $\lambda$ range (\AA)
  & Stellar tracks & Library of stellar
  spectra  \\
  \hline
  {\bf PD01}  & [$10^{-4}$, 16]  (51) &
  [0.0004,0.05] (5) &  [91,$9.5\cdot  10^{9}$] (1301) & 
  Pd94$^a$ and & Extended \citet{kurucz93}$^b$  \\
   & & & & analytical TP-AGBs &   \\
  \thinhline                             
               
  {\bf P2}  & [0, 20]  (516) &
  [0.0001,0.1] (7) &  [91,$16\cdot  10^{5}$] (1221) & 
  Modified Pd94$^c$ & BaSeL-2.0$^d$  \\
  \thinhline                                            

  {\bf BC99} & [$0$, 20]  (221) & [0.0001,0.1] (7) & [91,$16\cdot  10^{5}$] (1221) & Pd94 & Extended \citet{kurucz93}$^e$  \\
  \thinhline                        

  {\bf BC03}  & [$0$, 20]  (221) &
  [0.0001,0.05] (6) & [91,$16\cdot  10^{5}$] (1221) &
  Pd94 and an  &  STELIB$^f$, BaSeL-3.1 and \\
& & & &  observationally & spectra for carbon  stars  \\
& & & &  motivated & and TP-AGBs based on \\
& & & &  prescription & models and observations \\
& & & &  for TP-AGBs  & of the Galactic stars \\
  \thinhline                        
  {\bf MN05}  & [$10^{-6}$, 15]  (67) &
  [0.0001,0.07] (6)  & [91,$16\cdot  10^{5}$] (1221) & Fuel consumption
  &  BaSeL-3.0 and \\
  & & & & integration  & \citet{lancon02} \\
  & & & & (see text) & spectra for TP-AGBs\\
  \thinhline                        
  {\bf CB07}  & [$0$, 20]  (221) &
  [0.0001,0.05] (6) & [91,$3.6\cdot  10^{8}$] (1238) & Pd08$^g$  & Similar to BC03 \\
& & & &  & with a different pres- \\
& & & &  & cription for TP-AGBs \\
  \thinhline                        
  {\bf CW09} & [$3\cdot  10^{-4}$, 15]  (188) &
  [0.0002,0.03] (22) &  [91,$10^{8}$] (1963) &  Pd08
  and models & BaSeL-3.1 and \\
  & & & & for stars with   & \citet{lancon02} \\
  & & & &  $0.1\leq M(M_{\odot})<0.15$ & spectra for TP-AGBs \\

  \hline
\end{tabular}
  \caption{The first column contains the labels given here to the SPS
    models used in this study: PD01 \citep{grasil}, P2 \citep[{\sc P\'EGASE.2}, ][]{pegase,pegase2}, BC99 \citep[an updated version of][]{bc99}, BC03 \citep{bc03},
  MN05 \citep{mn05,maraston11}, CB07 \citep{cb07} and
  CW09 \citep[FSPS v2.3, ][]{conroy1,conroy2,conroy3}. Columns two, three and four summarise the age,
    metallicity and wavelength
    ranges, with the number of values given by the SPS models in
    parentheses. Column five gives the stellar
    evolutionary tracks. Column six lists the libraries of stellar
    spectra used by the SPS models.
    $^a$ Most of the Pd94 stellar
    evolutionary tracks are described by \citet{padova94}. 
    $^b$ The Pd01 model includes dust emission from AGB envelopes and thermal continuum emission from HII regions and synchroton radiation, for further details see \citet{bressan98}. 
    $^c$ P2 uses Pd94 isochrones with modifications
    for stars undergoing a helium flash, the mass loss during the
    early asymptotic giant branch phase, the TP-AGB phase, post-AGB
    and helium white dwarfs. 
    $^d$ BaSeL is a semi-empirical library of stellar spectra
    \citep{lejeune97,lejeune98,westera02}. 
    $^e$ See \citet{bc99}.
    $^f$ STELIB is a stellar spectral library covering the range from
  3200 \AA\ to 9500 \AA\ \citep{stelib}. 
    $^g$ Pd08 are isochrones from the Padova group which include a prescription for the
TP-AGB evolution of low and intermediate mass stars \citep{marigo07,marigo08}.
}
\label{tab:sps}
\end{table*}

Stellar population synthesis (SPS) models provide a link
between the predictions of galaxy formation and evolution models and observations. Models like {\sc galform}
predict the intrinsic properties of galaxies, such as their stellar masses, star formation and metal enrichment
histories, gas content, etc. Given this information, the SPS
models provide a way to estimate the spectral energy distribution (SED) of the model galaxies. In order to make this link, the SED of simple stellar
populations, i.e. of a group of coeval stars with the same metallicity, must be known.

The SPS models compute the characteristics of simple stellar
populations by first adopting a stellar evolutionary model. For a given initial stellar mass and metallicity, these models
calculate the stellar isochrones, i.e. the luminosity of a star at
different ages, from the main sequence zero-point to its
death. Then stellar spectral libraries are used to assign a full
spectrum to each set of values for the initial stellar mass, metallicity and
age. The SED of a simple stellar population is then obtained by
integrating the light from coeval stars weighted by the chosen stellar
initial mass function (IMF). The IMF describes
the initial number of stars formed in a given stellar mass bin. The IMF is usually approximated by one or more power
laws \citep[though, other parametrisations have also been used in the
  literature, in particular for masses below 1$M_{\odot}$, see ][]{chabrier03,vandokkum08}. This approximation is such that, in a given stellar mass
interval, the IMF can be expressed as: d$N(m)/$dln$m\propto m^{-x}$.

The SED of a galaxy is found by convolving the star formation history,
$\dot{m}_*(t)$, with the SED of a single stellar population, $\phi_{\lambda}$ (which includes the convolution with the IMF):
\begin{equation}
S_{\lambda}(t) = \int\limits_0^t
\phi_{\lambda}\left(t-t',Z(t')\right)\dot{m}_*(t'){\rm d}t',
\end{equation}
where $S_{\lambda}(t)$ is the resulting SED at time $t$ and $Z(t)$ is
the metallicity of the stars formed at time $t$. In
evaluating this integral, $\phi_{\lambda}(t-t',Z(t'))$ is obtained by linearly interpolating the tables provided by the SPS models to the appropriate time and metallicity. The SED of a galaxy obtained in this way can be convolved with a filter response function in order to obtain broadband luminosities.

\subsection{Comparing SPS models}\label{sec:ssp}
\begin{figure}
\begin{minipage}{8.5cm}
\includegraphics[width=4.3cm]{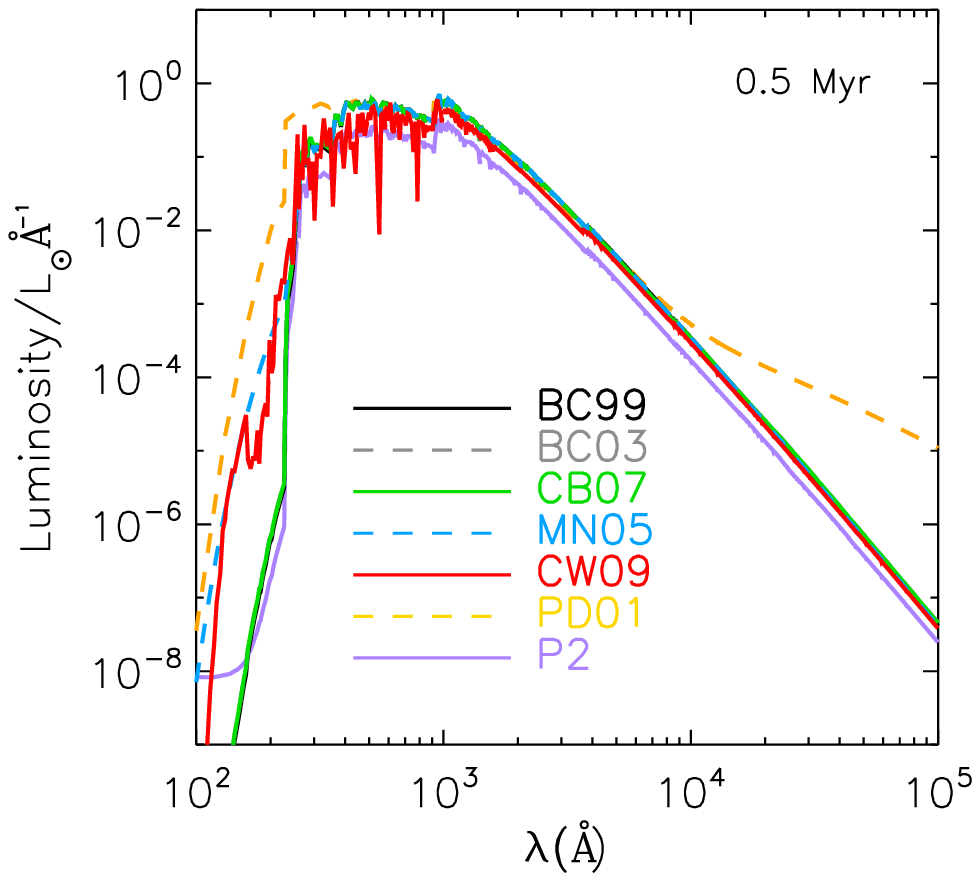}
\includegraphics[width=4.3cm]{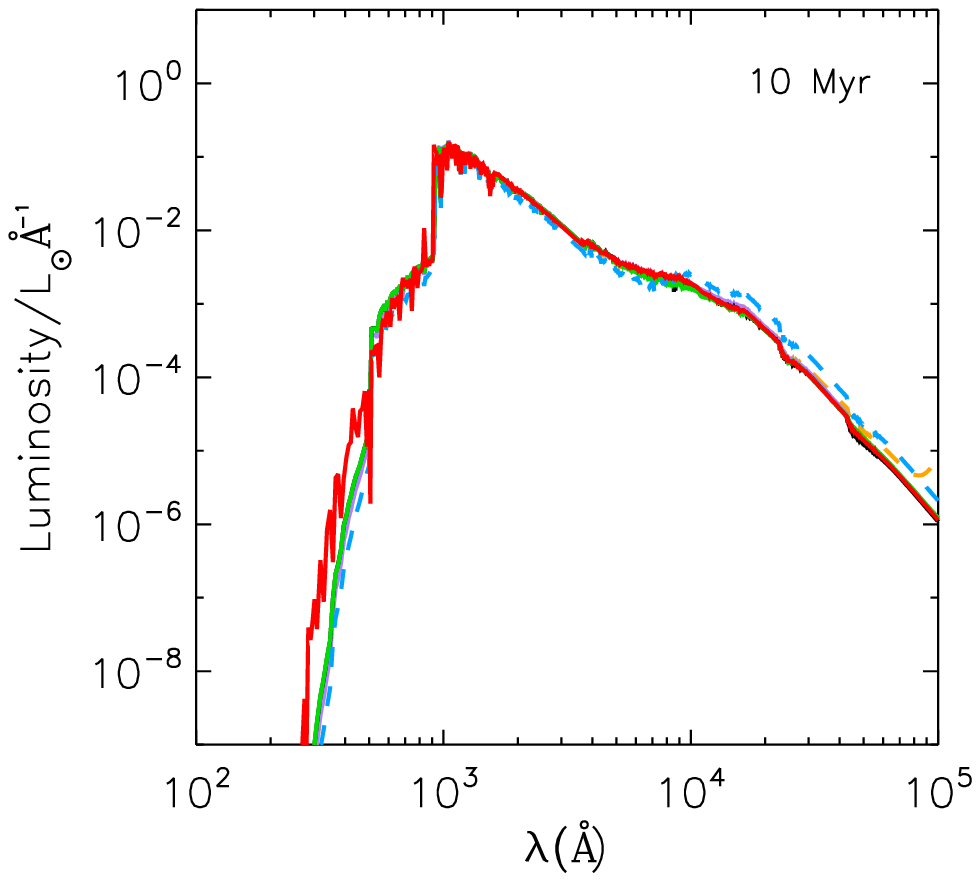}
\end{minipage}

\begin{minipage}{8.5cm}
\includegraphics[width=4.3cm]{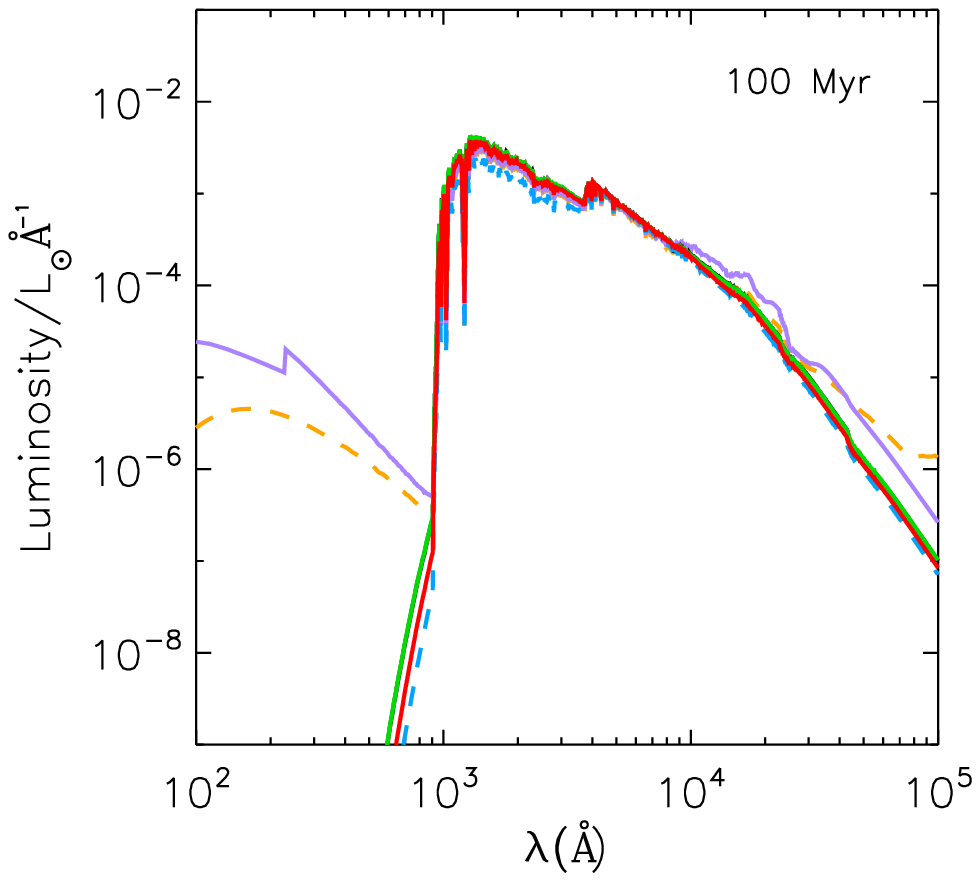}
\includegraphics[width=4.3cm]{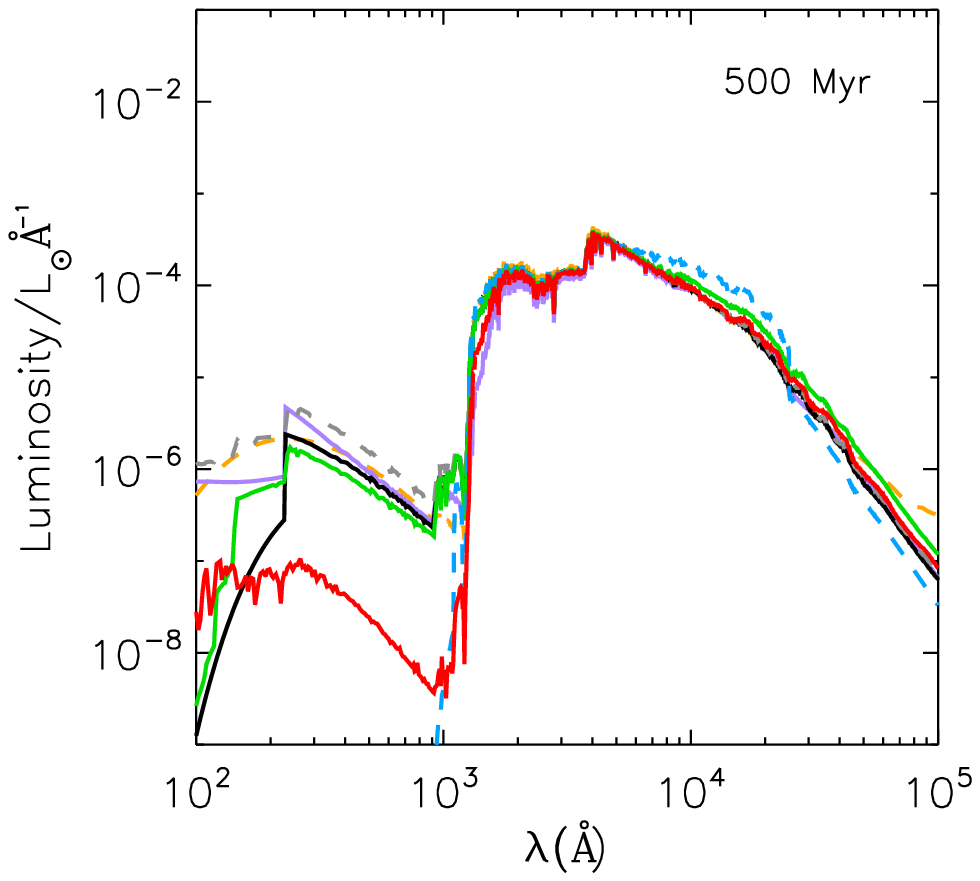}
\end{minipage}

\begin{minipage}{8.55cm}
\includegraphics[width=4.3cm]{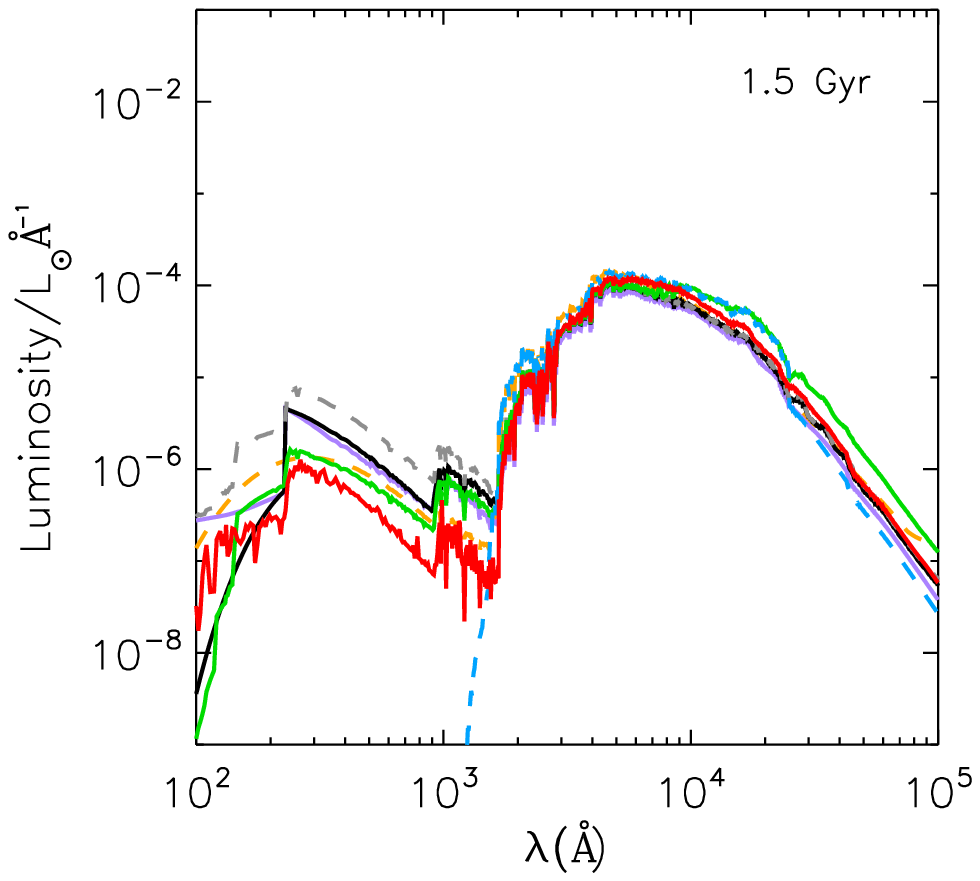}
\includegraphics[width=4.3cm]{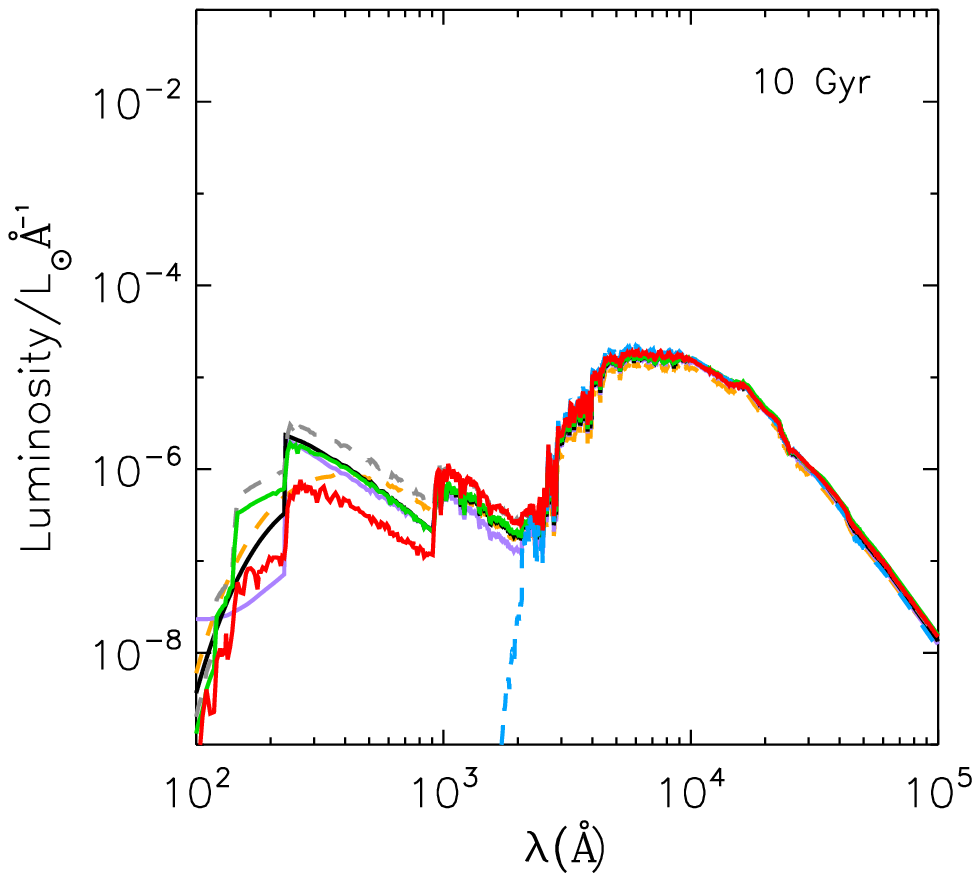}
\end{minipage}    

\caption{An illustration of the differences between the single stellar population luminosity
  obtained for a Salpeter IMF and solar metallicity, normalized to 1 M$_{\sun}$, as a function of
  wavelength for fixed ages using the BC99,
  BC03, MN05, CB07, CW09, PD01 and P2 SPS models as indicated in the
  legend. From left to
  right and top to bottom, the panels show the spectra of the simple stellar populations at ages: 0.5 Myr, 10
   Myr, 500 Myr and 1.5 Gyr. Note that the y-axis limits change from
   10 to 0.01, going from the first to second row.
}
\label{fig:sspage}
\end{figure}

As described above, the main ingredients of a SPS model
are: (i) the stellar evolutionary tracks, (ii) stellar spectral
libraries and (iii) the IMF. By default, {\sc
  galform} assumes a \citet{kennicutt_imf} IMF for which d$N(m)/$dln$m\propto
m^{-0.4}$ if $m\leq 1M_{\odot}$ and d$N(m)/$dln$m\propto
m^{-1.5}$ otherwise. We will be using
the different SPS models in combination with the
\citet{kennicutt_imf} IMF with
a mass range set to $0.1 \leq m(M_{\sun}) \leq 100$, except
when this IMF is not provided in the public release of the SPS model (only for
Figs. \ref{fig:sspage}, \ref{fig:ssp} and \ref{fig:cb07}). In such cases we will
use a Salpeter IMF for which d$N(m)/$dln$m\propto m^{-1.35}$, at all masses.

Table \ref{tab:sps} summarises the main ingredients of the following SPS
models: PD01 \citep{grasil}, P2\footnote{There is a new
  implementation of the P2 model, the {\sc P\'EGASE.3} model
  \citep{p3a,p3}, which is not yet public. This model extends into the far infrared
  region by calculating self-consistently the re-emission by dust of
  the UV/optical continuum. This new implementation currently also
  uses the Pd94 stellar tracks and does not include a modified
  treatment of the TP-AGB stellar phase. Therefore, in the studied
  wavelength range, we expect that we would obtain similar results using either version of the P2 model.} \citep{pegase,pegase2}, BC99 \citep[an updated version of][]{bc99}, BC03 \citep{bc03},
  MN05 \citep{mn05}, CB07 \citep{cb07} and
  CW09 \citep[FSPS v2.3, ][]{conroy1,conroy2,conroy3}. BC99 is the default SPS
  model used in the present model and also in other {\sc galform}
  developments such as: \citet{cole00}, \citet{bower06} and
  \citet{lagos12}. PD01 is the SPS model used in the version of
  {\sc galform} described by \citet{baugh05} and \citet{lacey11}. 

As can be seen in Table \ref{tab:sps} all of these SPS models cover
similar ranges in stellar metallicities and ages. In general, these SPS
models can be used in combination with a variety of low and high
resolution stellar spectral libraries. In order to cover the largest
possible wavelength and metallicity range, here we use the SPS models
combined with a low resolution stellar spectra library. The PD01 and
BC99 models are run in combination with an extended version of the spectra
from \citet{kurucz93} \citep[see
also][]{bressan98}. The other models make use of BaSeL, a semi-empirical
library of stellar spectra \citep{lejeune97,lejeune98,westera02}, with
extensions and modifications as indicated in Table \ref{tab:sps}.

All the SPS models used here, except the MN05 one, use isochrones to
compute the simple stellar populations for stars in different
evolutionary stages. The PD01, P2, BC99 and BC03 models are based on the
stellar tracks from the Padova group \citep[mostly described
in][]{padova94}, including modifications that differ from model to
model, as indicated in Table \ref{tab:sps}. Both the CB07 and CW09
models are based on a later version of the stellar tracks from the
Padova group which include a prescription for the
TP-AGB evolution of low and intermediate mass stars
\citep{marigo07,marigo08}. The CW09 SPS model also includes a
modification for stars with very low masses with respect to the Pd08 stellar tracks. 

The MN05 SPS model is the only one among those used in this study that
computes the characteristics of the simple stellar populations of post
main sequence stars by integrating the amount of hydrogen and/or
helium that is consumed during a given post main sequence phase. The fuel at a given
age depends on the mass of stars completing their hydrogen burning
phase. Thus, this integration neglects the dispersion of stellar
masses in post main sequence phases, an aspect supported by
observations. This approach has the advantage of minimising the use
of uncertain theoretical models for describing post main sequence phases of star
evolution. For main sequence
stars, this model uses
the tracks and isochrones from
  \citet{cassini97,cassini97b,cassini00} complemented with those from
  the Geneva group \citep{schaller92,meynet94} for very young
  evolutionary stages and from the Padova group for systems with high
  metallicities. The MN05 model allows for blue and red populations on the
  horizontal branch, recommending the use of red populations for
  metallicities $Z\geq 0.01$.

\subsubsection{Comparing simple stellar populations}
\begin{figure*}

\hspace{-0.5cm}
\begin{minipage}{5.8cm}
\includegraphics[width=6.1cm]{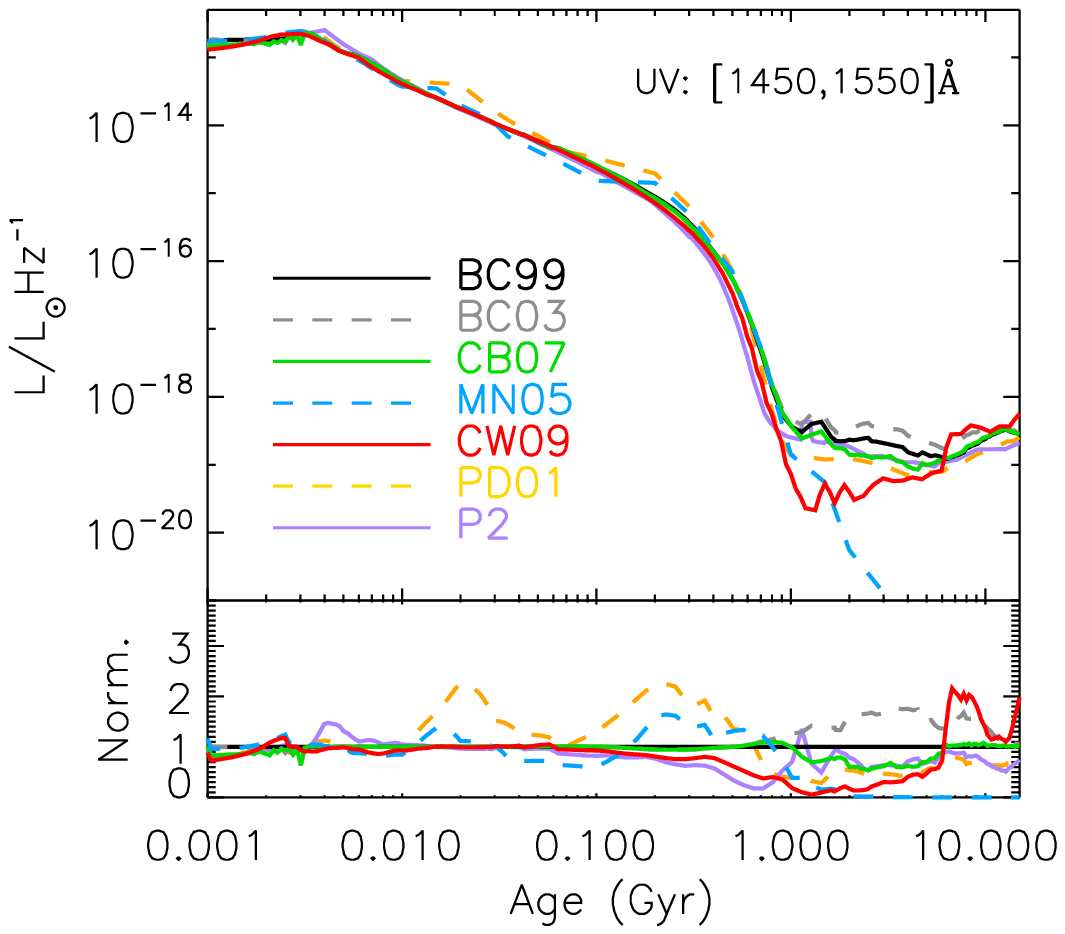}
\end{minipage}
\begin{minipage}{5.8cm}
\includegraphics[width=6.1cm]{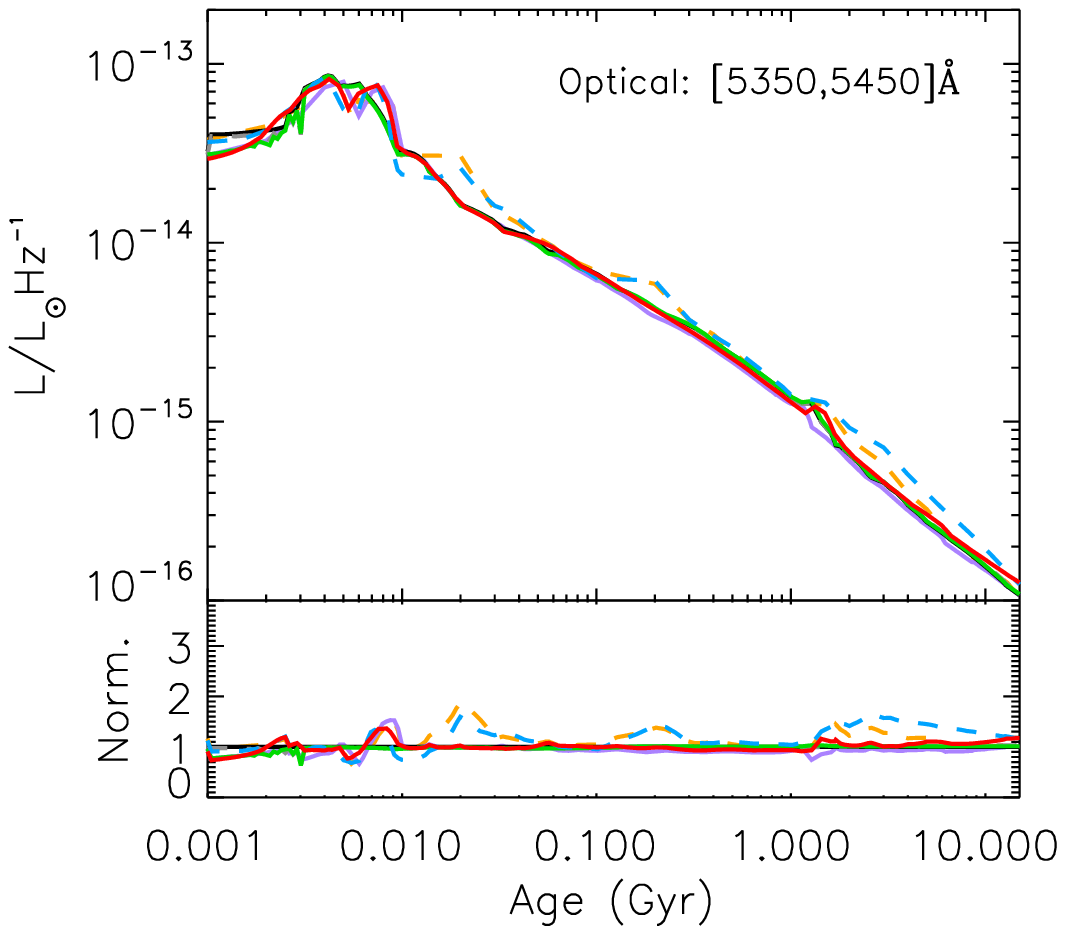}
\end{minipage}
\begin{minipage}{5.8cm}
\includegraphics[width=6.1cm]{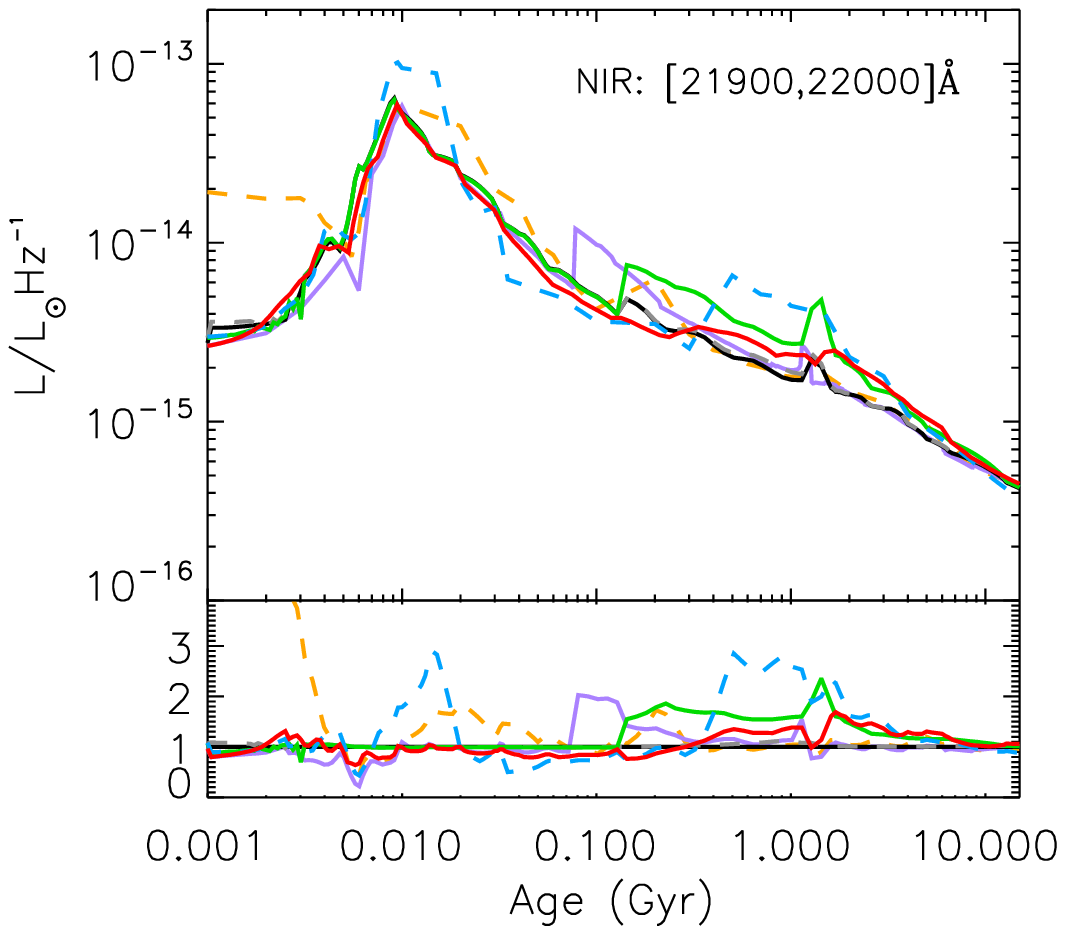}
\end{minipage}   

\caption{An illustration of the evolution with age of different single stellar population luminosity obtained for a Salpeter IMF
   with solar metallicity, normalized to 1 M$_{\sun}$, using the BC99,
  BC03, MN05, CB07, CW09, PD01 and P2 SPS models, as indicated in the
  legend. The luminosities are shown as a function of  age convolved
    with a top-hat filter covering a fixed range in wavelength:
   [1450,1550]\AA\ (left), [5350,5450]\AA\ (centre) and [21900,22000]\AA\
   (right). Note that the scale of the y-axis changes from the UV
   panel to the other two. The lower panels show the single stellar
   population luminosity for each SPS model, divided by that from the BC99
   model on a linear scale. 
}
\label{fig:ssp}
\end{figure*}

In order to gain some insight into the differences between the
SPS models summarised in Table \ref{tab:sps}, in this section we compare the
luminosities of simple stellar populations (SSPs). These are groups of coeval
stars with a fixed metallicity born in a
single instantaneous or short burst, with their luminosities weighted by the IMF.

To produce SSP spectra, the SPS models can be
run in most cases with a small number of different IMFs. However, the IMF is not a free parameter for
most of them. We have run the SPS models with a Salpeter IMF since it is one of
the few IMFs available in all cases.

Fig. \ref{fig:sspage} shows the SSPs computed with the
SPS models listed in Table \ref{tab:sps} for a
\citeauthor{salpeter} IMF and solar metallicity. Each panel shows a
fixed age. The SSPs plotted in Fig. \ref{fig:sspage} show similar global trends. At very early ages, the SSP
luminosity is dominated by the strong UV emission from hot, massive
stars. The emission at $\lambda<1000$ \AA\ declines with time, becoming
almost nonexistent after about 10 Myr. At this point in time, the emission in the infrared range starts
to increase. After a few hundred Myr there is again feeble emission at
$\lambda \leq 1000$ \AA\ coming from UV-bright old stars, such as hot
stars on the horizontal branch (HB) or the post asymptotic giant
branch stars \citep[e.g.][]{renzini}. After 1 Gyr the NIR
light is dominated by red giant branch stars and then the luminosity
in the optical and infrared ranges declines with time.

The trends seen in Fig. \ref{fig:sspage} are found to be similar for both
subsolar and supersolar metallicities. 

Though the general trends described above are followed by SSPs from all the SPS
models, it is worth noting  the cases where they differ. Fig. \ref{fig:sspage} shows that at ages below 1 Myr, the PD01 model
produces more luminous SSPs at $\lambda >10^4$
\AA\ than the other SPS models. The PD01 model also predicts a larger fraction of ionising
photons shortwards of 912\AA\ for young simple stellar
populations. These two differences
decrease rapidly with age. The difference seen for NIR wavelengths is related to thermal continuum emission from HII regions, whose contribution is included in the PD01 model \citep{bressan02} but not in the other models.

After $\sim 100$ Myr, the SSP spectra become very
different at wavelengths $\lambda < 10^3$
\AA. At these ages, this wavelength region is mostly sensitive to the light coming from stars
in a hot post red giant branch phase, which is far less well
understood than earlier stellar phases, in particular at ages
above 5 Gyr \citep[e.g.][]{brown08,conroy1,smith12}. This region of
ages and wavelengths is the only one where the BC99 and BC03 models
appear to differ noticeably from one another and, thus, the
  predicted evolution of the UV ($\sim$1500 \AA), optical and IR luminosity is
  practically  indistinguishable for these two SPS models.

For the MN05 SPS model, Fig. \ref{fig:sspage} shows the SSPs of
red HB stars, since we are considering the case of solar
metallicity. Above 1.5 Gyr, the luminosity of the SSPs for blue HB stars from MN05 model extends bluewards, compared
with the red HB ones shown in Fig. \ref{fig:sspage}, though this is
still far from the spectra obtained with the other SPS models. Nevertheless,
we will compute luminosity function at wavelengths longer than $10^3$ \AA, and
thus these strong differences will not affect our later
discussion. It is worth noting that for younger ages than $\sim$1.5 Gyr, the SSPs derived from the MN05 SPS
model with blue HB stars are much fainter, by about an order of
magnitude in the optical range, than the SSPs for the rest of the models. The use of blue
HB stars within the MN05 SPS model is recommended for very metal poor
cases.

The left panel in Fig. \ref{fig:ssp} compares the simple stellar population (Salpeter
IMF, solar metallicity)
luminosities obtained as a function of age and convolved with a top
hat filter defined over the wavelength range 1450$\leq \lambda$(\AA)
$\leq$1550. In this figure it can be seen clearly that the UV
luminosities for the different SSPs are in good agreement up to 1 Gyr, beyond which noticable
differences arise which are likely to be related to the poorly constrained advanced phases
of stellar evolution, as mentioned before.

Both Fig. \ref{fig:sspage} and the middle panel of Fig. \ref{fig:ssp} show that the different SSPs
are in remarkably good agreement in the optical range, with
differences of less than a factor of two in luminosity at $\lambda \approx 5400$
\AA. The optical luminosity is dominated by main-sequence, subgiant
and red giant branch stars \citep[e.g.][]{bc99}, stellar evolutionary phases which are well
understood.

The right panel in Fig. \ref{fig:ssp} compares the SSP luminosities as a function of age, convolved with a top
hat filter defined between 21900$\leq \lambda$(\AA)
$\leq$22000. Around 10 Myr both the MN05 and PD01 models produce SSPs with stronger
emission in the infrared than in the other models. This difference is probably
related to the modelling of supergiant stars. The difference lasts
less than 50 Myr and thus will have a neglegible effect on the 
predicted global luminosity functions.

At ages between 0.1 and 1 Gyr, the P2, CB07 and MN05 models produce SSPs
with a noticeably higher NIR luminosity than the rest of the models, due to
stars that are in a TP-AGB phase. Stars with initial
masses from 0.8 $M_{\sun}$ to 8 $M_{\sun}$ evolve onto the asymptotic giant branch and can enter the TP-AGB
phase, though only those with $M>2 M_{\sun}$ will contribute
significantly to integrated NIR luminosities
\citep[e.g.][]{frogel90}. Stars in the TP-AGB phase can be very luminous at wavelengths above
6000 \AA, as can be seen in the right panels of 
Fig. \ref{fig:sspage}. However, observed TP-AGB stars have unkown ages,
metallicities and mass losses rates, which makes it very difficult to model them
and to constrain their characteristics
\citep[e.g.][]{mn05,conroy1}. As can also be seen in
Fig. \ref{fig:ssp}, the time and duration at which the TP-AGB stars becomes dominant depends on the SPS
model.

Between 1 and 2 Gyrs, all SPS present an increase in
NIR emission related to the evolution of low-mass stars, $M_*<2M_{\odot}$, through
the helium flash \citep[e.g.][]{bc03}. 

\section{The evolution of the luminosity function}\label{sec:lf}

In this section we present the predicted galaxy luminosity function computed using different
SPS models within {\sc galform}, at three rest-frame wavelengths: UV
($\sim$1500\AA), optical ($\sim$5400\AA) and NIR
($\sim$22000\AA). As shown in the previous section, over the ranges in which
we are interested, both the BC99 and BC03 models give  very similar
simple stellar population spectra
and thus, we will show results only for the BC99 model. The CB07
model differs from the previous two mainly in the infrared and
thus, we will limit the comparison for this model to that wavelength range.

\subsection{The rest-frame UV wavelength range}
\begin{figure}
{\epsfxsize=8.5truecm
\epsfbox[57 382 364 769]{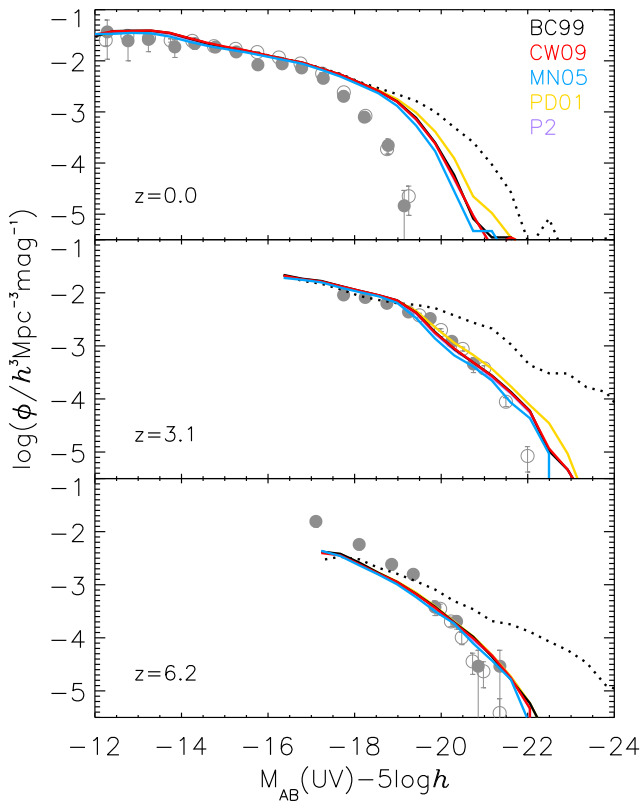}}
\caption
{The predicted rest-frame far-UV ($\sim$1500\AA) luminosity function at $z=$0 (top
  panel), $z\sim 3$ (central panel) and $z\sim 6$ (bottom panel),
  obtained with the SPS models from BC99 (black lines), CW09 (red
  lines), MN05 (blue lines), PD01 (yellow lines) and P2 (purple lines). The dotted lines show the luminosity function for the default SSP model (BC99) without
  attenuation by dust. The observational data is as follows: $z=$0 -
  \citet{galex05} (filled circles, 1500 \AA) and \citet{driver12}
  (empty circles, 1500 \AA); $z=$3.1 - \citet{sawicki06} (filled circles, 1700 \AA), and
  \citet{reddy09}
  (empty circles, 1700 \AA); $z=$6.2 - \citet{bou07} (filled circles, 1350\AA)
  and \citet{mclure09}
  (empty circles, 1500\AA). }
\label{fig:lfuv}
\end{figure}

Fig. \ref{fig:lfuv} shows the evolution of the predicted luminosity function in
the rest-frame UV (1500 \AA) for all galaxies at $z=0$, 3.1 and 6.2. The predicted luminosity functions obtained starting from different SPS models are very
similar (we note that this is also the case when making the
  comparison for a Salpeter IMF, including the BC03 and CB07 SPS models). At $z=$0, the model overpredicts the number of UV bright
galaxies. The same trend is found when increasing the resolution of the dark matter merger trees by using a Monte Carlo approach \citep{lagos14}.

At $z=$3.1 and $z=$6.2 the predicted luminosity function turns down
at luminosities fainter than those plotted in Fig. \ref{fig:lfuv}, due
to the finite halo mass resolution of the MS-W7 simulation. For brighter
magnitudes, those shown in Fig. \ref{fig:lfuv}, the match between predictions and observations is very good
at $z=$3.1, while at $z=$6.2 the model underpredicts the number of UV faint
galaxies. At these high redshifts there is a significant contribution to the bright end
of the rest-frame UV luminosity function from galaxies experiencing a starburst. 
The mismatch seen at $z=$6.2 could be alleviated if longer duration starbursts
are allowed. However, increasing the burst
duration increases the tension between the model and observations for
the bright end of the K-band luminosity function at high redshift. 

Similar trends to those shown in Fig. \ref{fig:lfuv} are expected
for colour selected Lyman Break Galaxies (LBGs), since, as shown in
\citet{drops}, the predicted far-UV luminosity function in this case 
is almost identical to the total UV luminosity function up to the corresponding observational limits.

The rest-frame UV luminosity is sensitive mainly to both the recent star formation
history and dust attenuation in a galaxy \citep[see][for a discussion of the impact that the dust attenuation
modelling has on the rest-frame UV luminosity]{drops}. However, these two aspects
depend on our modelling of the formation and evolution of galaxies
and are treated independently of the particular SPS model adopted. The
main uncertainties in the SPS models affecting the UV luminosity 
function are the modelling of the horizontal branch stars and the blue stragglers,
which contribute to the galaxy light after about 5 Gyr
\citep{conroy1}. While the characteristics and fractions of these two
types of stars can have a strong impact on the colours of optically
red galaxies \citep{brown08,smith12}, \citeauthor{conroy1} reported that changing the
fraction of stars in the blue horizontal branch has a negligible
effect on the UV luminosity function.

\subsection{The rest-frame optical wavelength range}
\begin{figure}
{\epsfxsize=8.5truecm
\epsfbox[55 394 353 849]{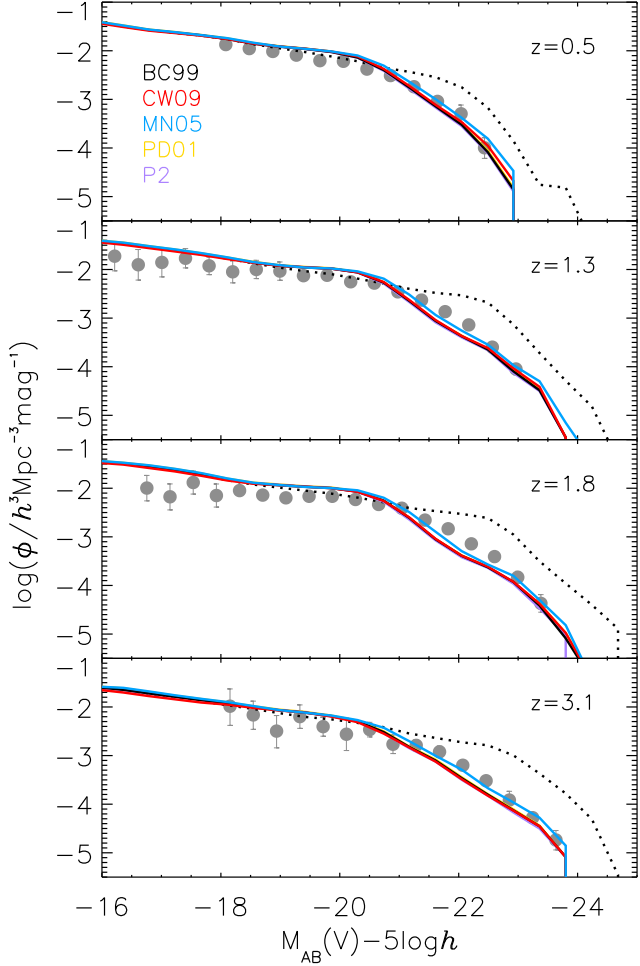}}
\caption
{From top to bottom, the predicted rest-frame V-band ($\lambda _{\rm eff}=5404$ \AA) luminosity
  function at $z=0.5$, 1.3, 1.8 and $z= 3.1$, obtained using the
  default SPS model (BC99, black lines)  and the SPS models from CW09 (red
  lines), MN05 (blue lines), PD01 (yellow lines) and P2 (purple
  lines). The dotted lines show the BC99 luminosity function without
  attenuation by dust. The filled circles show the observations from \citet{marchesini12}.
}
\label{fig:lfv}
\end{figure}

Fig. \ref{fig:lfv} shows the evolution of the luminosity function in
the rest-frame V-band. This is remarkably similar for all of the SPS
models considered, including the BC03 and CB07 SPS models. In
fact, the predicted luminosity function obtained with the PD01 and P2
SPS models are
practically indistinguisible from that with the BC99 SPS model. This
was expected since the SPS models are mainly constrained by
observations in the optical and, thus, mainly by stars on the main
sequence \citep[e.g.][]{conroy3}.

The predicted faint end normalisation increases by less than a factor 2 from $z=$3
to $z=$1.3, and then remains practically constant. The model predicts stronger
evolution at the bright end, where the break steepens at lower
redshifts since at later times there are fewer bright stars on the main
sequence. Both trends are in good agreement with the observations from \citet{marchesini12}.

\subsection{The rest-frame NIR wavelength range}\label{sec:nir}
\begin{figure}
{\epsfxsize=8.5truecm
\epsfbox[8 30 303 449]{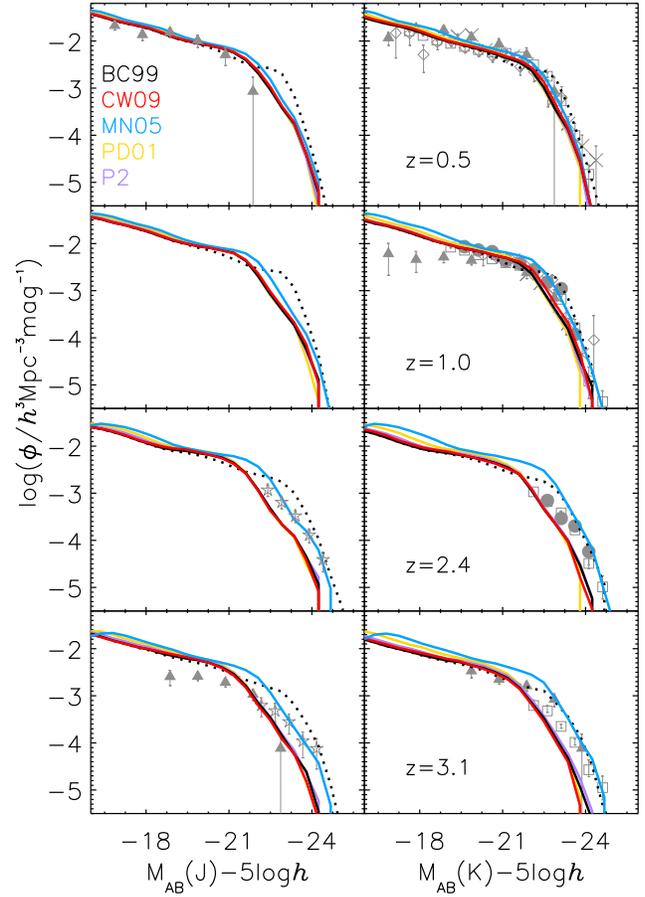}}
\caption
{The predicted rest-frame J-band ($\lambda_{\rm eff}=1.2\mu$m, left
  panels) and K-band ($\lambda_{\rm eff}=2.2\mu$m, right panels)
  luminosity function at $z=$0.5, 1, 2.4 and 3.1 (from top to bottom), compared to observational data. The lines show the predictions
  using a \citet{kennicutt_imf} IMF and the SPS models from BC99 (black lines), CW09 (red
  lines), MN05 (blue lines), PD01 (yellow lines) and P2 (purple
  lines). The dotted lines show the BC99 luminosity function without
  attenuation by dust. The symbols correspond to observational data
  from \citet{drory03} (crosses), \citet{pozzetti03} (diamonds),
  \citet{saracco06} (triangles), \citet{caputi06} (circles),
  \citet{cirasuolo10} (squares) and \citet{stefanon13} (stars). 
}
\label{fig:nirlf}
\end{figure}

The predicted rest-frame NIR luminosity function shows little evolution from $z=$2.5 to $z=$0, for all
the implemented SPS models except the MN05 and CB07 ones. At higher redshifts, $z=$3.1,
the rest-frame NIR luminosity function normalisation declines. These trends can be seen in
Fig. \ref{fig:nirlf} for the J and K bands and similar results are
found for the H-band. As for the rest-frame optical luminosity
functions, the predicted rest-frame NIR luminosity function
obtained with the PD01 and P2 SPS models are practically
indistinguishable from the default SPS one, BC99. 

At $z=1$ and for all the SPS models studied, there are too many
NIR faint galaxies (i.e. around $\sim2$ magnitudes fainter than L$_*$). \citet{lagos13} argued that this could be related
to the need for a more realistic modelling of the mass loading in SNe
outflows. Alternatively, \citet{henriques13} proposed solving this
problem by changing the reincorporation timescale of gas heated by SNe and ejected from haloes.

The predicted rest-frame NIR luminosity functions without taking into account the
attenuation by dust are also shown in Fig. \ref{fig:nirlf} for the
BC99 SPS model. Comparing these with the attenuated ones, we can see
that there is an appreciable attenuation and that it increases with
redshift, as noted by \citet{mitchell13}. For a model galaxy with a given extinction
optical depth and inclination, the attenuation is calculated in a
self-consistent way taking into account the results of a radiative
transfer model \citep{ferrara99}. In our model, the extinction optical depth
is assumed to be proportional to the mass of cold gas and inversely
proportional to the square of radial scalelength of the disc \citep{drops}. Thus, the increase
of attenuation with redshift is probably related to galaxies having
both larger
amounts of cold gas and smaller disk sizes at high redshifts.

\begin{figure}
{\epsfxsize=8.5truecm
\epsfbox[58 364 353 661]{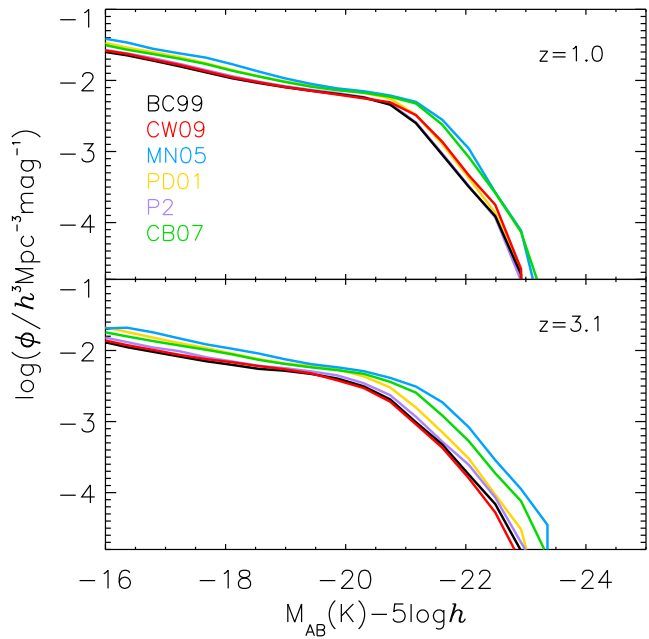}}
\caption
{The predicted rest-frame K-band luminosity function at $z=$1 (top
  panel) and $z= 3.1$ (bottom panel), using a Salpeter IMF combined
  with the default SPS model (BC99, black  lines) and the SPS models from
  CW09 (red  lines), MN05 (blue lines), PD01 (yellow lines), P2
  (purple lines) and CB07 (green lines).}
\label{fig:cb07}
\end{figure}

Fig. \ref{fig:nirlf} shows that at $z>1$, the rest-frame NIR
luminosity function using the MN05 SPS model agrees with the observed
bright end of the luminosity function, while overpredicting the number
of faint galaxies. Besides, at $z\ge 1$, the rest-frame NIR luminosity
  function predicted using this SPS model exhibits a global increase in normalisation  with respect to that at
$z=0$. This behaviour is also seen when the CB07 model is
used. In order to be able to compare the predictions obtained using the CB07 SPS model
with the others, we use the Salpeter IMF which is available
for all seven considered SPS models. Note that changing the IMF
affects the number of stars of a given mass present in a galaxy and
thus, differences in the predicted luminosity functions are expected
for the same SPS model. Fig. \ref{fig:cb07}
shows the predicted rest-frame luminosity function in the K-band (note that similar results are found in the J
and H-bands). This figure illustrates that, due to the
intrinsic uncertainties in the SPS models, the predicted luminosity
functions cannot match observations better than the spread due to the
use of different SPS models.

Recall that Fig. \ref{fig:ssp} showed how, compared with the other considered SPS
models, both MN05 and CB07
SPS models produce more luminous stars in the NIR for stellar populations with ages over 0.5
Gyr. The rest-frame K-band luminosity of stars with ages $0.3<$age(Gyr)$<2$ is
dominated by TP-AGB stars \citep{mn05}. For both the MN05 and CB07
models, this translates into both a shallower slope in the bright end and a change in the global normalisation of the luminosity function at
$z>1$ with respect to that at $z=0$. This clear change in
normalization is not seen in the other SPS models, even if they
explicitly consider the contribution of TP-AGB stars. We find
similar trends when using the Lacey et al. (in preparation) model, whose free
parameters are tuned using the MN05 SPS model.

We note that the shift in
  luminosity due to changing the BC99 SPS model by the MN05 one is larger for fainter
  model galaxies. The shallower slope of the luminosity function in
  the faint end makes this larger change far less visible than in the
  bright end.

\begin{figure}
{\epsfxsize=8.5truecm
\epsfbox[29 7 303 262]{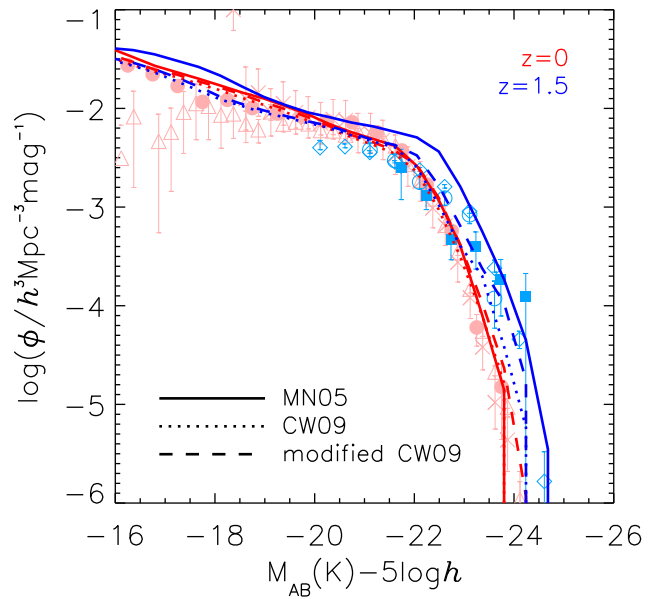}}
\caption
{The predicted rest-frame K-band luminosity function at $z=$0 (red
  lines) and $z=$1.5 (blue lines) compared to observations. The
  solid lines show the predictions using the MN05 SPS model, the
  dotted lines correspond to the predictions from the default CW09 SPS
  model and the dashed lines show the predictions when the luminosity of the
  TP-AGB stars is increased in the CW09 model. The observational
  data are as follows: $z=$0 - \citet{driver12} (filled circles), \citet{cole01} (triangles),
  \citet{kochanek01} (crosses); $z= 1.5$ - \citet{pozzetti03}
  (filled squares), \citet{caputi06} (open circles), \citet{cirasuolo10} (diamonds).
}
\label{fig:evz}
\end{figure}

In order to explore this point further we have run the CW09 SPS model varying the
properties of the TP-AGB stars. \citet{conroy1} modelled the uncertainties
associated with the TP-AGB phase by allowing a shift in both their
luminosity, $\Delta _L = \Delta {\rm log}(L_{\rm bol})$, and temperature,
$\Delta _T = \Delta {\rm log}(T_{\rm eff})$, with respect to the default tracks
(see Table \ref{tab:sps}). Their comparisson to star cluster colours and the fitting of broadband photometry of blue
$\sim L_*$ galaxies at $z\sim 0$ suggested that the allowed ranges are $-0.2 \le \Delta _L \le0.2$
and $-0.1 \le \Delta _T \le 0.1$. We have used these four extremes to
produce new simple stellar populations to be fed into the
semi-analytical model and find that
only the bright end of the predicted K-band luminosity function at $z>1$ is affected by
these changes. Fig. \ref{fig:evz} shows the K-band luminosity function at both $z=$0 and
z=1.5 for the default CW09 SPS model and that modified by increasing the
luminosity of the TP-AGB stars by $\Delta _L=0.2$. It is clear from
this plot that the CW09 SPS model modified in this way reproduces better the observed
evolution of the K-band luminosity function. However, \citet{conroy1} found that
a shift of $\Delta _L=-0.4$ was preferred in order to reproduce the
colours of star clusters with their SPS model. These different
preferred values depending on the data set used for comparison
illustrates the uncertainties derived from the difficulty modelling the TP-AGB phase.

Fig. \ref{fig:evz} also shows the evolution of the K-band luminosity function predicted
using the MN05 SPS model, which is similar to that predicted with the
CB07 SPS model, as seen in Fig. \ref{fig:cb07}. There is a clear increase in the luminosity function
normalization from $z=$0 to $z=$1.5 with these two SPS models. The other
models, including the modified CW09 SPS one, predict a much smaller
change in normalisation; for these models most of the evolution occurs at the bright end of the luminosity function. The observations appear to agree with the
latter type of evolution. 

As discussed in \S\ref{sec:ssp}, the TP-AGB stars dominate the NIR light of
stars with ages between 0.2 Gyr and 4 Gyr, depending on the
SPS model. We have explored the predicted mean ages of galaxies when the SPS
model is varied. Only small variations are found when we compare the
distribution of both stellar mass-weighted ages for galaxies with
$M_*>10^8h^{-1}M_{\odot}$ and V-luminosity weighted ages for galaxies with $M_{AB}
(V) - 5$log$h\le -18$. At  $z=$0, when the luminosity functions from
both BC99 and MN05 SPS models agree very well, galaxies with $M_{AB}
(K) - 5$log$h\le -18$ have median K-weighted ages about 25 per cent younger when using the MN05
SPS model, compared to the BC99 model. This difference increases to 40 per cent
for galaxies with $M_{AB}(K) - 5$log$h\le -22$. However, the offset is reduced at higher redshifts. At $z=$1.5, the
median K-weighted age obtained with the MN05 model is less than 20 per
cent younger than for the BC99 model. These differences are likely to
be related to the different contribution of intermediate age stars to
the integrated NIR luminosity in the different SPS models. The
difference in ages cannot completely explain the variation in the NIR
luminosity function found at $z>1$.

Constraining the TP-AGB phase using the predicted rest-frame K-band luminosity function
evolution is not possible due to the uncertainties in other
physical processes which can also change this prediction. However, it
is still
interesting to note the difference between the observed evolution and
that predicted with either the MN05 or the CB07 SPS models.
 
Due to the uncertainties in the modelling of the TP-AGB phase, our results advice against using the rest-frame K-band luminosity function at $z>1$ to tune the free parameters in models of galaxy formation and evolution. We have found that, at $z>1$, the rest-frame optical luminosity functions are robust against uncertainties in the SSPs models and thus, favored for tuning the model free parameters.

\begin{figure*}

\hspace{-0.5cm}
\begin{minipage}{5.8cm}
\includegraphics[width=6.2cm]{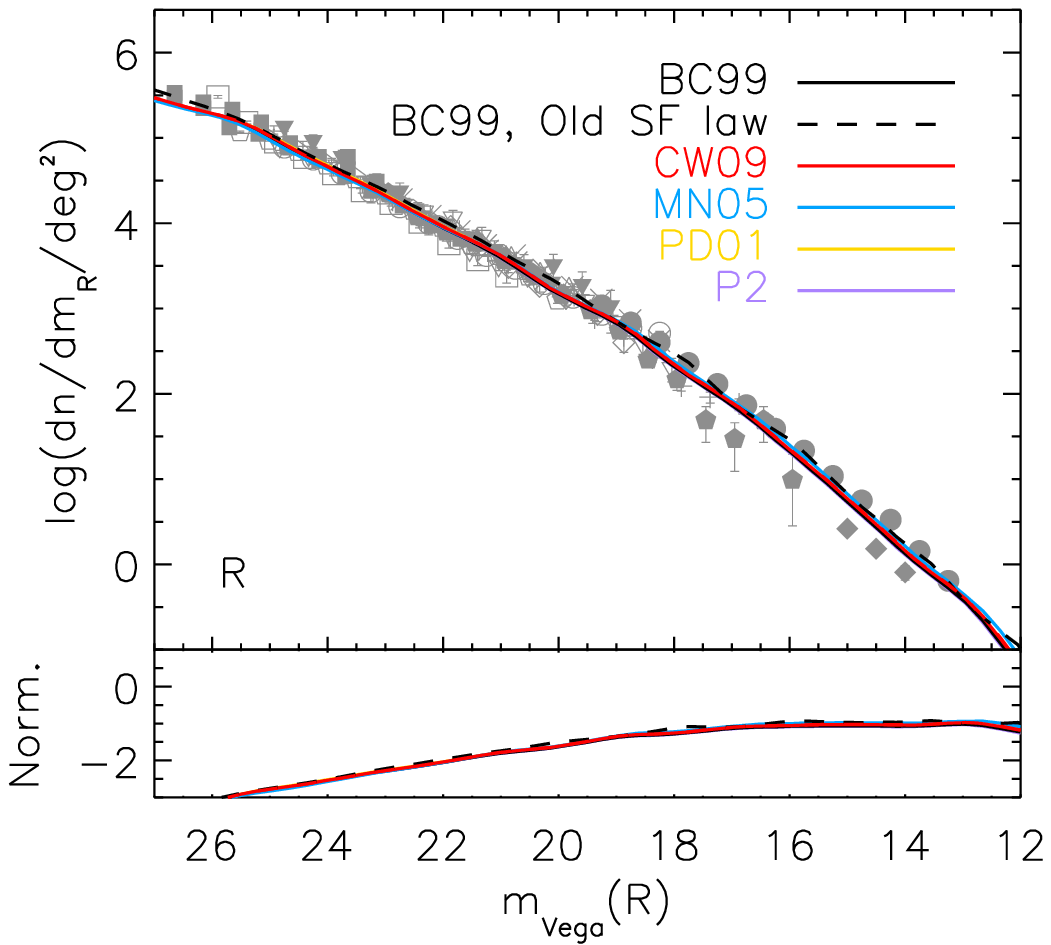}
\end{minipage}
\begin{minipage}{5.8cm}
\includegraphics[width=6.2cm]{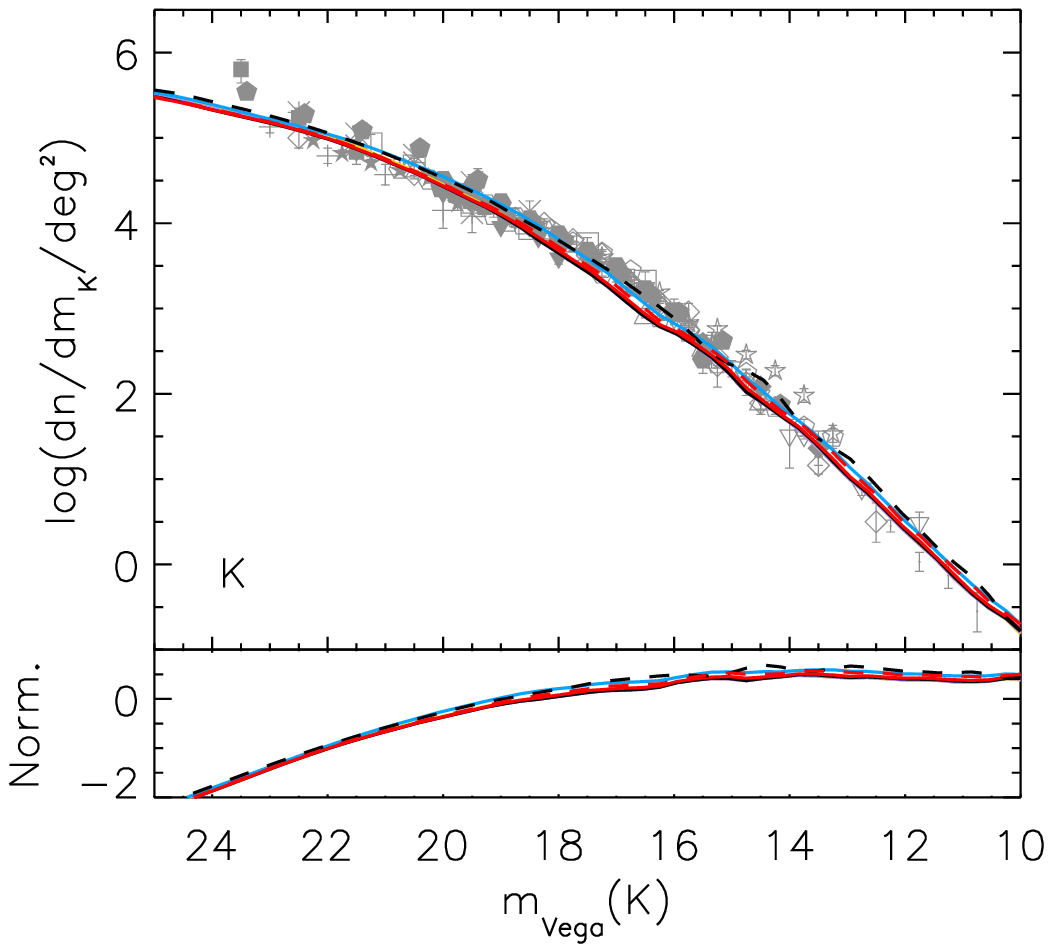}
\end{minipage}    
\begin{minipage}{5.8cm}
\includegraphics[width=6.2cm]{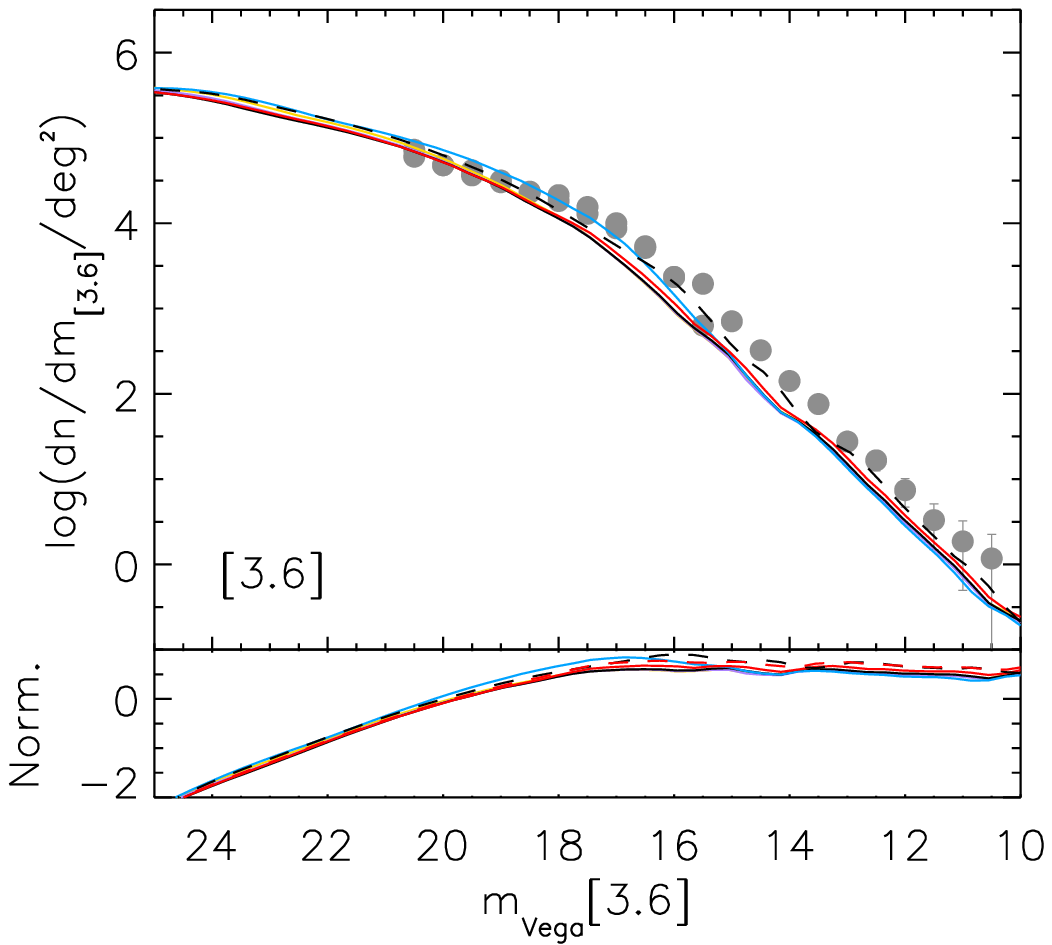}
\end{minipage}    

\caption{Main panels: Differential number counts in R-band
  ($\lambda_{\rm eff}=6490$ \AA, left), K-band  ($\lambda_{\rm eff}=2.2\mu$m, center) and [3.6]-band
  ($\lambda_{\rm eff}=3.6\mu$m, right). Bottom panels: Differential
  number counts normalized by the expected counts in a Euclidean
  universe, i.e. log$($d$n/$d$m/$deg$^2$)$-0.6(m-12.)$, with $m$ being
  the corresponding magnitude. The symbols in the
  left and central panel correspond to the compilation from Nigel
  Metcalfe (\url{http://astro.dur.ac.uk/~nm/pubhtml/counts/counts.html}).  
  The symbols in the right panel correspond to observations from the three fields
  in \citet{fazio04}, the error bars show the Poisson
  uncertainties. K-band observations: \citet{mobasher} (dots), \citet{gardner93} (diamonds),
  \citet{soifer94} (crosses), \citet{djor95} (plus signs),
  \citet{mcleod95} (squares), \citet{gardner96} (rotated triangles),
  \citet{moustakas97} (filled rotated triangles), \citet{bershady98}
  (filled squares), \citet{szokoly98} (dilled diamonds),
  \citet{mccracken} (asterisks), \citet{vaisanen00} (stars),
  \citet{huang01} (inverted triangles), \citet{kummel01} (triangles),
  \citet{maihara01} (filled triangles), \citet{saracco01} (filled
  stars), \citet{iovino05} (hexagons), \citet{metcalfe06} (filled
  inverted triangles),  \citet{imai07} (pentagons), \citet{keenan09}
  (filled pentagons) and \citet{ch09} (filled hexagons). 
  R-band observations: 
  \citet{koo86} (empty diamonds), \citet{infante86} (rotated empty
  triangle), \citet{yee87} (filled triangles), \citet{tyson88} (empty
  squares), \citet{metcalfe91} (asteriks), \citet{picard91} (empty
  pentagons), \citet{aat91} (empty triangles), \citet{steidel93}
  (rotated filled triangle), \citet{couch93} (upsided-down filled triangles),  
\citet{driver94} (filled pentagons), \citet{metcalfe95} (dots), \citet{metcalfe01} (filled squares),
\citet{huang01} (empty stars), \citet{kummel01} (crosses), 
\citet{sdss01} (filled circles),
\citet{mccracken03} (empty circles), \citet{capak04} (filled stars).
}
\label{fig:nc}
\end{figure*}
\section{Number counts}\label{sec:nc}

The number counts of galaxies depend on how the galaxy luminosity
function varies with redshift. Thus, although they are a simple quantity
 to measure, the number counts contain information about the processes
that affect the evolution of galaxies \citep[e.g.][]{guiderdoni90,barro09}. 

Fig. \ref{fig:nc} shows the predicted number counts in R, K and [3.6]-bands
compared with observations The optical and NIR number counts predicted using different SPS models are
practically indistinguishable and give a very good
match to the observations in the R and K bands (similar results are also
found for the i-band, $\lambda_{\rm eff}=7481$ \AA), though they are
underpredicting by less than a factor of 2 the bright [3.6] number counts. 

The faint end of the number counts is expected to be dominated by high redshift galaxies
while the bright end is determined by nearby ones. Thus,
the differences seen in the rest-frame
K-band luminosity function at $z>1$ will start to be appreciable at
longer wavelenths in the observer frame.

The faint [3.6]-band
number counts are expected to be dominated by the contribution of
galaxies at high redshifts, $z \sim 1$, and thus, by their rest-frame NIR light. The
right panel in Fig. \ref{fig:nc} shows that around [3.6]$_{\rm Vega}$=18, there is
a difference of a factor $\sim1.6$ between the number counts using the
MN05 model and the BC99 one. This is in agreement with the different evolution 
seen in the rest-frame K-band luminosity function at $z\ge 0.8$,
i.e. where the TP-AGB stars start to dominate the global
luminosity. 

\subsection{The number counts of EROs}
\begin{figure}
{\epsfxsize=8.5truecm
\epsfbox[53 41 504 490]{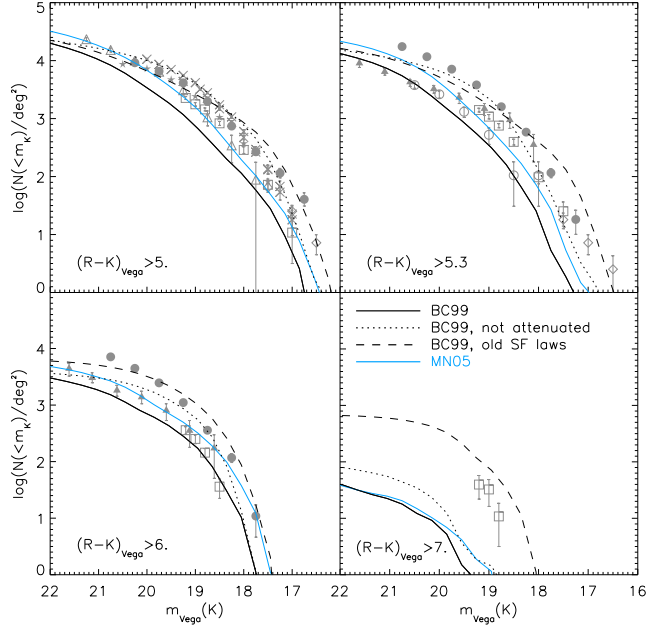}}
\caption
{The predicted cumulative number counts of EROs selected by their
  $(R-K)$ colour, as indicated in each panel. The solid black lines
  correspond to the BC99 SSP model and the blue solid lines to the
  MN05 one. The dotted lines show the BC99 SPS model counts without
  attenuation by dust. The dashed black lines show the predicted
  number counts when the BC99 SPS model is used in
  combination with the old prescription for the star formation
  law. The symbols correspond to
  observations from \citet{daddi00} (squares), \citet{smith02}
  (filled triangles), \citet{smail02}
  (empty circles), \citet{roche02} (empty triangles),
  \citet{vaisanen04} (diamonds), \citet{kong06} (crosses), the UKIDSS survey \citep[][filled
  circles]{simpson06,lawrence07} and \citet{palamara13} (stars). The errors shown are Poisson.
}
\label{fig:eros}
\end{figure}

Since the predicted number counts of K-selected galaxies are in
excellent agreement with observations, we further explore the nature of
galaxies that are bright in the NIR and that have been claimed to be
particularly sensitive to the modelling of the TP-AGB phase \citep{fontanot10,henriques11}: the Extremely Red Objects (EROs).

EROs are K-selected galaxies with very red optical to NIR colours, for example
$(R-K)_{\rm Vega}>5$. Bright EROs with $m_{\rm Vega}(K)<20$ have
inferred stellar masses above $10^{10.5}\, h^{-1}M_{\odot}$
\citep{conselice08}, they appear to be in place at $z\sim 1$ \citep{k201} and
about half of them are dominated by an old stellar component \citep{vaisanen04}. Such
characteristics posed a problem to hierarchical models of galaxy
formation and evolution for about a decade \citep{smith02}. In
hierarchical models, more massive haloes tend to form more
recently. However, the formation of galaxy mass does not follow that of the host dark matter haloes, since processes
affecting the star formation, gas cooling and galaxy merging can also
have an impact on the mass assembly of galaxies. 

In \citet{eros1} and \citet{xieros}, we showed that the predicted numbers and characteristics of
the EROs depend strongly on the feedback processes which suppress the
formation of massive galaxies, in particular the AGN feedback, and on the
star formation timescale. We presented a model, \citet{bower06} with
the BC99 SPS, that matched the
observed ERO numbers. Both \citet{fontanot10} and \citet{henriques11}
found that a galaxy formation model coupled with the MN05 SPS model
could also reproduce the observed number counts of EROs.

Fig. \ref{fig:eros} shows the predicted cumulative K-band number counts
for EROs selected by their $(R-K)$ colour, compared with different
observations. The ERO number counts predicted using the BC99, PD01, P2
or CW09 SPS models are very similar (for clarity we have included
in the figure only the default case, the BC99 SPS model). The
semi-analytical model coupled with all of these SPS models underpredicts the
observed number counts by a factor of $\sim 2.5$ for EROs with
$(R-K)_{\rm Vega}>5$ and by more than an order of magnitude for EROs with
$(R-K)_{\rm Vega}>7$, though these differences can be alleviated by
modifying the colour cut bluewards by 0.2 and 0.5 magnitudes respectively. Note
that a large fraction of the dispersion found among observations is likely
related to the sensitivity of the colour cut to the choice of filters and to aperture effects when estimating colours \citep{simpson06}. For the intermediate colour cuts, these models match some
of the observational data sets, though not those covering the largest
observed areas \citep{lawrence07,palamara13}. 

The mismatch between the default predictions and observations is larger
than expected, considering that the AGN feedback in the model
presented in this work is very similar to that in the \citet{bower06}
model. The timescale of starbursts has been increased from the
\citeauthor{bower06} model to the one presented here. Nevertheless,
this change has a minimal impact on the EROs number counts.

One of the main changes introduced in the model by
\citet{lagos11} is an improved treatment of star
formation in disks. In our work we make use of this calculation of the star
formation rate in disks, in which the star formation rate surface density is assumed to be
proportional to the molecular hydrogen surface density in galaxies. When we run the model using the old implementation for calculating the
star formation rate in disks, we obtain number counts that adequately reproduce 
the observational data and even overpredict the number of EROs with $(R-K)_{\rm Vega}>7$. These results are closer to those reported in
\citeauthor{eros1} The K-band luminosity function predicted at $z=0$
with the old star formation law in disks although consistent with
observation, it has a break at fainter luminosities than observed.

As in \citet{eros1}, we find that using the old star formation law most of the predicted EROs are passively evolving and had their last burst of star formation more
than 1 Gyr ago for both star formation implementations (the exact split varies with mangitude). However,
with the new star formation implementation the fraction of EROs that have experienced
a recent burst of star formation increases. 

The comparison between the number counts calculated with the two
implementations of the star formation rate law shows that, besides the AGN feedback, the ERO counts
depend on the way star formation is modelled, since this has a strong
impact on the numbers of galaxies that experience significant bursts of star
formation at different cosmic times. A basic reason for the
difference might be the average slower depletion of gas by quiescent star formation
in the new star formation prescription compared with the old prescription. It is important to note that there are compelling observations from $z=$0 to $z\sim 3$ supporting the new star formation implementation \citep{bigiel11,tacconi13}. Moreover, some observations indicate that massive and passively evolving galaxies (as most of the predicted EROs are) could have gas depletion timescales even longer than the 2 Gyr assumed in our model \citep{saintonge11,martig13} and, thus, even further away from the old one.

When using the MN05 SPS model, the match between the predicted and
observed ERO number counts is somewhat improved, with respect to the other SPS models, except for the most
extreme case of EROs with $(R-K)_{\rm Vega}>7$. This is in agreement with
previous results from \citet{fontanot10} and
\citet{henriques11}. We have also explored the
prediction obtained when using a modified version of the CW09 SPS model similar
to that described in \S\ref{sec:nir}, for which we have increased the
luminosity of the TP-AGB stars. This change has a visible impact increasing the predicted
cumulative number counts of EROs. However, this increase is still too
small to match the predictions using the MN05 SPS model, in agreement
with the different evolution of the K-band luminosity function seen in Fig. \ref{fig:evz} using the MN05 and modified CW09 SPS models.

The predicted number counts without including the attenuation by
interstellar dust agree remarkably well with observations, except for
EROs  with $(R-K)_{\rm Vega}>7$. Thus, uncertainties in the treatment of dust
attenuation might also be partially responsible for the mismatch
between predictions and observations.

\section{Conclusions}\label{sec:conclusions}

We have presented a new hierarchical model of galaxy formation and evolution based
on the MS-W7 simulation (Lacey et al. in preparation). This run has the same specifications as the Millennium Simulation but uses a WMAP7 cosmology \citep{wmap7}, which is close to Planck cosmology
 for the scales relevant to galaxy evolution \citep{planck}. This new model
is a development of the {\sc galform} semi-analytical model, which
includes physical treatments of the hierarchical assembly of dark
matter haloes, shock-heating and cooling of gas, star formation,
feedback from SNe, AGN and photoionization of the IGM, galaxy mergers and chemical enrichment. The luminosities of
galaxies are calculated from a stellar population synthesis (SPS)
model and dust attenuation is then included using a self-consistent
theoretical model based on the results of a radiative transfer calculation. 

The model presented here is an extension of the \citet{lagos11} model
which introduced a  new treatment of star formation which follows the
atomic and molecular hydrogen in the ISM. The free parameters in the
model have been chosen in order to reproduce the rest-frame
luminosity functions in b$_J$ and K-bands at $z=$0 with the WMAP7
cosmology, and to predict a reasonable
evolution of the luminosity function in both the rest-frame UV and K-bands.

The variation in the model predictions arising from the choice of SPS
model gives an indication of how accurately different properties can be predicted. Thus, with this new model we have explored how sensitive the predicted
luminosity function evolution from the rest-frame UV to the NIR wavelength range is
to the particular choice of SPS
model. In order to do this, we have run our galaxy formation and
evolution model coupled with seven different SPS models: PD01
\citep{grasil}, P2 \citep{pegase,pegase2}, BC99
\citep[an updated version of][]{bc99}, BC03 \citep{bc03},
  MN05 \citep{mn05}, CB07 \citep{cb07} and
  CW09 \citep{conroy1,conroy2,conroy3}. Our default model uses the
  BC99 SPS model and a
\citeauthor{kennicutt_imf} IMF. All of the studied SPS models cover a large
range in wavelength, at least from the UV to the NIR, and most are publicly
available. In the wavelength range of interest here, and for stellar ages above 10 Myr, the main differences between
the spectra of simple stellar populations obtained with different SPS models arise from stars in post-main sequence phases, in
particular, those in a thermally pulsating asymptotic
giant branch (TP-AGB) phase. This is a phase which is particularly difficult to
model due to its variability, including uncertain mass losses by
winds. There are observations both supporting and disfavouring the
modelling of a strong TP-AGB
phase \citep[e.g.][]{lyu12,zibetti13}. From the SPS models explored, the
MN05 and CB07 ones include a strong TP-AGB phase and the CW09 one
allows changes in the intensity of this phase. 

We found that the predicted rest-frame UV luminosity function is insensitive
to the choice of SPS model. Compared with observations, we
have found that our model overpredicts the number of bright rest-frame UV
galaxies at $z=0$. At higher
redshifts, $z\le 6$, the model reproduces the observed rest-frame UV luminosity
function reasonably well.

The predicted rest-frame optical luminosity function is also found to be insensitive to the
choice of SPS model. In this case, the model agrees remarkably well
with observations up to $z\sim 3$.

We have found that the evolution of the rest-frame NIR luminosity function
strongly depends on the particular treatment of the TP-AGB phase done in the SPS models. When we couple the galaxy formation and
evolution model with the two SPS models which include the strongest
contribution from TP-AGB stars, the MN05 and CB07 models, the
predicted luminosity function strongly evolves from $z=$0 to $z=$1.5, with a
change in the magnitude corresponding to $L_*$ galaxies. For the other SPS models,
even when we artificially increase the rest-frame luminosity of TP-AGB stars in
the CW09 model, we only find a change at the bright end of the
rest-frame NIR
luminosity function. Similar trends are found when using the Lacey et
al. (in preparation) model, whose free
parameters are tuned using the MN05 SPS model. The observations of the
luminosity function evolution are in better agreement with an
evolution of the rest-frame NIR luminosity function mainly affecting
the its bright end. When
using the MN05 model, we predict galaxies at
z=1.5 to have slightly smaller K-weighted ages. The difference in K-weighted ages increases towards lower redshifts,
being 20 per cent at $z=$0 for galaxies brighter than $M_K-5$log$h<-18$
(the percentage increases for brighter cuts). At $z=0$, despite this
difference in K-weighted ages, the NIR luminosity functions predicted with the
different SPS models agree. Thus, the difference in ages appears to be
related with the specific constribution of intermediate age stars to
the global NIR luminosity when using either the BC99 or the MN05 SPS
models, rather than the origin of the difference of the predicted rest-frame NIR luminosity
function. Our results suggest that, once other set of observations are also used to
constrain the model, the predicted evolution of the rest-frame NIR
luminosity function might help in constraining the strength of the
TP-AGB phase.

The number counts of galaxies depends on the evolution of the observed
luminosity function. Thus, as a further exploration we have compared
the predicted number counts from the optical to the NIR wavelength
range. We find very good agreement between observations and
predictions. Only at wavelengths as long as 3.6$\mu$m we start to see
the impact of TP-AGB stars on the global galactic number counts.

Galaxies selected by their red optical to NIR colours are particularly
sensitive to the implementation of the TP-AGB in the SPS
models. However, we find that the number counts of galaxies with
$(R-K)>5$ are more sensitive to the treatment of the star formation in
disks than to the strength of the TP-AGB phase.

In conclusion, we find that we can use the predicted rest-frame UV to
optical luminosity functions without being hindered by uncertainties
in the particular choice of SPS model done within the galaxy formation and
evolution model. The rest-frame NIR luminosity function is more problematic, since it depends on the particular treatment done in the SPS model to account for TP-AGB stars.

\subsection*{ACKNOWLEDGEMENTS}

The authors would like to thank Andrew Benson, Richard Bower, Shaun
Cole and Carlos Frenk for their help developing {\sc galform};
Stephane Charlot, Charlie Conroy, Michel Fioc, Claudia Maraston and
Brigitte Rocca-Volmerange for their helpful comments on the SPS
models; Olivier Ilbert and David Palamara for providing us with
tabulated observational data; Peder Norberg for making the model runs
for the Millennium Archive Data Base. This work used the DiRAC Data
Centric system at Durham University, operated by the Institute for
Computational Cosmology on behalf of the STFC DiRAC HPC Facility
(www.dirac.ac.uk). This equipment was funded by BIS National
E-infrastructure capital grant ST/K00042X/1, STFC capital grant
ST/H008519/1, and STFC DiRAC Operations grant ST/K003267/1 and Durham
University. DiRAC is part of the National E-Infrastructure. We
acknowledge support from the Durham STFC rolling grant in theoretical cosmology. VGP acknowledges financial
support from the Agence Nationale de la Recherche OMEGA grant ANR-11-JS56-003-01, past support from the UK Space
Agency and logistic support from St Andrews University. DJRC
acknowledges support from the Royal Astronomical Society Grant for a
summer studentship. VGP and DJRC thank Alastair Edge for his help getting this studentship.


\bibliographystyle{mn2e}
\bibliography{biblio}

\begin{thebibliography}{204}
\expandafter\ifx\csname natexlab\endcsname\relax\def\natexlab#1{#1}\fi

\bibitem[{{Almeida} {et~al}\mbox{.}(2007){Almeida}, {Baugh}, \&
  {Lacey}}]{almeida07}
{Almeida} C., {Baugh} C.~M., {Lacey} C.~G., 2007, \mnras, 376, 1711

\bibitem[{{Andreon}(2013)}]{andreon13}
{Andreon} S., 2013, \aap, 554, A79

\bibitem[{{Baldry} {et~al}\mbox{.}(2012){Baldry}, {Driver}, {Loveday},
  {Taylor}, {Kelvin}, {Liske}, {Norberg}, {Robotham}, {Brough}, {Hopkins},
  {Bamford}, {Peacock}, {Bland-Hawthorn}, {Conselice}, {Croom}, {Jones},
  {Parkinson}, {Popescu}, {Prescott}, {Sharp}, \& {Tuffs}}]{baldry12}
{Baldry} I.~K. {et~al.}, 2012, \mnras, 421, 621

\bibitem[{{Baldry} \& {Glazebrook}(2003)}]{BG}
{Baldry} I.~K., {Glazebrook} K., 2003, \apj, 593, 258

\bibitem[{{Barro} {et~al}\mbox{.}(2009){Barro}, {Gallego},
  {P{\'e}rez-Gonz{\'a}lez}, {Eliche-Moral}, {Balcells}, {Villar}, {Cardiel},
  {Cristobal-Hornillos}, {Gil de Paz}, {Guzm{\'a}n}, {Pell{\'o}}, {Prieto}, \&
  {Zamorano}}]{barro09}
{Barro} G. {et~al.}, 2009, \aap, 494, 63

\bibitem[{{Baugh}(2006)}]{baugh06}
{Baugh} C.~M., 2006, Reports of Progress in Physics, 69, 3101

\bibitem[{{Baugh} {et~al}\mbox{.}(2005){Baugh}, {Lacey}, {Frenk}, {Granato},
  {Silva}, {Bressan}, {Benson}, \& {Cole}}]{baugh05}
{Baugh} C.~M., {Lacey} C.~G., {Frenk} C.~S., {Granato} G.~L., {Silva} L.,
  {Bressan} A., {Benson} A.~J., {Cole} S., 2005, \mnras, 356, 1191

\bibitem[{{Benson}(2010)}]{benson10}
{Benson} A.~J., 2010, \physrep, 495, 33

\bibitem[{{Benson} \& {Bower}(2010)}]{bb10}
{Benson} A.~J., {Bower} R., 2010, \mnras, 405, 1573

\bibitem[{{Benson} {et~al}\mbox{.}(2003){Benson}, {Bower}, {Frenk}, {Lacey},
  {Baugh}, \& {Cole}}]{benson03}
{Benson} A.~J., {Bower} R.~G., {Frenk} C.~S., {Lacey} C.~G., {Baugh} C.~M.,
  {Cole} S., 2003, \apj, 599, 38

\bibitem[{{Benson} {et~al}\mbox{.}(2002){Benson}, {Lacey}, {Baugh}, {Cole}, \&
  {Frenk}}]{benson02}
{Benson} A.~J., {Lacey} C.~G., {Baugh} C.~M., {Cole} S., {Frenk} C.~S., 2002,
  \mnras, 333, 156

\bibitem[{{Bershady} {et~al}\mbox{.}(1998){Bershady}, {Lowenthal}, \&
  {Koo}}]{bershady98}
{Bershady} M.~A., {Lowenthal} J.~D., {Koo} D.~C., 1998, \apj, 505, 50

\bibitem[{{Bertelli} {et~al}\mbox{.}(1994){Bertelli}, {Bressan}, {Chiosi},
  {Fagotto}, \& {Nasi}}]{padova94}
{Bertelli} G., {Bressan} A., {Chiosi} C., {Fagotto} F., {Nasi} E., 1994, \aaps,
  106, 275

\bibitem[{{Bigiel} {et~al}\mbox{.}(2008){Bigiel}, {Leroy}, {Walter}, {Brinks},
  {de Blok}, {Madore}, \& {Thornley}}]{bigiel08}
{Bigiel} F., {Leroy} A., {Walter} F., {Brinks} E., {de Blok} W.~J.~G., {Madore}
  B., {Thornley} M.~D., 2008, \aj, 136, 2846

\bibitem[{{Bigiel} {et~al}\mbox{.}(2011){Bigiel}, {Leroy}, {Walter}, {Brinks},
  {de Blok}, {Kramer}, {Rix}, {Schruba}, {Schuster}, {Usero}, \&
  {Wiesemeyer}}]{bigiel11}
{Bigiel} F. {et~al.}, 2011, \apjl, 730, L13

\bibitem[{{Blanton} {et~al}\mbox{.}(2005){Blanton}, {Schlegel}, {Strauss},
  {Brinkmann}, {Finkbeiner}, {Fukugita}, {Gunn}, {Hogg}, {Ivezi{\'c}}, {Knapp},
  {Lupton}, {Munn}, {Schneider}, {Tegmark}, \& {Zehavi}}]{blanton05}
{Blanton} M.~R. {et~al.}, 2005, \aj, 129, 2562

\bibitem[{{Blitz} \& {Rosolowsky}(2006)}]{blitz06}
{Blitz} L., {Rosolowsky} E., 2006, \apj, 650, 933

\bibitem[{{Bouwens} {et~al}\mbox{.}(2007){Bouwens}, {Illingworth}, {Franx}, \&
  {Ford}}]{bou07}
{Bouwens} R.~J., {Illingworth} G.~D., {Franx} M., {Ford} H., 2007, \apj, 670,
  928

\bibitem[{{Bower} {et~al}\mbox{.}(2012){Bower}, {Benson}, \& {Crain}}]{bower12}
{Bower} R.~G., {Benson} A.~J., {Crain} R.~A., 2012, \mnras, 422, 2816

\bibitem[{{Bower} {et~al}\mbox{.}(2006){Bower}, {Benson}, {Malbon}, {Helly},
  {Frenk}, {Baugh}, {Cole}, \& {Lacey}}]{bower06}
{Bower} R.~G., {Benson} A.~J., {Malbon} R., {Helly} J.~C., {Frenk} C.~S.,
  {Baugh} C.~M., {Cole} S., {Lacey} C.~G., 2006, \mnras, 370, 645

\bibitem[{{Bressan} {et~al}\mbox{.}(1998){Bressan}, {Granato}, \&
  {Silva}}]{bressan98}
{Bressan} A., {Granato} G.~L., {Silva} L., 1998, \aap, 332, 135

\bibitem[{{Bressan} {et~al}\mbox{.}(2002){Bressan}, {Silva}, \&
  {Granato}}]{bressan02}
{Bressan} A., {Silva} L., {Granato} G.~L., 2002, \aap, 392, 377

\bibitem[{{Brown} {et~al}\mbox{.}(2008){Brown}, {Smith}, {Ferguson},
  {Sweigart}, {Kimble}, \& {Bowers}}]{brown08}
{Brown} T.~M., {Smith} E., {Ferguson} H.~C., {Sweigart} A.~V., {Kimble} R.~A.,
  {Bowers} C.~W., 2008, \apj, 682, 319

\bibitem[{{Bruzual}(2007)}]{cb07}
{Bruzual} A.~G., 2007, in IAU Symposium, Vol. 241, IAU Symposium, {Vazdekis}
  A., {Peletier} R., eds., pp. 125--132

\bibitem[{{Bruzual} \& {Charlot}(1993)}]{bc99}
{Bruzual} G., {Charlot} S., 1993, \apj, 405, 538

\bibitem[{{Bruzual} \& {Charlot}(2003)}]{bc03}
{Bruzual} G., {Charlot} S., 2003, \mnras, 344, 1000

\bibitem[{{Burgarella} {et~al}\mbox{.}(2013){Burgarella}, {Buat}, {Gruppioni},
  {Cucciati}, {Heinis}, {Berta}, {B{\'e}thermin}, {Bock}, {Cooray}, {Dunlop},
  {Farrah}, {Franceschini}, {Le Floc'h}, {Lutz}, {Magnelli}, {Nordon},
  {Oliver}, {Page}, {Popesso}, {Pozzi}, {Riguccini}, {Vaccari}, \&
  {Viero}}]{burgarella13}
{Burgarella} D. {et~al.}, 2013, \aap, 554, A70

\bibitem[{{Capak} {et~al}\mbox{.}(2004){Capak}, {Cowie}, {Hu}, {Barger},
  {Dickinson}, {Fernandez}, {Giavalisco}, {Komiyama}, {Kretchmer}, {McNally},
  {Miyazaki}, {Okamura}, \& {Stern}}]{capak04}
{Capak} P. {et~al.}, 2004, \aj, 127, 180

\bibitem[{{Cappellari} {et~al}\mbox{.}(2012){Cappellari}, {McDermid},
  {Alatalo}, {Blitz}, {Bois}, {Bournaud}, {Bureau}, {Crocker}, {Davies},
  {Davis}, {de Zeeuw}, {Duc}, {Emsellem}, {Khochfar}, {Krajnovi{\'c}},
  {Kuntschner}, {Lablanche}, {Morganti}, {Naab}, {Oosterloo}, {Sarzi}, {Scott},
  {Serra}, {Weijmans}, \& {Young}}]{cappellari12}
{Cappellari} M. {et~al.}, 2012, \nat, 484, 485

\bibitem[{{Caputi} {et~al}\mbox{.}(2006){Caputi}, {McLure}, {Dunlop},
  {Cirasuolo}, \& {Schael}}]{caputi06}
{Caputi} K.~I., {McLure} R.~J., {Dunlop} J.~S., {Cirasuolo} M., {Schael} A.~M.,
  2006, \mnras, 366, 609

\bibitem[{{Cassisi} {et~al}\mbox{.}(1997{\natexlab{a}}){Cassisi}, {Castellani},
  \& {Castellani}}]{cassini97}
{Cassisi} S., {Castellani} M., {Castellani} V., 1997{\natexlab{a}}, \aap, 317,
  108

\bibitem[{{Cassisi} {et~al}\mbox{.}(2000){Cassisi}, {Castellani},
  {Ciarcelluti}, {Piotto}, \& {Zoccali}}]{cassini00}
{Cassisi} S., {Castellani} V., {Ciarcelluti} P., {Piotto} G., {Zoccali} M.,
  2000, \mnras, 315, 679

\bibitem[{{Cassisi} {et~al}\mbox{.}(1997{\natexlab{b}}){Cassisi},
  {degl'Innocenti}, \& {Salaris}}]{cassini97b}
{Cassisi} S., {degl'Innocenti} S., {Salaris} M., 1997{\natexlab{b}}, \mnras,
  290, 515

\bibitem[{{Chabrier}(2003)}]{chabrier03}
{Chabrier} G., 2003, \apjl, 586, L133

\bibitem[{{Charlot} {et~al}\mbox{.}(1996){Charlot}, {Worthey}, \&
  {Bressan}}]{charlot96}
{Charlot} S., {Worthey} G., {Bressan} A., 1996, \apj, 457, 625

\bibitem[{{Chen} {et~al}\mbox{.}(2010){Chen}, {Liang}, {Hammer}, {Prugniel},
  {Zhong}, {Rodrigues}, {Zhao}, \& {Flores}}]{chen10}
{Chen} X.~Y., {Liang} Y.~C., {Hammer} F., {Prugniel} P., {Zhong} G.~H.,
  {Rodrigues} M., {Zhao} Y.~H., {Flores} H., 2010, \aap, 515, A101

\bibitem[{{Cimatti} {et~al}\mbox{.}(2002){Cimatti}, {Daddi}, {Mignoli},
  {Pozzetti}, {Renzini}, {Zamorani}, {Broadhurst}, {Fontana}, {Saracco},
  {Poli}, {Cristiani}, {D'Odorico}, {Giallongo}, {Gilmozzi}, \& {Menci}}]{k201}
{Cimatti} A. {et~al.}, 2002, \aap, 381, L68

\bibitem[{{Cirasuolo} {et~al}\mbox{.}(2010){Cirasuolo}, {McLure}, {Dunlop},
  {Almaini}, {Foucaud}, \& {Simpson}}]{cirasuolo10}
{Cirasuolo} M., {McLure} R.~J., {Dunlop} J.~S., {Almaini} O., {Foucaud} S.,
  {Simpson} C., 2010, \mnras, 401, 1166

\bibitem[{{Clements} {et~al}\mbox{.}(2010){Clements}, {Rigby}, {Maddox},
  {Dunne}, {Mortier}, {Pearson}, {Amblard}, {Auld}, {Baes}, {Bonfield},
  {Burgarella}, {Buttiglione}, {Cava}, {Cooray}, {Dariush}, {de Zotti}, {Dye},
  {Eales}, {Frayer}, {Fritz}, {Gardner}, {Gonzalez-Nuevo}, {Herranz}, {Ibar},
  {Ivison}, {Jarvis}, {Lagache}, {Leeuw}, {Lopez-Caniego}, {Negrello},
  {Pascale}, {Pohlen}, {Rodighiero}, {Samui}, {Serjeant}, {Sibthorpe}, {Scott},
  {Smith}, {Temi}, {Thompson}, {Valtchanov}, {van der Werf}, \&
  {Verma}}]{hatlas}
{Clements} D.~L. {et~al.}, 2010, \aap, 518, L8

\bibitem[{{Cole} {et~al}\mbox{.}(1994){Cole}, {Aragon-Salamanca}, {Frenk},
  {Navarro}, \& {Zepf}}]{cole94}
{Cole} S., {Aragon-Salamanca} A., {Frenk} C.~S., {Navarro} J.~F., {Zepf} S.~E.,
  1994, \mnras, 271, 781

\bibitem[{{Cole} {et~al}\mbox{.}(2000){Cole}, {Lacey}, {Baugh}, \&
  {Frenk}}]{cole00}
{Cole} S., {Lacey} C.~G., {Baugh} C.~M., {Frenk} C.~S., 2000, \mnras, 319, 168

\bibitem[{{Cole} {et~al}\mbox{.}(2001){Cole}, {Norberg}, {Baugh}, {Frenk},
  {Bland-Hawthorn}, {Bridges}, {Cannon}, {Colless}, {Collins}, {Couch},
  {Cross}, {Dalton}, {De Propris}, {Driver}, {Efstathiou}, {Ellis},
  {Glazebrook}, {Jackson}, {Lahav}, {Lewis}, {Lumsden}, {Maddox}, {Madgwick},
  {Peacock}, {Peterson}, {Sutherland}, \& {Taylor}}]{cole01}
{Cole} S. {et~al.}, 2001, \mnras, 326, 255

\bibitem[{{Conroy} \& {Gunn}(2010)}]{conroy3}
{Conroy} C., {Gunn} J.~E., 2010, \apj, 712, 833

\bibitem[{{Conroy} {et~al}\mbox{.}(2009){Conroy}, {Gunn}, \& {White}}]{conroy1}
{Conroy} C., {Gunn} J.~E., {White} M., 2009, \apj, 699, 486

\bibitem[{{Conroy} {et~al}\mbox{.}(2010){Conroy}, {White}, \& {Gunn}}]{conroy2}
{Conroy} C., {White} M., {Gunn} J.~E., 2010, \apj, 708, 58

\bibitem[{{Conselice} {et~al}\mbox{.}(2008){Conselice}, {Bundy}, {U},
  {Eisenhardt}, {Lotz}, \& {Newman}}]{conselice08}
{Conselice} C.~J., {Bundy} K., {U} V., {Eisenhardt} P., {Lotz} J., {Newman} J.,
  2008, \mnras, 383, 1366

\bibitem[{{Coppin} {et~al}\mbox{.}(2006){Coppin}, {Chapin}, {Mortier}, {Scott},
  {Borys}, {Dunlop}, {Halpern}, {Hughes}, {Pope}, {Scott}, {Serjeant}, {Wagg},
  {Alexander}, {Almaini}, {Aretxaga}, {Babbedge}, {Best}, {Blain}, {Chapman},
  {Clements}, {Crawford}, {Dunne}, {Eales}, {Edge}, {Farrah}, {Gazta{\~n}aga},
  {Gear}, {Granato}, {Greve}, {Fox}, {Ivison}, {Jarvis}, {Jenness}, {Lacey},
  {Lepage}, {Mann}, {Marsden}, {Martinez-Sansigre}, {Oliver}, {Page},
  {Peacock}, {Pearson}, {Percival}, {Priddey}, {Rawlings}, {Rowan-Robinson},
  {Savage}, {Seigar}, {Sekiguchi}, {Silva}, {Simpson}, {Smail}, {Stevens},
  {Takagi}, {Vaccari}, {van Kampen}, \& {Willott}}]{coppin06}
{Coppin} K. {et~al.}, 2006, \mnras, 372, 1621

\bibitem[{{Couch} {et~al}\mbox{.}(1993){Couch}, {Jurcevic}, \&
  {Boyle}}]{couch93}
{Couch} W.~J., {Jurcevic} J.~S., {Boyle} B.~J., 1993, \mnras, 260, 241

\bibitem[{{Crist{\'o}bal-Hornillos}
  {et~al}\mbox{.}(2009){Crist{\'o}bal-Hornillos}, {Aguerri}, {Moles}, {Perea},
  {Castander}, {Broadhurst}, {Alfaro}, {Ben{\'{\i}}tez}, {Cabrera-Ca{\~n}o},
  {Cepa}, {Cervi{\~n}o}, {Fern{\'a}ndez-Soto}, {Gonz{\'a}lez Delgado},
  {Husillos}, {Infante}, {M{\'a}rquez}, {Mart{\'{\i}}nez}, {Masegosa}, {del
  Olmo}, {Prada}, {Quintana}, \& {S{\'a}nchez}}]{ch09}
{Crist{\'o}bal-Hornillos} D. {et~al.}, 2009, \apj, 696, 1554

\bibitem[{{Croton} {et~al}\mbox{.}(2006){Croton}, {Springel}, {White}, {De
  Lucia}, {Frenk}, {Gao}, {Jenkins}, {Kauffmann}, {Navarro}, \&
  {Yoshida}}]{croton06}
{Croton} D.~J. {et~al.}, 2006, \mnras, 365, 11

\bibitem[{{Cucciati} {et~al}\mbox{.}(2012){Cucciati}, {Tresse}, {Ilbert}, {Le
  F{\`e}vre}, {Garilli}, {Le Brun}, {Cassata}, {Franzetti}, {Maccagni},
  {Scodeggio}, {Zucca}, {Zamorani}, {Bardelli}, {Bolzonella}, {Bielby},
  {McCracken}, {Zanichelli}, \& {Vergani}}]{cucciati12}
{Cucciati} O. {et~al.}, 2012, \aap, 539, A31

\bibitem[{{Daddi} {et~al}\mbox{.}(2000){Daddi}, {Cimatti}, \&
  {Renzini}}]{daddi00}
{Daddi} E., {Cimatti} A., {Renzini} A., 2000, \aap, 362, L45

\bibitem[{{Davis} {et~al}\mbox{.}(1985){Davis}, {Efstathiou}, {Frenk}, \&
  {White}}]{fof}
{Davis} M., {Efstathiou} G., {Frenk} C.~S., {White} S.~D.~M., 1985, \apj, 292,
  371

\bibitem[{{De Lucia} {et~al}\mbox{.}(2010){De Lucia}, {Boylan-Kolchin},
  {Benson}, {Fontanot}, \& {Monaco}}]{delucia10}
{De Lucia} G., {Boylan-Kolchin} M., {Benson} A.~J., {Fontanot} F., {Monaco} P.,
  2010, \mnras, 406, 1533

\bibitem[{{Djorgovski} {et~al}\mbox{.}(1995){Djorgovski}, {Soifer}, {Pahre},
  {Larkin}, {Smith}, {Neugebauer}, {Smail}, {Matthews}, {Hogg}, {Blandford},
  {Cohen}, {Harrison}, \& {Nelson}}]{djor95}
{Djorgovski} S. {et~al.}, 1995, \apjl, 438, L13

\bibitem[{{Driver} {et~al}\mbox{.}(1994){Driver}, {Phillipps}, {Davies},
  {Morgan}, \& {Disney}}]{driver94}
{Driver} S.~P., {Phillipps} S., {Davies} J.~I., {Morgan} I., {Disney} M.~J.,
  1994, \mnras, 268, 393

\bibitem[{{Driver} {et~al}\mbox{.}(2012){Driver}, {Robotham}, {Kelvin},
  {Alpaslan}, {Baldry}, {Bamford}, {Brough}, {Brown}, {Hopkins}, {Liske},
  {Loveday}, {Norberg}, {Peacock}, {Andrae}, {Bland-Hawthorn}, {Bourne},
  {Cameron}, {Colless}, {Conselice}, {Croom}, {Dunne}, {Frenk}, {Graham},
  {Gunawardhana}, {Hill}, {Jones}, {Kuijken}, {Madore}, {Nichol}, {Parkinson},
  {Pimbblet}, {Phillipps}, {Popescu}, {Prescott}, {Seibert}, {Sharp},
  {Sutherland}, {Taylor}, {Thomas}, {Tuffs}, {van Kampen}, {Wijesinghe}, \&
  {Wilkins}}]{driver12}
{Driver} S.~P. {et~al.}, 2012, \mnras, 427, 3244

\bibitem[{{Drory} {et~al}\mbox{.}(2003){Drory}, {Bender}, {Feulner}, {Hopp},
  {Maraston}, {Snigula}, \& {Hill}}]{drory03}
{Drory} N., {Bender} R., {Feulner} G., {Hopp} U., {Maraston} C., {Snigula} J.,
  {Hill} G.~J., 2003, \apj, 595, 698

\bibitem[{{Dutton} {et~al}\mbox{.}(2011){Dutton}, {Conroy}, {van den Bosch},
  {Simard}, {Mendel}, {Courteau}, {Dekel}, {More}, \& {Prada}}]{dutton11}
{Dutton} A.~A. {et~al.}, 2011, \mnras, 416, 322

\bibitem[{{Eke} {et~al}\mbox{.}(1996){Eke}, {Cole}, \& {Frenk}}]{Eke96}
{Eke} V.~R., {Cole} S., {Frenk} C.~S., 1996, \mnras, 282, 263

\bibitem[{{Eminian} {et~al}\mbox{.}(2008){Eminian}, {Kauffmann}, {Charlot},
  {Wild}, {Bruzual}, {Rettura}, \& {Loveday}}]{eminian08}
{Eminian} C., {Kauffmann} G., {Charlot} S., {Wild} V., {Bruzual} G., {Rettura}
  A., {Loveday} J., 2008, \mnras, 384, 930

\bibitem[{{Fanidakis} {et~al}\mbox{.}(2011){Fanidakis}, {Baugh}, {Benson},
  {Bower}, {Cole}, {Done}, \& {Frenk}}]{fanidakis11}
{Fanidakis} N., {Baugh} C.~M., {Benson} A.~J., {Bower} R.~G., {Cole} S., {Done}
  C., {Frenk} C.~S., 2011, \mnras, 410, 53

\bibitem[{{Fazio} {et~al}\mbox{.}(2004){Fazio}, {Ashby}, {Barmby}, {Hora},
  {Huang}, {Pahre}, {Wang}, {Willner}, {Arendt}, {Moseley}, {Brodwin},
  {Eisenhardt}, {Stern}, {Tollestrup}, \& {Wright}}]{fazio04}
{Fazio} G.~G. {et~al.}, 2004, \apjs, 154, 39

\bibitem[{{Ferrara} {et~al}\mbox{.}(1999){Ferrara}, {Bianchi}, {Cimatti}, \&
  {Giovanardi}}]{ferrara99}
{Ferrara} A., {Bianchi} S., {Cimatti} A., {Giovanardi} C., 1999, \apjs, 123,
  437

\bibitem[{{Fioc} \& {Rocca-Volmerange}(1997)}]{pegase}
{Fioc} M., {Rocca-Volmerange} B., 1997, A\&A, 326, 950

\bibitem[{{Fioc} \& {Rocca-Volmerange}(1999)}]{pegase2}
{Fioc} M., {Rocca-Volmerange} B., 1999, arXiv:astro-ph/9912179

\bibitem[{{Font} {et~al}\mbox{.}(2008){Font}, {Bower}, {McCarthy}, {Benson},
  {Frenk}, {Helly}, {Lacey}, {Baugh}, \& {Cole}}]{font08}
{Font} A.~S. {et~al.}, 2008, \mnras, 389, 1619

\bibitem[{{Fontanot} \& {Monaco}(2010)}]{fontanot10}
{Fontanot} F., {Monaco} P., 2010, \mnras, 405, 705

\bibitem[{{Frogel} {et~al}\mbox{.}(1990){Frogel}, {Mould}, \&
  {Blanco}}]{frogel90}
{Frogel} J.~A., {Mould} J., {Blanco} V.~M., 1990, \apj, 352, 96

\bibitem[{{Gardner} {et~al}\mbox{.}(1993){Gardner}, {Cowie}, \&
  {Wainscoat}}]{gardner93}
{Gardner} J.~P., {Cowie} L.~L., {Wainscoat} R.~J., 1993, \apjl, 415, L9

\bibitem[{{Gardner} {et~al}\mbox{.}(1996){Gardner}, {Sharples}, {Carrasco}, \&
  {Frenk}}]{gardner96}
{Gardner} J.~P., {Sharples} R.~M., {Carrasco} B.~E., {Frenk} C.~S., 1996,
  \mnras, 282, L1

\bibitem[{{Geha} {et~al}\mbox{.}(2013){Geha}, {Brown}, {Tumlinson}, {Kalirai},
  {Simon}, {Kirby}, {VandenBerg}, {Mu{\~n}oz}, {Avila}, {Guhathakurta}, \&
  {Ferguson}}]{geha13}
{Geha} M. {et~al.}, 2013, \apj, 771, 29

\bibitem[{{Gilbank} {et~al}\mbox{.}(2011){Gilbank}, {Bower}, {Glazebrook},
  {Balogh}, {Baldry}, {Davies}, {Hau}, {Li}, {McCarthy}, \&
  {Sawicki}}]{gilbank11}
{Gilbank} D.~G. {et~al.}, 2011, \mnras, 414, 304

\bibitem[{{Gonz{\'a}lez} {et~al}\mbox{.}(2009){Gonz{\'a}lez}, {Lacey}, {Baugh},
  {Frenk}, \& {Benson}}]{juan09}
{Gonz{\'a}lez} J.~E., {Lacey} C.~G., {Baugh} C.~M., {Frenk} C.~S., {Benson}
  A.~J., 2009, \mnras, 397, 1254

\bibitem[{{Gonzalez-Perez} {et~al}\mbox{.}(2009){Gonzalez-Perez}, {Baugh},
  {Lacey}, \& {Almeida}}]{eros1}
{Gonzalez-Perez} V., {Baugh} C.~M., {Lacey} C.~G., {Almeida} C., 2009, \mnras,
  398, 497

\bibitem[{{Gonzalez-Perez} {et~al}\mbox{.}(2011){Gonzalez-Perez}, {Baugh},
  {Lacey}, \& {Kim}}]{xieros}
{Gonzalez-Perez} V., {Baugh} C.~M., {Lacey} C.~G., {Kim} J.-W., 2011, \mnras,
  417, 517

\bibitem[{{Gonzalez-Perez} {et~al}\mbox{.}(2013){Gonzalez-Perez}, {Lacey},
  {Baugh}, {Frenk}, \& {Wilkins}}]{drops}
{Gonzalez-Perez} V., {Lacey} C.~G., {Baugh} C.~M., {Frenk} C.~S., {Wilkins}
  S.~M., 2013, \mnras, 429, 1609

\bibitem[{{Goudfrooij} {et~al}\mbox{.}(2006){Goudfrooij}, {Gilmore},
  {Kissler-Patig}, \& {Maraston}}]{goud06}
{Goudfrooij} P., {Gilmore} D., {Kissler-Patig} M., {Maraston} C., 2006, \mnras,
  369, 697

\bibitem[{{Guiderdoni} \& {Rocca-Volmerange}(1990)}]{guiderdoni90}
{Guiderdoni} B., {Rocca-Volmerange} B., 1990, \aap, 227, 362

\bibitem[{{Gunawardhana} {et~al}\mbox{.}(2013){Gunawardhana}, {Hopkins},
  {Bland-Hawthorn}, {Brough}, {Sharp}, {Loveday}, {Taylor}, {Jones},
  {Lara-L{\'o}pez}, {Bauer}, {Colless}, {Owers}, {Baldry},
  {L{\'o}pez-S{\'a}nchez}, {Foster}, {Bamford}, {Brown}, {Driver},
  {Drinkwater}, {Liske}, {Meyer}, {Norberg}, {Robotham}, {Ching}, {Cluver},
  {Croom}, {Kelvin}, {Prescott}, {Steele}, {Thomas}, \& {Wang}}]{gamasfrd}
{Gunawardhana} M.~L.~P. {et~al.}, 2013, \mnras, 433, 2764

\bibitem[{{Guo} {et~al}\mbox{.}(2013){Guo}, {White}, {Angulo}, {Henriques},
  {Lemson}, {Boylan-Kolchin}, {Thomas}, \& {Short}}]{qi12}
{Guo} Q., {White} S., {Angulo} R.~E., {Henriques} B., {Lemson} G.,
  {Boylan-Kolchin} M., {Thomas} P., {Short} C., 2013, \mnras, 428, 1351

\bibitem[{{Guo} {et~al}\mbox{.}(2011){Guo}, {White}, {Boylan-Kolchin}, {De
  Lucia}, {Kauffmann}, {Lemson}, {Li}, {Springel}, \& {Weinmann}}]{qi11}
{Guo} Q. {et~al.}, 2011, \mnras, 413, 101

\bibitem[{{Henriques} {et~al}\mbox{.}(2011){Henriques}, {Maraston}, {Monaco},
  {Fontanot}, {Menci}, {De Lucia}, \& {Tonini}}]{henriques11}
{Henriques} B., {Maraston} C., {Monaco} P., {Fontanot} F., {Menci} N., {De
  Lucia} G., {Tonini} C., 2011, \mnras, 415, 3571

\bibitem[{{Henriques} {et~al}\mbox{.}(2012){Henriques}, {White}, {Lemson},
  {Thomas}, {Guo}, {Marleau}, \& {Overzier}}]{henriques12}
{Henriques} B.~M.~B., {White} S.~D.~M., {Lemson} G., {Thomas} P.~A., {Guo} Q.,
  {Marleau} G.-D., {Overzier} R.~A., 2012, \mnras, 421, 2904

\bibitem[{{Henriques} {et~al}\mbox{.}(2013){Henriques}, {White}, {Thomas},
  {Angulo}, {Guo}, {Lemson}, \& {Springel}}]{henriques13}
{Henriques} B.~M.~B., {White} S.~D.~M., {Thomas} P.~A., {Angulo} R.~E., {Guo}
  Q., {Lemson} G., {Springel} V., 2013, \mnras, 431, 3373

\bibitem[{{Hopkins} \& {Beacom}(2006)}]{hopkins06}
{Hopkins} A.~M., {Beacom} J.~F., 2006, \apj, 651, 142

\bibitem[{{Huang} {et~al}\mbox{.}(2001){Huang}, {Thompson}, {K{\"u}mmel},
  {Meisenheimer}, {Wolf}, {Beckwith}, {Fockenbrock}, {Fried}, {Hippelein}, {von
  Kuhlmann}, {Phleps}, {R{\"o}ser}, \& {Thommes}}]{huang01}
{Huang} J.-S. {et~al.}, 2001, \aap, 368, 787

\bibitem[{{Ilbert} {et~al}\mbox{.}(2013){Ilbert}, {McCracken}, {Le F{\`e}vre},
  {Capak}, {Dunlop}, {Karim}, {Renzini}, {Caputi}, {Boissier}, {Arnouts},
  {Aussel}, {Comparat}, {Guo}, {Hudelot}, {Kartaltepe}, {Kneib}, {Krogager},
  {Le Floc'h}, {Lilly}, {Mellier}, {Milvang-Jensen}, {Moutard}, {Onodera},
  {Richard}, {Salvato}, {Sanders}, {Scoville}, {Silverman}, {Taniguchi},
  {Tasca}, {Thomas}, {Toft}, {Tresse}, {Vergani}, {Wolk}, \& {Zirm}}]{ilbert13}
{Ilbert} O. {et~al.}, 2013, \aap, 556, A55

\bibitem[{{Ilbert} {et~al}\mbox{.}(2010){Ilbert}, {Salvato}, {Le Floc'h},
  {Aussel}, {Capak}, {McCracken}, {Mobasher}, {Kartaltepe}, {Scoville},
  {Sanders}, {Arnouts}, {Bundy}, {Cassata}, {Kneib}, {Koekemoer}, {Le
  F{\`e}vre}, {Lilly}, {Surace}, {Taniguchi}, {Tasca}, {Thompson}, {Tresse},
  {Zamojski}, {Zamorani}, \& {Zucca}}]{ilbert10}
{Ilbert} O. {et~al.}, 2010, \apj, 709, 644

\bibitem[{{Imai} {et~al}\mbox{.}(2007){Imai}, {Matsuhara}, {Oyabu}, {Wada},
  {Takagi}, {Fujishiro}, {Hanami}, \& {Pearson}}]{imai07}
{Imai} K., {Matsuhara} H., {Oyabu} S., {Wada} T., {Takagi} T., {Fujishiro} N.,
  {Hanami} H., {Pearson} C.~P., 2007, \aj, 133, 2418

\bibitem[{{Infante} {et~al}\mbox{.}(1986){Infante}, {Pritchet}, \&
  {Quintana}}]{infante86}
{Infante} L., {Pritchet} C., {Quintana} H., 1986, \aj, 91, 217

\bibitem[{{Iovino} {et~al}\mbox{.}(2005){Iovino}, {McCracken}, {Garilli},
  {Foucaud}, {Le F{\`e}vre}, {Maccagni}, {Saracco}, {Bardelli}, {Busarello},
  {Scodeggio}, {Zanichelli}, {Paioro}, {Bottini}, {Le Brun}, {Picat},
  {Scaramella}, {Tresse}, {Vettolani}, {Adami}, {Arnaboldi}, {Arnouts},
  {Bolzonella}, {Cappi}, {Charlot}, {Ciliegi}, {Contini}, {Franzetti},
  {Gavignaud}, {Guzzo}, {Ilbert}, {Marano}, {Marinoni}, {Mazure}, {Meneux},
  {Merighi}, {Paltani}, {Pell{\`o}}, {Pollo}, {Pozzetti}, {Radovich},
  {Zamorani}, {Zucca}, {Bertin}, {Bondi}, {Bongiorno}, {Cucciati}, {Gregorini},
  {Mathez}, {Mellier}, {Merluzzi}, {Ripepi}, \& {Rizzo}}]{iovino05}
{Iovino} A. {et~al.}, 2005, \aap, 442, 423

\bibitem[{{Jiang} {et~al}\mbox{.}(2013){Jiang}, {Helly}, {Cole}, \&
  {Frenk}}]{jiang13}
{Jiang} L., {Helly} J.~C., {Cole} S., {Frenk} C.~S., 2013, ArXiv e-prints

\bibitem[{{Jones} {et~al}\mbox{.}(1991){Jones}, {Fong}, {Shanks}, {Ellis}, \&
  {Peterson}}]{aat91}
{Jones} L.~R., {Fong} R., {Shanks} T., {Ellis} R.~S., {Peterson} B.~A., 1991,
  \mnras, 249, 481

\bibitem[{{Karim} {et~al}\mbox{.}(2011){Karim}, {Schinnerer},
  {Mart{\'{\i}}nez-Sansigre}, {Sargent}, {van der Wel}, {Rix}, {Ilbert},
  {Smol{\v c}i{\'c}}, {Carilli}, {Pannella}, {Koekemoer}, {Bell}, \&
  {Salvato}}]{karim11}
{Karim} A. {et~al.}, 2011, \apj, 730, 61

\bibitem[{{Karim} {et~al}\mbox{.}(2013){Karim}, {Swinbank}, {Hodge}, {Smail},
  {Walter}, {Biggs}, {Simpson}, {Danielson}, {Alexander}, {Bertoldi}, {de
  Breuck}, {Chapman}, {Coppin}, {Dannerbauer}, {Edge}, {Greve}, {Ivison},
  {Knudsen}, {Menten}, {Schinnerer}, {Wardlow}, {Wei{\ss}}, \& {van der
  Werf}}]{karim13}
{Karim} A. {et~al.}, 2013, \mnras, 432, 2

\bibitem[{{Kauffmann} {et~al}\mbox{.}(1993){Kauffmann}, {White}, \&
  {Guiderdoni}}]{kauffmann93}
{Kauffmann} G., {White} S.~D.~M., {Guiderdoni} B., 1993, \mnras, 264, 201

\bibitem[{{Keenan} {et~al}\mbox{.}(2009){Keenan}, {Crockett}, {Aggarwal},
  {Jess}, \& {Mathioudakis}}]{keenan09}
{Keenan} F.~P., {Crockett} P.~J., {Aggarwal} K.~M., {Jess} D.~B.,
  {Mathioudakis} M., 2009, \aap, 495, 359

\bibitem[{{Kelson} \& {Holden}(2010)}]{kelson10}
{Kelson} D.~D., {Holden} B.~P., 2010, \apjl, 713, L28

\bibitem[{{Kennicutt}(1983)}]{kennicutt_imf}
{Kennicutt}, Jr. R.~C., 1983, \apj, 272, 54

\bibitem[{{Kim} {et~al}\mbox{.}(2012){Kim}, {Lacey}, {Cole}, {Baugh}, {Frenk},
  \& {Efstathiou}}]{kim12}
{Kim} H.-S., {Lacey} C.~G., {Cole} S., {Baugh} C.~M., {Frenk} C.~S.,
  {Efstathiou} G., 2012, \mnras, 425, 2674

\bibitem[{{Kim} {et~al}\mbox{.}(2013){Kim}, {Power}, {Baugh}, {Wyithe},
  {Lacey}, {Lagos}, \& {Frenk}}]{kim13}
{Kim} H.-S., {Power} C., {Baugh} C.~M., {Wyithe} J.~S.~B., {Lacey} C.~G.,
  {Lagos} C.~D.~P., {Frenk} C.~S., 2013, \mnras, 428, 3366

\bibitem[{{Knudsen} {et~al}\mbox{.}(2008){Knudsen}, {van der Werf}, \&
  {Kneib}}]{knudsen08}
{Knudsen} K.~K., {van der Werf} P.~P., {Kneib} J.-P., 2008, \mnras, 384, 1611

\bibitem[{{Kochanek} {et~al}\mbox{.}(2001){Kochanek}, {Pahre}, {Falco},
  {Huchra}, {Mader}, {Jarrett}, {Chester}, {Cutri}, \&
  {Schneider}}]{kochanek01}
{Kochanek} C.~S. {et~al.}, 2001, \apj, 560, 566

\bibitem[{{Komatsu} {et~al}\mbox{.}(2011){Komatsu}, {Smith}, {Dunkley},
  {Bennett}, {Gold}, {Hinshaw}, {Jarosik}, {Larson}, {Nolta}, {Page},
  {Spergel}, {Halpern}, {Hill}, {Kogut}, {Limon}, {Meyer}, {Odegard}, {Tucker},
  {Weiland}, {Wollack}, \& {Wright}}]{wmap7}
{Komatsu} E. {et~al.}, 2011, \apjs, 192, 18

\bibitem[{{Kong} {et~al}\mbox{.}(2006){Kong}, ~, {Arimoto}, {Renzini},
  {Broadhurst}, {Cimatti}, {Ikuta}, {Ohta}, {da Costa}, {Olsen}, {Onodera}, \&
  {Tamura}}]{kong06}
{Kong} X. {et~al.}, 2006, \apj, 638, 72

\bibitem[{{Koo}(1986)}]{koo86}
{Koo} D.~C., 1986, \apj, 311, 651

\bibitem[{{Kriek} {et~al}\mbox{.}(2010){Kriek}, {Labb{\'e}}, {Conroy},
  {Whitaker}, {van Dokkum}, {Brammer}, {Franx}, {Illingworth}, {Marchesini},
  {Muzzin}, {Quadri}, \& {Rudnick}}]{kriek10}
{Kriek} M. {et~al.}, 2010, \apjl, 722, L64

\bibitem[{{K{\"u}mmel} \& {Wagner}(2001)}]{kummel01}
{K{\"u}mmel} M.~W., {Wagner} S.~J., 2001, \aap, 370, 384

\bibitem[{{Kurucz}(1993)}]{kurucz93}
{Kurucz} R., 1993, Atomic data for opacity calculations.~Kurucz CD-ROM No.~1.~
  Cambridge, Mass.: Smithsonian Astrophysical Observatory, 1993., 1

\bibitem[{{La Barbera} {et~al}\mbox{.}(2013){La Barbera}, {Ferreras},
  {Vazdekis}, {de la Rosa}, {de Carvalho}, {Trevisan}, {Falc{\'o}n-Barroso}, \&
  {Ricciardelli}}]{labarbera13}
{La Barbera} F., {Ferreras} I., {Vazdekis} A., {de la Rosa} I.~G., {de
  Carvalho} R.~R., {Trevisan} M., {Falc{\'o}n-Barroso} J., {Ricciardelli} E.,
  2013, \mnras, 433, 3017

\bibitem[{{Lacey} {et~al}\mbox{.}(2011){Lacey}, {Baugh}, {Frenk}, \&
  {Benson}}]{lacey11}
{Lacey} C.~G., {Baugh} C.~M., {Frenk} C.~S., {Benson} A.~J., 2011, \mnras, 412,
  1828

\bibitem[{{Lagos} {et~al}\mbox{.}(2011{\natexlab{a}}){Lagos}, {Baugh}, {Lacey},
  {Benson}, {Kim}, \& {Power}}]{lagos11}
{Lagos} C.~D.~P., {Baugh} C.~M., {Lacey} C.~G., {Benson} A.~J., {Kim} H.-S.,
  {Power} C., 2011{\natexlab{a}}, \mnras, 418, 1649

\bibitem[{{Lagos} {et~al}\mbox{.}(2013{\natexlab{a}}){Lagos}, {Baugh}, {Zwaan},
  {Lacey}, {Gonzalez-Perez}, {Power}, {Swinbank}, \& {van Kampen}}]{lagos14}
{Lagos} C.~d.~P., {Baugh} C.~M., {Zwaan} M.~A., {Lacey} C.~G., {Gonzalez-Perez}
  V., {Power} C., {Swinbank} A.~M., {van Kampen} E., 2013{\natexlab{a}},
  arXiv:astro-ph/1310.4178

\bibitem[{{Lagos} {et~al}\mbox{.}(2012){Lagos}, {Bayet}, {Baugh}, {Lacey},
  {Bell}, {Fanidakis}, \& {Geach}}]{lagos12}
{Lagos} C.~d.~P., {Bayet} E., {Baugh} C.~M., {Lacey} C.~G., {Bell} T.~A.,
  {Fanidakis} N., {Geach} J.~E., 2012, \mnras, 426, 2142

\bibitem[{{Lagos} {et~al}\mbox{.}(2013{\natexlab{b}}){Lagos}, {Lacey}, \&
  {Baugh}}]{lagos13}
{Lagos} C.~d.~P., {Lacey} C.~G., {Baugh} C.~M., 2013{\natexlab{b}}, \mnras,
  436, 1787

\bibitem[{{Lagos} {et~al}\mbox{.}(2011{\natexlab{b}}){Lagos}, {Lacey}, {Baugh},
  {Bower}, \& {Benson}}]{lagos10}
{Lagos} C.~D.~P., {Lacey} C.~G., {Baugh} C.~M., {Bower} R.~G., {Benson} A.~J.,
  2011{\natexlab{b}}, \mnras, 416, 1566

\bibitem[{{Lan{\c c}on} \& {Mouhcine}(2002)}]{lancon02}
{Lan{\c c}on} A., {Mouhcine} M., 2002, \aap, 393, 167

\bibitem[{{Lawrence} {et~al}\mbox{.}(2007){Lawrence}, {Almaini}, {Edge},
  {Hambly}, {Jameson}, {Lucas}, {Casali}, {Adamson}, {Dye}, {Emerson},
  {Foucaud}, {Hewett}, {Hirst}, {Hodgkin}, {Irwin}, {Lodieu}, {McMahon},
  {Simpson}, {Smail}, {Mortlock}, \& {Folger}}]{lawrence07}
{Lawrence}, A.~{Warren} S.~J. {et~al.}, 2007, \mnras, 379, 1599

\bibitem[{{Le Borgne} {et~al}\mbox{.}(2003){Le Borgne}, {Bruzual}, {Pell{\'o}},
  {Lan{\c c}on}, {Rocca-Volmerange}, {Sanahuja}, {Schaerer}, {Soubiran}, \&
  {V{\'{\i}}lchez-G{\'o}mez}}]{stelib}
{Le Borgne} J.-F. {et~al.}, 2003, \aap, 402, 433

\bibitem[{{Lejeune} {et~al}\mbox{.}(1997){Lejeune}, {Cuisinier}, \&
  {Buser}}]{lejeune97}
{Lejeune} T., {Cuisinier} F., {Buser} R., 1997, \aaps, 125, 229

\bibitem[{{Lejeune} {et~al}\mbox{.}(1998){Lejeune}, {Cuisinier}, \&
  {Buser}}]{lejeune98}
{Lejeune} T., {Cuisinier} F., {Buser} R., 1998, \aaps, 130, 65

\bibitem[{{Leroy} {et~al}\mbox{.}(2008){Leroy}, {Walter}, {Brinks}, {Bigiel},
  {de Blok}, {Madore}, \& {Thornley}}]{leroy08}
{Leroy} A.~K., {Walter} F., {Brinks} E., {Bigiel} F., {de Blok} W.~J.~G.,
  {Madore} B., {Thornley} M.~D., 2008, \aj, 136, 2782

\bibitem[{{Linder}(2005)}]{Linder05}
{Linder} E.~V., 2005, \prd, 72, 043529

\bibitem[{{Lyubenova} {et~al}\mbox{.}(2012){Lyubenova}, {Kuntschner},
  {Rejkuba}, {Silva}, {Kissler-Patig}, \& {Tacconi-Garman}}]{lyu12}
{Lyubenova} M., {Kuntschner} H., {Rejkuba} M., {Silva} D.~R., {Kissler-Patig}
  M., {Tacconi-Garman} L.~E., 2012, \aap, 543, A75

\bibitem[{{MacArthur} {et~al}\mbox{.}(2010){MacArthur}, {McDonald}, {Courteau},
  \& {Jes{\'u}s Gonz{\'a}lez}}]{macarthur10}
{MacArthur} L.~A., {McDonald} M., {Courteau} S., {Jes{\'u}s Gonz{\'a}lez} J.,
  2010, \apj, 718, 768

\bibitem[{{Maihara} {et~al}\mbox{.}(2001){Maihara}, {Iwamuro}, {Tanabe},
  {Taguchi}, {Hata}, {Oya}, {Kashikawa}, {Iye}, {Miyazaki}, {Karoji},
  {Yoshida}, {Totani}, {Yoshii}, {Okamura}, {Shimasaku}, {Saito}, {Ando},
  {Goto}, {Hayashi}, {Kaifu}, {Kobayashi}, {Kosugi}, {Motohara}, {Nishimura},
  {Noumaru}, {Ogasawara}, {Sasaki}, {Sekiguchi}, {Takata}, {Terada},
  {Yamashita}, {Usuda}, \& {Tokunaga}}]{maihara01}
{Maihara} T. {et~al.}, 2001, \pasj, 53, 25

\bibitem[{{Mancone} \& {Gonzalez}(2012)}]{mancone12}
{Mancone} C.~L., {Gonzalez} A.~H., 2012, \pasp, 124, 606

\bibitem[{{Maraston}(2005)}]{mn05}
{Maraston} C., 2005, \mnras, 362, 799

\bibitem[{{Maraston} \& {Str{\"o}mb{\"a}ck}(2011)}]{maraston11}
{Maraston} C., {Str{\"o}mb{\"a}ck} G., 2011, \mnras, 418, 2785

\bibitem[{{Marchesini} {et~al}\mbox{.}(2012){Marchesini}, {Stefanon},
  {Brammer}, \& {Whitaker}}]{marchesini12}
{Marchesini} D., {Stefanon} M., {Brammer} G.~B., {Whitaker} K.~E., 2012, \apj,
  748, 126

\bibitem[{{Marigo} \& {Girardi}(2007)}]{marigo07}
{Marigo} P., {Girardi} L., 2007, in Astronomical Society of the Pacific
  Conference Series, Vol. 374, From Stars to Galaxies: Building the Pieces to
  Build Up the Universe, {Vallenari} A., {Tantalo} R., {Portinari} L.,
  {Moretti} A., eds., p.~33

\bibitem[{{Marigo} {et~al}\mbox{.}(2008){Marigo}, {Girardi}, {Bressan},
  {Groenewegen}, {Silva}, \& {Granato}}]{marigo08}
{Marigo} P., {Girardi} L., {Bressan} A., {Groenewegen} M.~A.~T., {Silva} L.,
  {Granato} G.~L., 2008, \aap, 482, 883

\bibitem[{{Martig} {et~al}\mbox{.}(2013){Martig}, {Crocker}, {Bournaud},
  {Emsellem}, {Gabor}, {Alatalo}, {Blitz}, {Bois}, {Bureau}, {Cappellari},
  {Davies}, {Davis}, {Dekel}, {de Zeeuw}, {Duc}, {Falc{\'o}n-Barroso},
  {Khochfar}, {Krajnovi{\'c}}, {Kuntschner}, {Morganti}, {McDermid}, {Naab},
  {Oosterloo}, {Sarzi}, {Scott}, {Serra}, {Griffin}, {Teyssier}, {Weijmans}, \&
  {Young}}]{martig13}
{Martig} M. {et~al.}, 2013, \mnras, 432, 1914

\bibitem[{{Martin} {et~al}\mbox{.}(2010){Martin}, {Papastergis}, {Giovanelli},
  {Haynes}, {Springob}, \& {Stierwalt}}]{martin10}
{Martin} A.~M., {Papastergis} E., {Giovanelli} R., {Haynes} M.~P., {Springob}
  C.~M., {Stierwalt} S., 2010, \apj, 723, 1359

\bibitem[{{Mathewson} \& {Ford}(1996)}]{math}
{Mathewson} D.~S., {Ford} V.~L., 1996, \apjs, 107, 97

\bibitem[{{McCracken} {et~al}\mbox{.}(2000){McCracken}, {Metcalfe}, {Shanks},
  {Campos}, {Gardner}, \& {Fong}}]{mccracken}
{McCracken} H.~J., {Metcalfe} N., {Shanks} T., {Campos} A., {Gardner} J.~P.,
  {Fong} R., 2000, \mnras, 311, 707

\bibitem[{{McCracken} {et~al}\mbox{.}(2003){McCracken}, {Radovich}, {Bertin},
  {Mellier}, {Dantel-Fort}, {Le F{\`e}vre}, {Cuillandre}, {Gwyn}, {Foucaud}, \&
  {Zamorani}}]{mccracken03}
{McCracken} H.~J. {et~al.}, 2003, \aap, 410, 17

\bibitem[{{McGaugh} \& {Schombert}(2013)}]{mcgaugh13}
{McGaugh} S., {Schombert} J., 2013, ArXiv e-prints

\bibitem[{{McLeod} {et~al}\mbox{.}(1995){McLeod}, {Bernstein}, {Rieke},
  {Tollestrup}, \& {Fazio}}]{mcleod95}
{McLeod} B.~A., {Bernstein} G.~M., {Rieke} M.~J., {Tollestrup} E.~V., {Fazio}
  G.~G., 1995, \apjs, 96, 117

\bibitem[{{McLure} {et~al}\mbox{.}(2009){McLure}, {Cirasuolo}, {Dunlop},
  {Foucaud}, \& {Almaini}}]{mclure09}
{McLure} R.~J., {Cirasuolo} M., {Dunlop} J.~S., {Foucaud} S., {Almaini} O.,
  2009, \mnras, 395, 2196

\bibitem[{{Melbourne} {et~al}\mbox{.}(2012){Melbourne}, {Williams},
  {Dalcanton}, {Rosenfield}, {Girardi}, {Marigo}, {Weisz}, {Dolphin}, {Boyer},
  {Olsen}, {Skillman}, \& {Seth}}]{melbourne12}
{Melbourne} J. {et~al.}, 2012, \apj, 748, 47

\bibitem[{{Merson} {et~al}\mbox{.}(2013){Merson}, {Baugh}, {Helly},
  {Gonzalez-Perez}, {Cole}, {Bielby}, {Norberg}, {Frenk}, {Benson}, {Bower},
  {Lacey}, \& {Lagos}}]{merson13}
{Merson} A.~I. {et~al.}, 2013, \mnras, 429, 556

\bibitem[{{Metcalfe} {et~al}\mbox{.}(2001){Metcalfe}, {Shanks}, {Campos},
  {McCracken}, \& {Fong}}]{metcalfe01}
{Metcalfe} N., {Shanks} T., {Campos} A., {McCracken} H.~J., {Fong} R., 2001,
  \mnras, 323, 795

\bibitem[{{Metcalfe} {et~al}\mbox{.}(1991){Metcalfe}, {Shanks}, {Fong}, \&
  {Jones}}]{metcalfe91}
{Metcalfe} N., {Shanks} T., {Fong} R., {Jones} L.~R., 1991, \mnras, 249, 498

\bibitem[{{Metcalfe} {et~al}\mbox{.}(1995){Metcalfe}, {Shanks}, {Fong}, \&
  {Roche}}]{metcalfe95}
{Metcalfe} N., {Shanks} T., {Fong} R., {Roche} N., 1995, \mnras, 273, 257

\bibitem[{{Metcalfe} {et~al}\mbox{.}(2006){Metcalfe}, {Shanks}, {Weilbacher},
  {McCracken}, {Fong}, \& {Thompson}}]{metcalfe06}
{Metcalfe} N., {Shanks} T., {Weilbacher} P.~M., {McCracken} H.~J., {Fong} R.,
  {Thompson} D., 2006, \mnras, 370, 1257

\bibitem[{{Meynet} {et~al}\mbox{.}(1994){Meynet}, {Maeder}, {Schaller},
  {Schaerer}, \& {Charbonnel}}]{meynet94}
{Meynet} G., {Maeder} A., {Schaller} G., {Schaerer} D., {Charbonnel} C., 1994,
  \aaps, 103, 97

\bibitem[{{Mitchell} {et~al}\mbox{.}(2013){Mitchell}, {Lacey}, {Baugh}, \&
  {Cole}}]{mitchell13}
{Mitchell} P.~D., {Lacey} C.~G., {Baugh} C.~M., {Cole} S., 2013, \mnras, 435,
  87

\bibitem[{{Mobasher} {et~al}\mbox{.}(1986){Mobasher}, {Ellis}, \&
  {Sharples}}]{mobasher}
{Mobasher} B., {Ellis} R.~S., {Sharples} R.~M., 1986, \mnras, 223, 11

\bibitem[{{Moustakas} {et~al}\mbox{.}(1997){Moustakas}, {Davis}, {Graham},
  {Silk}, {Peterson}, \& {Yoshii}}]{moustakas97}
{Moustakas} L.~A., {Davis} M., {Graham} J.~R., {Silk} J., {Peterson} B.~A.,
  {Yoshii} Y., 1997, \apj, 475, 445

\bibitem[{{Muzzin} {et~al}\mbox{.}(2013){Muzzin}, {Marchesini}, {Stefanon},
  {Franx}, {McCracken}, {Milvang-Jensen}, {Dunlop}, {Fynbo}, {Brammer},
  {Labb{\'e}}, \& {van Dokkum}}]{muzzin13}
{Muzzin} A. {et~al.}, 2013, \apj, 777, 18

\bibitem[{{Norberg} {et~al}\mbox{.}(2002){Norberg}, {Cole}, {Baugh}, {Frenk},
  {Baldry}, {Bland-Hawthorn}, {Bridges}, {Cannon}, {Colless}, {Collins},
  {Couch}, {Cross}, {Dalton}, {De Propris}, {Driver}, {Efstathiou}, {Ellis},
  {Glazebrook}, {Jackson}, {Lahav}, {Lewis}, {Lumsden}, {Maddox}, {Madgwick},
  {Peacock}, {Peterson}, {Sutherland}, {Taylor}, \& {2DFGRS Team}}]{norberg02}
{Norberg} P. {et~al.}, 2002, \mnras, 336, 907

\bibitem[{{Okamoto} {et~al}\mbox{.}(2005){Okamoto}, {Eke}, {Frenk}, \&
  {Jenkins}}]{okamoto05}
{Okamoto} T., {Eke} V.~R., {Frenk} C.~S., {Jenkins} A., 2005, \mnras, 363, 1299

\bibitem[{{Okamoto} {et~al}\mbox{.}(2008){Okamoto}, {Gao}, \&
  {Theuns}}]{okamoto08}
{Okamoto} T., {Gao} L., {Theuns} T., 2008, \mnras, 390, 920

\bibitem[{{Oliver} {et~al}\mbox{.}(2010){Oliver}, {Wang}, {Smith}, {Altieri},
  {Amblard}, {Arumugam}, {Auld}, {Aussel}, {Babbedge}, {Blain}, {Bock},
  {Boselli}, {Buat}, {Burgarella}, {Castro-Rodr{\'{\i}}guez}, {Cava},
  {Chanial}, {Clements}, {Conley}, {Conversi}, {Cooray}, {Dowell}, {Dwek},
  {Eales}, {Elbaz}, {Fox}, {Franceschini}, {Gear}, {Glenn}, {Griffin},
  {Halpern}, {Hatziminaoglou}, {Ibar}, {Isaak}, {Ivison}, {Lagache},
  {Levenson}, {Lu}, {Madden}, {Maffei}, {Mainetti}, {Marchetti},
  {Mitchell-Wynne}, {Mortier}, {Nguyen}, {O'Halloran}, {Omont}, {Page},
  {Panuzzo}, {Papageorgiou}, {Pearson}, {P{\'e}rez-Fournon}, {Pohlen},
  {Rawlings}, {Raymond}, {Rigopoulou}, {Rizzo}, {Roseboom}, {Rowan-Robinson},
  {S{\'a}nchez Portal}, {Savage}, {Schulz}, {Scott}, {Seymour}, {Shupe},
  {Stevens}, {Symeonidis}, {Trichas}, {Tugwell}, {Vaccari}, {Valiante},
  {Valtchanov}, {Vieira}, {Vigroux}, {Ward}, {Wright}, {Xu}, \&
  {Zemcov}}]{hermes}
{Oliver} S.~J. {et~al.}, 2010, \aap, 518, L21

\bibitem[{{Palamara} {et~al}\mbox{.}(2013){Palamara}, {Brown}, {Jannuzi},
  {Dey}, {Stern}, {Pimbblet}, {Weiner}, {Ashby}, {Kochanek}, {Gonzalez},
  {Brodwin}, {Le Floc'h}, \& {Rieke}}]{palamara13}
{Palamara} D.~P. {et~al.}, 2013, \apj, 764, 31

\bibitem[{{Picard}(1991)}]{picard91}
{Picard} A., 1991, \aj, 102, 445

\bibitem[{{Planck Collaboration} {et~al}\mbox{.}(2013){Planck Collaboration},
  {Ade}, {Aghanim}, {Armitage-Caplan}, {Arnaud}, {Ashdown}, {Atrio-Barandela},
  {Aumont}, {Baccigalupi}, {Banday}, \& et~al.}]{planck}
{Planck Collaboration} {et~al.}, 2013, arXiv:astro-ph/1303.5076

\bibitem[{{Pozzetti} {et~al}\mbox{.}(2003){Pozzetti}, {Cimatti}, {Zamorani},
  {Daddi}, {Menci}, {Fontana}, {Renzini}, {Mignoli}, {Poli}, {Saracco},
  {Broadhurst}, {Cristiani}, {D'Odorico}, {Giallongo}, \&
  {Gilmozzi}}]{pozzetti03}
{Pozzetti} L. {et~al.}, 2003, \aap, 402, 837

\bibitem[{{Reddy} \& {Steidel}(2009)}]{reddy09}
{Reddy} N.~A., {Steidel} C.~C., 2009, \apj, 692, 778

\bibitem[{{Renzini} \& {Fusi Pecci}(1988)}]{renzini}
{Renzini} A., {Fusi Pecci} F., 1988, \araa, 26, 199

\bibitem[{{Riffel} {et~al}\mbox{.}(2011){Riffel}, {Ruschel-Dutra}, {Pastoriza},
  {Rodr{\'{\i}}guez-Ardila}, {Santos}, {Bonatto}, \& {Ducati}}]{riffel11}
{Riffel} R., {Ruschel-Dutra} D., {Pastoriza} M.~G., {Rodr{\'{\i}}guez-Ardila}
  A., {Santos}, Jr. J.~F.~C., {Bonatto} C.~J., {Ducati} J.~R., 2011, \mnras,
  410, 2714

\bibitem[{{Rocca-Volmerange} {et~al}\mbox{.}(2007){Rocca-Volmerange}, {de
  Lapparent}, {Seymour}, \& {Fioc}}]{p3a}
{Rocca-Volmerange} B., {de Lapparent} V., {Seymour} N., {Fioc} M., 2007, \aap,
  475, 801

\bibitem[{{Rocca-Volmerange} {et~al}\mbox{.}(2013){Rocca-Volmerange},
  {Drouart}, {De Breuck}, {Vernet}, {Seymour}, {Wylezalek}, {Lehnert},
  {Nesvadba}, \& {Fioc}}]{p3}
{Rocca-Volmerange} B. {et~al.}, 2013, \mnras, 429, 2780

\bibitem[{{Roche} {et~al}\mbox{.}(2002){Roche}, {Almaini}, {Dunlop}, {Ivison},
  \& {Willott}}]{roche02}
{Roche} N.~D., {Almaini} O., {Dunlop} J., {Ivison} R.~J., {Willott} C.~J.,
  2002, \mnras, 337, 1282

\bibitem[{{Saintonge} {et~al}\mbox{.}(2011){Saintonge}, {Kauffmann}, {Wang},
  {Kramer}, {Tacconi}, {Buchbender}, {Catinella}, {Graci{\'a}-Carpio},
  {Cortese}, {Fabello}, {Fu}, {Genzel}, {Giovanelli}, {Guo}, {Haynes},
  {Heckman}, {Krumholz}, {Lemonias}, {Li}, {Moran}, {Rodriguez-Fernandez},
  {Schiminovich}, {Schuster}, \& {Sievers}}]{saintonge11}
{Saintonge} A. {et~al.}, 2011, \mnras, 415, 61

\bibitem[{{Salpeter}(1955)}]{salpeter}
{Salpeter} E.~E., 1955, \apj, 121, 161

\bibitem[{{Santini} {et~al}\mbox{.}(2012){Santini}, {Fontana}, {Grazian},
  {Salimbeni}, {Fontanot}, {Paris}, {Boutsia}, {Castellano}, {Fiore},
  {Gallozzi}, {Giallongo}, {Koekemoer}, {Menci}, {Pentericci}, \&
  {Somerville}}]{santini12}
{Santini} P. {et~al.}, 2012, \aap, 538, A33

\bibitem[{{Saracco} {et~al}\mbox{.}(2006){Saracco}, {Fiano}, {Chincarini},
  {Vanzella}, {Longhetti}, {Cristiani}, {Fontana}, {Giallongo}, \&
  {Nonino}}]{saracco06}
{Saracco} P. {et~al.}, 2006, \mnras, 367, 349

\bibitem[{{Saracco} {et~al}\mbox{.}(2001){Saracco}, {Giallongo}, {Cristiani},
  {D'Odorico}, {Fontana}, {Iovino}, {Poli}, \& {Vanzella}}]{saracco01}
{Saracco} P., {Giallongo} E., {Cristiani} S., {D'Odorico} S., {Fontana} A.,
  {Iovino} A., {Poli} F., {Vanzella} E., 2001, \aap, 375, 1

\bibitem[{{Sawicki} \& {Thompson}(2006)}]{sawicki06}
{Sawicki} M., {Thompson} D., 2006, \apj, 642, 653

\bibitem[{{Schaller} {et~al}\mbox{.}(1992){Schaller}, {Schaerer}, {Meynet}, \&
  {Maeder}}]{schaller92}
{Schaller} G., {Schaerer} D., {Meynet} G., {Maeder} A., 1992, \aaps, 96, 269

\bibitem[{{Shen} {et~al}\mbox{.}(2003){Shen}, {Mo}, {White}, {Blanton},
  {Kauffmann}, {Voges}, {Brinkmann}, \& {Csabai}}]{shen03}
{Shen} S., {Mo} H.~J., {White} S.~D.~M., {Blanton} M.~R., {Kauffmann} G.,
  {Voges} W., {Brinkmann} J., {Csabai} I., 2003, \mnras, 343, 978

\bibitem[{{Silva} {et~al}\mbox{.}(1998){Silva}, {Granato}, {Bressan}, \&
  {Danese}}]{grasil}
{Silva} L., {Granato} G.~L., {Bressan} A., {Danese} L., 1998, \apj, 509, 103

\bibitem[{{Simpson} {et~al}\mbox{.}(2006){Simpson}, {Cirasuolo}, {Dunlop},
  {Foucaud}, {Hirst}, {Ivison}, {Page}, {Rawlings}, {Sekiguchi}, {Smail}, \&
  {Watson}}]{simpson06}
{Simpson}, C.~{Almaini} O. {et~al.}, 2006, \mnras, 373, L21

\bibitem[{{Smail} {et~al}\mbox{.}(2002){Smail}, {Owen}, {Morrison}, {Keel},
  {Ivison}, \& {Ledlow}}]{smail02}
{Smail} I., {Owen} F.~N., {Morrison} G.~E., {Keel} W.~C., {Ivison} R.~J.,
  {Ledlow} M.~J., 2002, \apj, 581, 844

\bibitem[{{Smith} {et~al}\mbox{.}(2002){Smith}, {Smail}, {Kneib}, {Davis},
  {Takamiya}, {Ebeling}, \& {Czoske}}]{smith02}
{Smith} G.~P., {Smail} I., {Kneib} J.-P., {Davis} C.~J., {Takamiya} M.,
  {Ebeling} H., {Czoske} O., 2002, \mnras, 333, L16

\bibitem[{{Smith} {et~al}\mbox{.}(2012){Smith}, {Lucey}, \& {Carter}}]{smith12}
{Smith} R.~J., {Lucey} J.~R., {Carter} D., 2012, \mnras, 421, 2982

\bibitem[{{Soifer} {et~al}\mbox{.}(1994){Soifer}, {Matthews}, {Djorgovski},
  {Larkin}, {Graham}, {Harrison}, {Jernigan}, {Lin}, {Nelson}, {Neugebauer},
  {Smith}, {Smith}, \& {Ziomkowski}}]{soifer94}
{Soifer} B.~T. {et~al.}, 1994, \apjl, 420, L1

\bibitem[{{Spergel} {et~al}\mbox{.}(2003){Spergel}, {Verde}, {Peiris},
  {Komatsu}, {Nolta}, {Bennett}, {Halpern}, {Hinshaw}, {Jarosik}, {Kogut},
  {Limon}, {Meyer}, {Page}, {Tucker}, {Weiland}, {Wollack}, \&
  {Wright}}]{wmap1}
{Spergel} D.~N. {et~al.}, 2003, \apjs, 148, 175

\bibitem[{{Springel} {et~al}\mbox{.}(2005){Springel}, {White}, {Jenkins},
  {Frenk}, {Yoshida}, {Gao}, {Navarro}, {Thacker}, {Croton}, {Helly},
  {Peacock}, {Cole}, {Thomas}, {Couchman}, {Evrard}, {Colberg}, \&
  {Pearce}}]{springel05}
{Springel} V. {et~al.}, 2005, \nat, 435, 629

\bibitem[{{Springel} {et~al}\mbox{.}(2001){Springel}, {White}, {Tormen}, \&
  {Kauffmann}}]{springel01}
{Springel} V., {White} S.~D.~M., {Tormen} G., {Kauffmann} G., 2001, \mnras,
  328, 726

\bibitem[{{Stefanon} \& {Marchesini}(2013)}]{stefanon13}
{Stefanon} M., {Marchesini} D., 2013, \mnras, 429, 881

\bibitem[{{Steidel} \& {Hamilton}(1993)}]{steidel93}
{Steidel} C.~C., {Hamilton} D., 1993, \aj, 105, 2017

\bibitem[{{Szokoly} {et~al}\mbox{.}(1998){Szokoly}, {Subbarao}, {Connolly}, \&
  {Mobasher}}]{szokoly98}
{Szokoly} G.~P., {Subbarao} M.~U., {Connolly} A.~J., {Mobasher} B., 1998, \apj,
  492, 452

\bibitem[{{Tacconi} {et~al}\mbox{.}(2013){Tacconi}, {Neri}, {Genzel}, {Combes},
  {Bolatto}, {Cooper}, {Wuyts}, {Bournaud}, {Burkert}, {Comerford}, {Cox},
  {Davis}, {F{\"o}rster Schreiber}, {Garc{\'{\i}}a-Burillo}, {Gracia-Carpio},
  {Lutz}, {Naab}, {Newman}, {Omont}, {Saintonge}, {Shapiro Griffin}, {Shapley},
  {Sternberg}, \& {Weiner}}]{tacconi13}
{Tacconi} L.~J. {et~al.}, 2013, \apj, 768, 74

\bibitem[{{Tonini} {et~al}\mbox{.}(2009){Tonini}, {Maraston}, {Devriendt},
  {Thomas}, \& {Silk}}]{tonini09}
{Tonini} C., {Maraston} C., {Devriendt} J., {Thomas} D., {Silk} J., 2009,
  \mnras, 396, L36

\bibitem[{{Tonini} {et~al}\mbox{.}(2010){Tonini}, {Maraston}, {Thomas},
  {Devriendt}, \& {Silk}}]{tonini10}
{Tonini} C., {Maraston} C., {Thomas} D., {Devriendt} J., {Silk} J., 2010,
  \mnras, 403, 1749

\bibitem[{{Tyson}(1988)}]{tyson88}
{Tyson} J.~A., 1988, \aj, 96, 1

\bibitem[{{V{\"a}is{\"a}nen} \& {Johansson}(2004)}]{vaisanen04}
{V{\"a}is{\"a}nen} P., {Johansson} P.~H., 2004, \aap, 421, 821

\bibitem[{{V{\"a}is{\"a}nen} {et~al}\mbox{.}(2000){V{\"a}is{\"a}nen},
  {Tollestrup}, {Willner}, \& {Cohen}}]{vaisanen00}
{V{\"a}is{\"a}nen} P., {Tollestrup} E.~V., {Willner} S.~P., {Cohen} M., 2000,
  \apj, 540, 593

\bibitem[{{van Dokkum}(2008)}]{vandokkum08}
{van Dokkum} P.~G., 2008, \apj, 674, 29

\bibitem[{{Wardlow} {et~al}\mbox{.}(2011){Wardlow}, {Smail}, {Coppin},
  {Alexander}, {Brandt}, {Danielson}, {Luo}, {Swinbank}, {Walter}, {Wei{\ss}},
  {Xue}, {Zibetti}, {Bertoldi}, {Biggs}, {Chapman}, {Dannerbauer}, {Dunlop},
  {Gawiser}, {Ivison}, {Knudsen}, {Kov{\'a}cs}, {Lacey}, {Menten}, {Padilla},
  {Rix}, \& {van der Werf}}]{wardlow11}
{Wardlow} J.~L. {et~al.}, 2011, \mnras, 415, 1479

\bibitem[{{Wei{\ss}} {et~al}\mbox{.}(2009){Wei{\ss}}, {Kov{\'a}cs}, {Coppin},
  {Greve}, {Walter}, {Smail}, {Dunlop}, {Knudsen}, {Alexander}, {Bertoldi},
  {Brandt}, {Chapman}, {Cox}, {Dannerbauer}, {De Breuck}, {Gawiser}, {Ivison},
  {Lutz}, {Menten}, {Koekemoer}, {Kreysa}, {Kurczynski}, {Rix}, {Schinnerer},
  \& {van der Werf}}]{weiss09}
{Wei{\ss}} A. {et~al.}, 2009, \apj, 707, 1201

\bibitem[{{Westera} {et~al}\mbox{.}(2002){Westera}, {Lejeune}, {Buser},
  {Cuisinier}, \& {Bruzual}}]{westera02}
{Westera} P., {Lejeune} T., {Buser} R., {Cuisinier} F., {Bruzual} G., 2002,
  \aap, 381, 524

\bibitem[{{White} \& {Rees}(1978)}]{whiterees78}
{White} S.~D.~M., {Rees} M.~J., 1978, \mnras, 183, 341

\bibitem[{{Wilkins} {et~al}\mbox{.}(2012){Wilkins}, {Gonzalez-Perez}, {Lacey},
  \& {Baugh}}]{wilkins12}
{Wilkins} S.~M., {Gonzalez-Perez} V., {Lacey} C.~G., {Baugh} C.~M., 2012,
  \mnras, 427, 1490

\bibitem[{{Wilkins} {et~al}\mbox{.}(2008){Wilkins}, {Hopkins}, {Trentham}, \&
  {Tojeiro}}]{sw08}
{Wilkins} S.~M., {Hopkins} A.~M., {Trentham} N., {Tojeiro} R., 2008, \mnras,
  391, 363

\bibitem[{{Wyder} {et~al}\mbox{.}(2005){Wyder}, {Treyer}, {Milliard},
  {Schiminovich}, {Arnouts}, {Budav{\'a}ri}, {Barlow}, {Bianchi}, {Byun},
  {Donas}, {Forster}, {Friedman}, {Heckman}, {Jelinsky}, {Lee}, {Madore},
  {Malina}, {Martin}, {Morrissey}, {Neff}, {Rich}, {Siegmund}, {Small},
  {Szalay}, \& {Welsh}}]{galex05}
{Wyder} T.~K. {et~al.}, 2005, \apjl, 619, L15

\bibitem[{{Yasuda} {et~al}\mbox{.}(2001){Yasuda}, {Fukugita}, {Narayanan},
  {Lupton}, {Strateva}, {Strauss}, {Ivezi{\'c}}, {Kim}, {Hogg}, {Weinberg},
  {Shimasaku}, {Loveday}, {Annis}, {Bahcall}, {Blanton}, {Brinkmann},
  {Brunner}, {Connolly}, {Csabai}, {Doi}, {Hamabe}, {Ichikawa}, {Ichikawa},
  {Johnston}, {Knapp}, {Kunszt}, {Lamb}, {McKay}, {Munn}, {Nichol}, {Okamura},
  {Schneider}, {Szokoly}, {Vogeley}, {Watanabe}, \& {York}}]{sdss01}
{Yasuda} N. {et~al.}, 2001, \aj, 122, 1104

\bibitem[{{Yee} \& {Green}(1987)}]{yee87}
{Yee} H.~K.~C., {Green} R.~F., 1987, \apj, 319, 28

\bibitem[{{Zibetti} {et~al}\mbox{.}(2013){Zibetti}, {Gallazzi}, {Charlot},
  {Pierini}, \& {Pasquali}}]{zibetti13}
{Zibetti} S., {Gallazzi} A., {Charlot} S., {Pierini} D., {Pasquali} A., 2013,
  \mnras, 428, 1479

\bibitem[{{Zwaan} {et~al}\mbox{.}(2005){Zwaan}, {Meyer}, {Staveley-Smith}, \&
  {Webster}}]{zwaan05}
{Zwaan} M.~A., {Meyer} M.~J., {Staveley-Smith} L., {Webster} R.~L., 2005,
  \mnras, 359, L30

\end{thebibliography}


\appendix

\section{Other model predictions}

Here we present some further predictions of the new model of galaxy
formation and evolution using solely the
BC99 SPS model. These properties have been previously used to
constrain and/or test earlier versions of the {\sc galform} model, and are presented here for comparison.

\begin{table}
  \begin{tabular}{ c | c | c | c | c}
  \hline                       
  IMF & L$_{1500 \AA}$ & L$_{\rm H_{\alpha}}$   & $\dot{N}_{SN}$ & L$_{FIR}$\\ 
   & erg s$^{-1}$ Hz$^{-1}$ & 10$^{40}$ erg s$^{-1}$ & yr$^{-1}$ & 10$^{44}$ ergs$^{-1}$\\ 
  \hline                        
 Ken & 1.12$\times 10^{28}$ &  13.04   & 0.011 & 0.308 \\
 Sal & 8.82$\times 10^{27}$ &  12.20   & 0.009 & 0.249 \\
 Cha & 1.41$\times 10^{28}$ &  20.53   & 0.014 & 0.399 \\
 BG  & 1.74$\times 10^{28}$ &  29.48   & 0.017 & 0.504 \\
  \hline                        
\end{tabular}
  \caption{L$_{1500 \AA}$, L$_{\rm H_{\alpha}}$ and the ratio of
    core-collapsed SNe, $\dot{N}_{SN}$, produced by a constant SFR=1
    M$_{\odot}$yr$^{-1}$ after 1 Gyr obtained with the P2 SPS model for a fixed solar metallicity and
    without dust attenuation, assuming different IMFs. The L$_{FIR}$ (10 to
    1000 $\mu$m) luminosities are calculated with the P2 SPS model for a
    solar metallicity from the bolometric luminosity at 100 Myr. The
    different IMFs used are, from top to bottom:
    \citet{kennicutt_imf} (Ken), \citet{salpeter} (Sal),
    \citet{chabrier03} (Cha) and
    \citet{BG} (BG) IMFs.
}
\label{tab:sfrv}
\end{table}
\begin{figure}
{\epsfxsize=8.5truecm
\epsfbox[28 10 292 264]{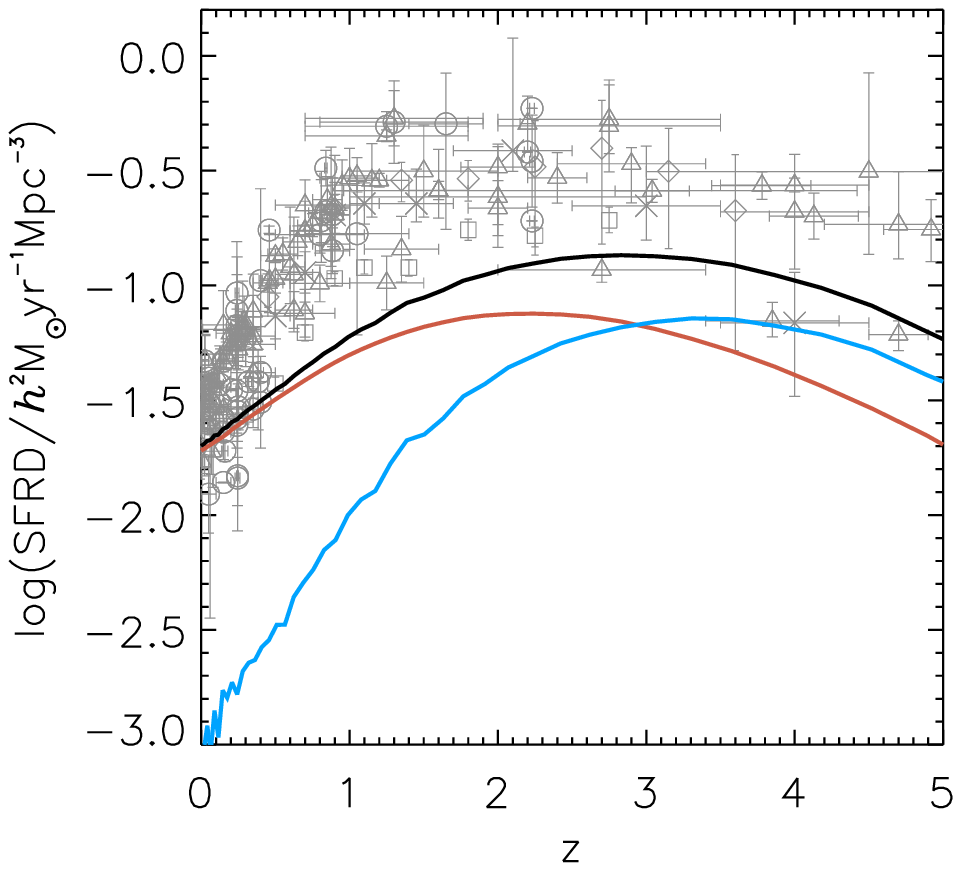}}
\caption
{The evolution of the predicted cosmic star formation ratio per unit
  comoving volume (SFRD, black line). The blue line shows the
  predicted contribution
  from galaxies experiencing a burst of star formation. The red line
  shows that for quiescent model galaxies. The
  crosses show the SFRD estimated by \citet{cucciati12} from the the
  intrinsic non-ionising UV stellar continuum of galaxies. The triangles present the
  compilation from \citet{hopkins06} for UV, H$_{\alpha}$, FIR and radio
  tracers ($\sim$1.4 GHz). The squares show the
  results from radio observations by \citet{karim11}. The circles show the compilation
done by \citet{gamasfrd} based on H$_{\alpha}$ and H$_{\beta}$
emission-line SFR density measurements. The diamonds show the total
SFRD estimated by \citet{burgarella13} using both UV and IR observations. The observational data sets
have been corrected to be compared to a Kennicutt IMF.
}
\label{fig:sfrv}
\end{figure}

Fig. \ref{fig:sfrv} shows the evolution of the predicted cosmic star
formation rate per unit of comoving volume (SFRD). The observational
data plotted in Fig. \ref{fig:sfrv} is corrected to a Kennicutt IMF
(we set the IMF integration limits to $0.1
\leq m(M_{\sun}) \leq 100$) by
using the values tabulated in Table \ref{tab:sfrv}. These values are derived
using the P2 SPS model assuming a constant star formation rate, a fixed
solar metallicity and without dust attenuation. The UV luminosities are obtained by
convolving with a top-hat filter of 400\AA\ width the spectral
distribution at 1 Gyr obtained with the P2 SPS model \citep[see][for a
discussion of the uncertainties of using the UV luminosity as a tracer
of the SFR]{wilkins12}. The
H$_{\alpha}$ luminosity is also obtained at 1 Gyr with the P2 SPS
model, while the FIR luminosity (10 to 1000 $\mu$m) is calculated at 100 Myr. For the case of measurements inferred from
radio observations ($\sim 1.4$ GHz) we use the ratio of core-collapsed
SNe calculated with the P2 SPS model and, thus, we
are neglecting the thermal contribution to the radio emission
\citep{bressan02}. 

The predicted SFRD is below the observationally inferred one, particularly 
around $z\sim 1$. This comparison is sensitive to the extrapolation of the 
observed luminosity function to low luminosities in order to obtain the total 
luminosity density and to the contribution of dust attenuation. Both 
corrections are uncertain. \citet{lagos14} have shown that the 
model presented here agrees remarkably well with the observed H-$\alpha$ 
luminosity function, whereas the SFRD inferred from the same observations 
is substantially higher than the model prediction, and in fact is one 
of the most discrepant points in Fig.~\ref{fig:sfrv}. This illustrates 
the difficulties in using the SFRD to constrain galaxy formation models 
and favours the use of the more directly observed luminosity function.

Fig. \ref{fig:sfrv} also shows that, at $z>3$ the
predicted SFRD is dominated by the contribution from galaxies
experiencing a burst of star formation. The strong decline predicted
at $z>4$ is an artefact of the fixed mass resolution limit of the N-body
simulation. This limit is not sufficient to account for the low mass
halos that contribute the most to the SFRD at these high redshifts. 
In their calculations, \citet{lagos14} increase the resolution of 
the dark matter merger trees which redshift, reaching values as small as
$4\times 10^6\, h^{-1}$Mpc at $z=$4,
in order to sample the dominant star-forming galaxies at high
redshifts. By doing so, the predicted decline at
$z>4$ seen in Fig.\ref{fig:sfrv} becomes much weaker and the model SFRD agrees better with the observational inferences.

\begin{figure}
{\epsfxsize=8.5truecm
\epsfbox[8 35 301 448]{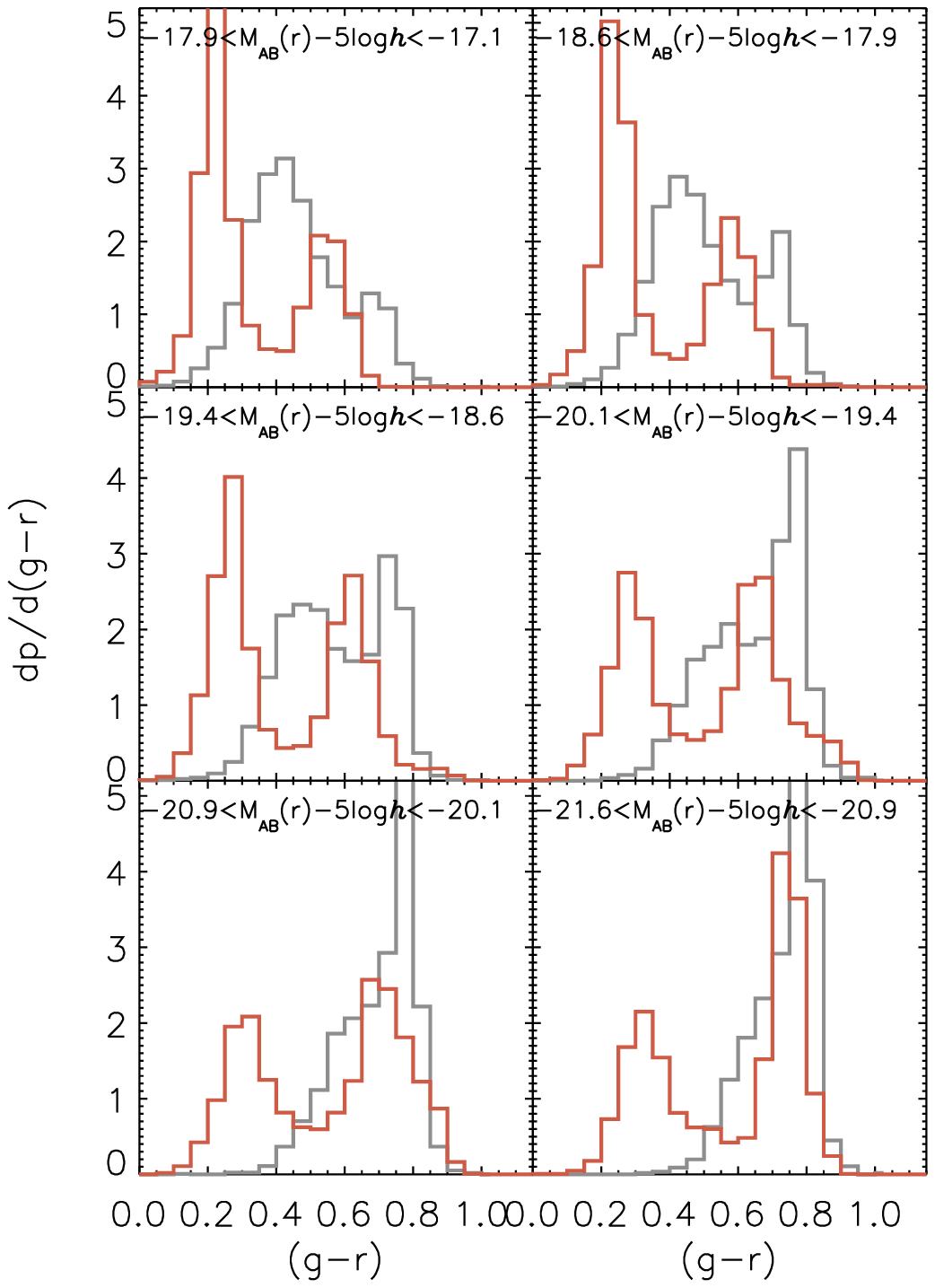}}
\caption
{The red histograms show the predicted (g-r) colour distribution at $z=$0 for
  decreasing absolute magnitude from top to bottom, as indicated in
  each panel. The grey histograms show the corresponding observational
  results from the SDSS data \citep{blanton05,juan09}. All histograms are normalised to have
  unit area.
}
\label{fig:gr}
\end{figure}

Fig. \ref{fig:gr} shows the predicted (g-r) colour distribution at $z=$0
compared with SDSS observations \citep{blanton05,juan09}. The model predicts
galaxies with colours covering a similar range to the
observations. However, the predicted colour
distribution is more strongly bimodal than is observed, as was previously
found for the \citet{bower06} model by \citet{juan09}. This mismatch is likely related to the simplistic treatment
of the gas stripping done for modelling satellite galaxies \citep{font08}.

\begin{figure}
{\epsfxsize=8.5truecm
\epsfbox[7 10 299 327]{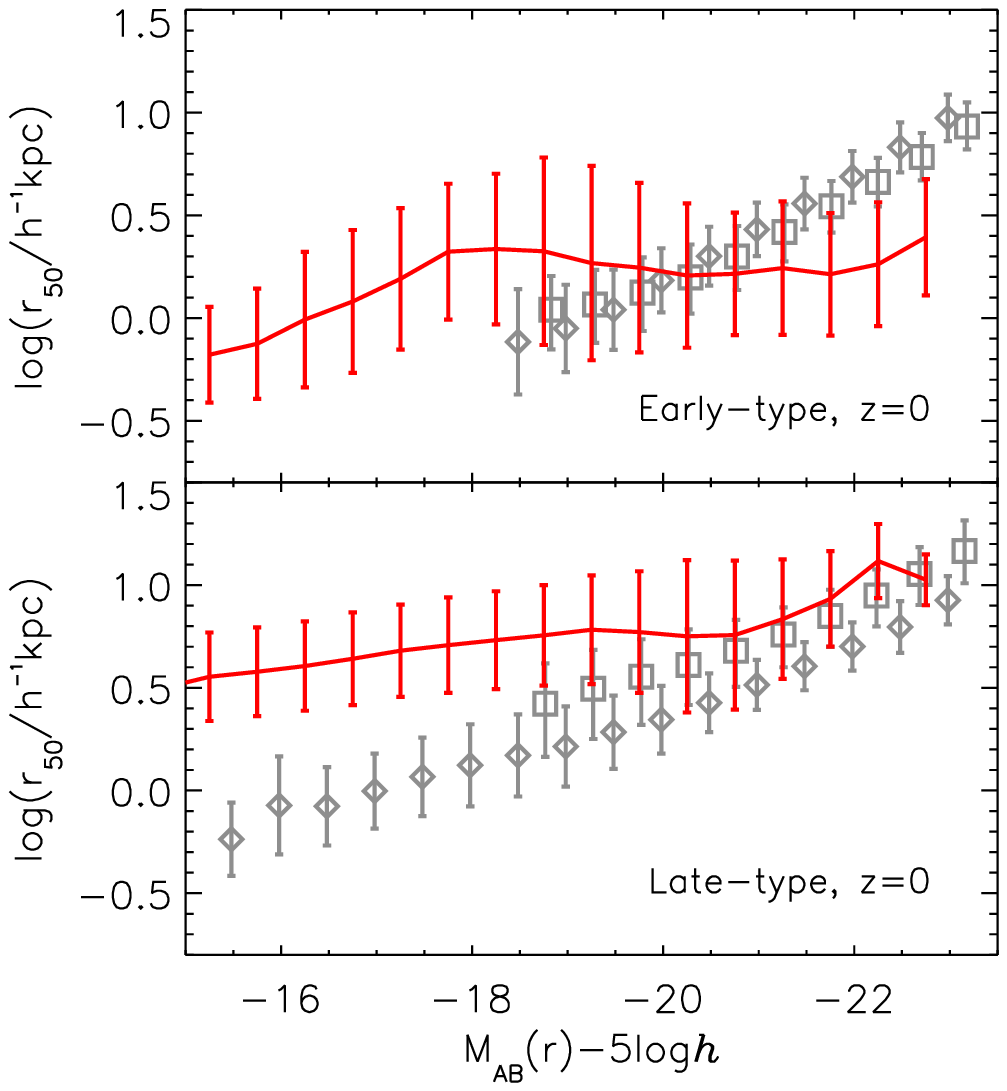}}
\caption{The solid lines show the predicted median half-light radii as a function or
  r-band magnitude for early (top) and late-type galaxies (bottom), compared with the observations from
  \citet{shen03} (diamonds) and \citet{dutton11} (squares). The error
  bars show the 1$\sigma$ range of the distributions.
}
\label{fig:size}
\end{figure}
Fig. \ref{fig:size} compares the predicted and observed median
half-light radii of early and late-type galaxies at
$z=0$ \citep{shen03,dutton11}. Model early-type galaxies are defined as those with a dust attenuated
bulge-to-total luminosity ratio in the r-band exceeding 0.5
\citep{juan09}. In {\sc galform}, the sizes of discs and bulges are predicted by
tracking the angular momentum of the gas cooling
from the halo and, in the case of bulges formed in mergers, by
applying the conservation of energy and the virial theorem
\citep{cole00,benson10}. Bright early-type model galaxies are smaller
than observed. The predicted sizes of galaxies with
M$_{AB}(r)-5$log$h<-19$ are consistent with the observations. Fainter late-type model galaxies are too large
compared with observed ones. Note that, nevertheless, the agreement is better than in \citet{bower06}
model. The calculation of galaxy sizes is complicated by
the pull of baryons on the dark matter halo \citep{almeida07}. Also,
there might be angular momentum losses of the infalling gas that are
not taken into account in our model \citep[e.g.][]{okamoto05}. Reproducing observed
galaxy sizes is a long standing problem of semi-analytical models \citep[e.g.][]{bb10}.

\begin{figure}
\includegraphics[width=8.5cm]{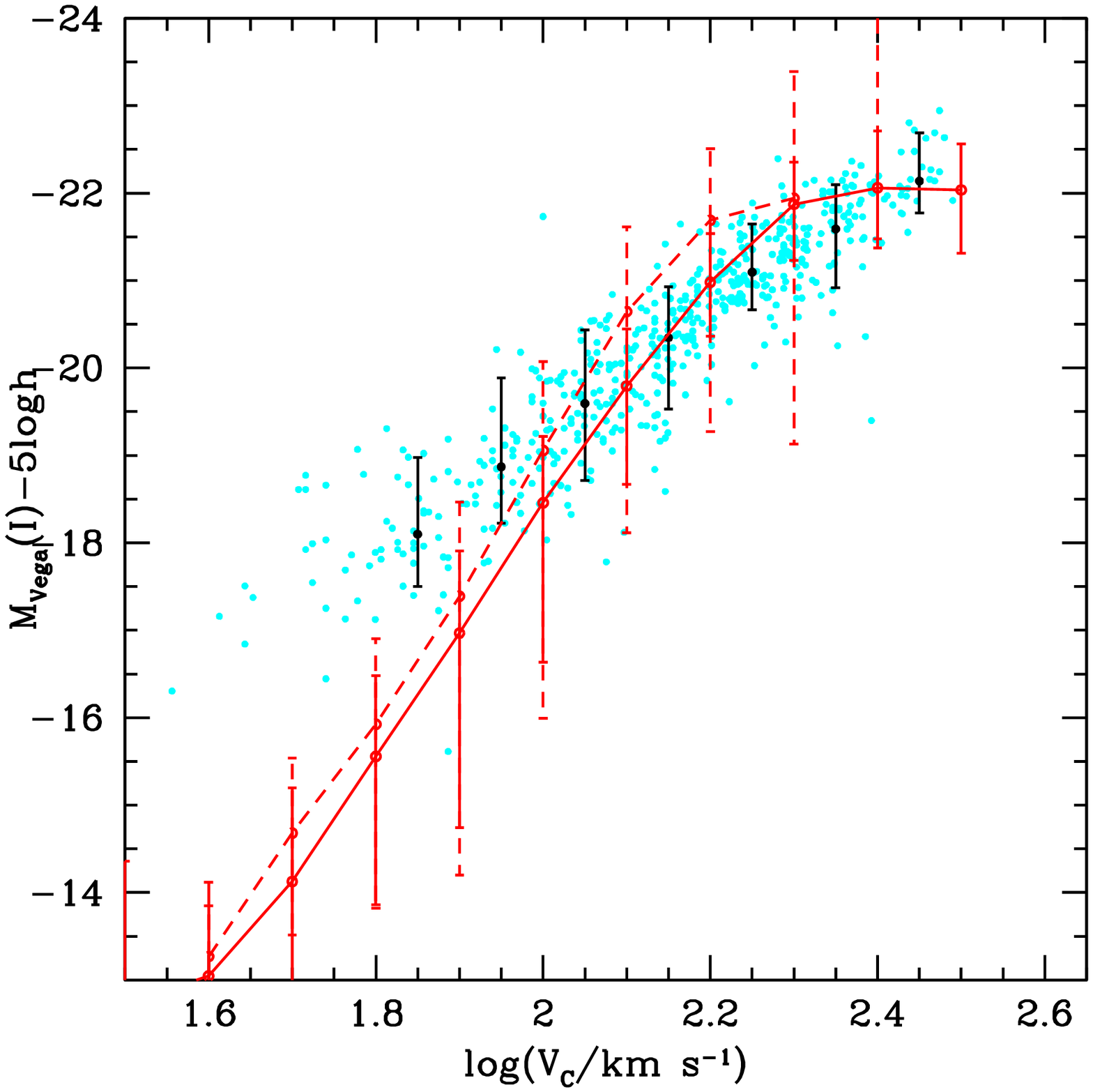}
\caption{The solid red line shows the predicted I-band Tully-Fisher relation at
  $z=$0 for galaxies with bulge-to-total B-band light ratios smaller than
  0.2. The dashed line corresponds to the case in which the host dark
  matter halo circular velocity is used. Observations of Sb-Sdm galaxies by
  \citet{math} are shown as cyan symbols. The black symbols
  show the medians and 10 to 90 percentile range for the same observational
  data in bins of Vc. 
}
\label{fig:tf}
\end{figure}
Fig. \ref{fig:tf} compares observations with the predicted I-band Tully-Fisher relation
at $z=$0. The observed Tully-Fisher relation is obtained from a sample
of undisturbed late-type galaxies and thus, in order to mimic the
observational selection, the model prediction is only shown for
galaxies with bulge-to-total light ratios smaller than
0.24 \citep[for more details see][]{cole00}. Whilst the agreement between the observations and model
predictions is good for bright galaxies, the
predicted slope differs from that
observed. Matching simultaneously and with the same level of agreement the
observed optical luminosity function and the Tully-Fisher relation at
z=0 remains difficult for semi-analytical models \citep[see also][]{croton06}.

\begin{figure}
\includegraphics[width=8.5cm]{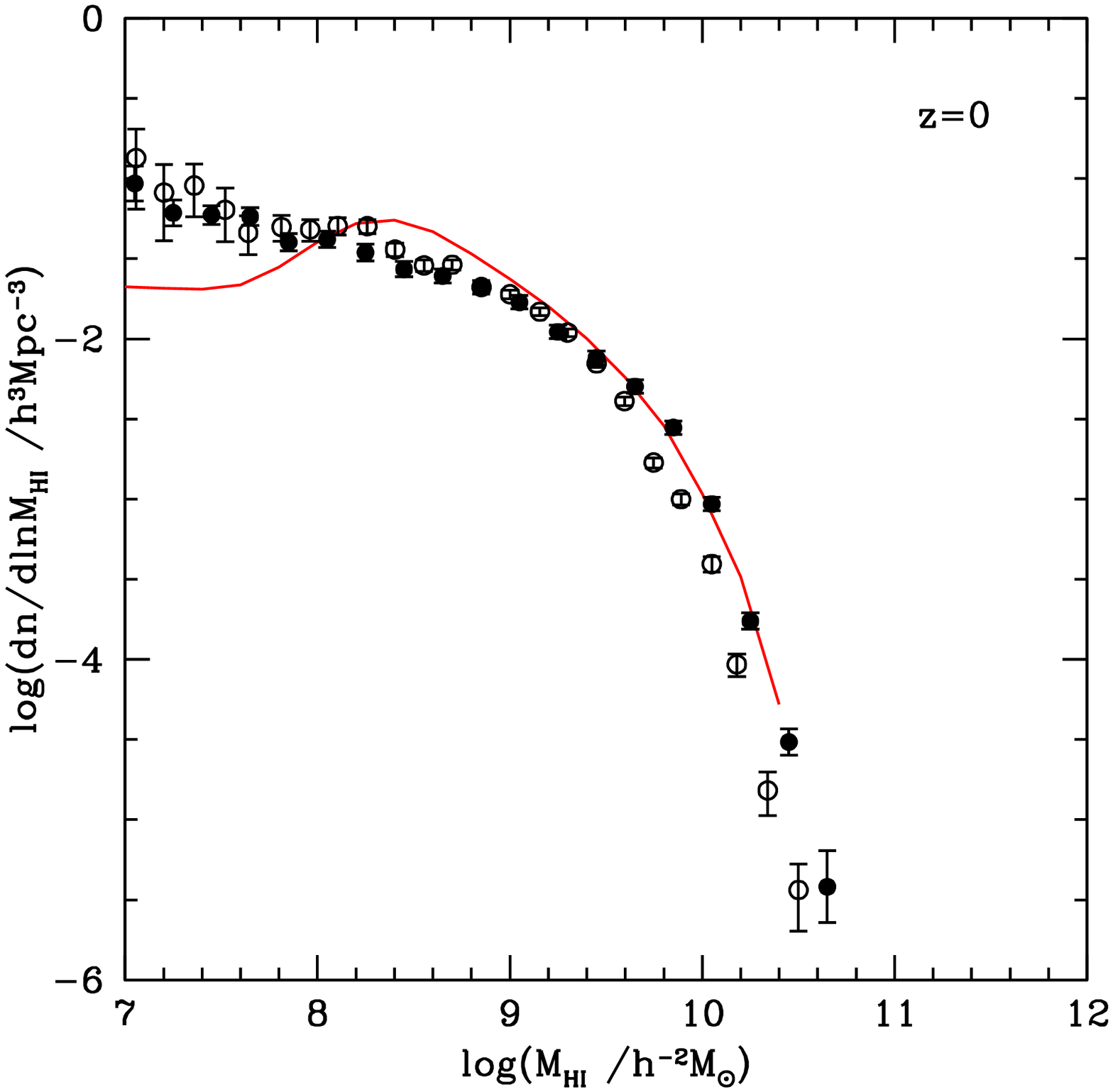}
\caption{The solid line shows the predicted HI
  mass function at $z=$0. The empty circles show observational
  results from \citet{zwaan05} (HIPASS survey) and the filled circles from
  \citet{martin10} (ALFALFA survey).
}
\label{fig:mHI}
\end{figure}
Fig. \ref{fig:mHI} shows the HI mass function at $z=$0. Following \citet{lagos10}, in our model
the star formation law explicitly distinguishes between the atomic and molecular
phases of the neutral hydrogen in the inter stellar medium. The model
prediction agrees remarkably well with the observations, except at the
lowest gas masses. The predicted turn over is mainly due to the
resolution of the simulation. Nevertheless, a turn over is also
expected for galaxies modelled using certain
implementations of the photoionization \citep{kim13}. The HI traces the cold gas content in galaxies. In
turn, the cold gas mass content of galaxies is affected
by the rate at which accreted gas is converted into stars, and thus
can be a useful observable to distinguish between different star
formation laws. Therefore, the good match between the model
predictions and observations provides encouraging support for a star
formation rate following the empirical relation deduced by \citet{blitz06}.

\begin{figure}
{\epsfxsize=8.5truecm
\epsfbox[31 7 296 263]{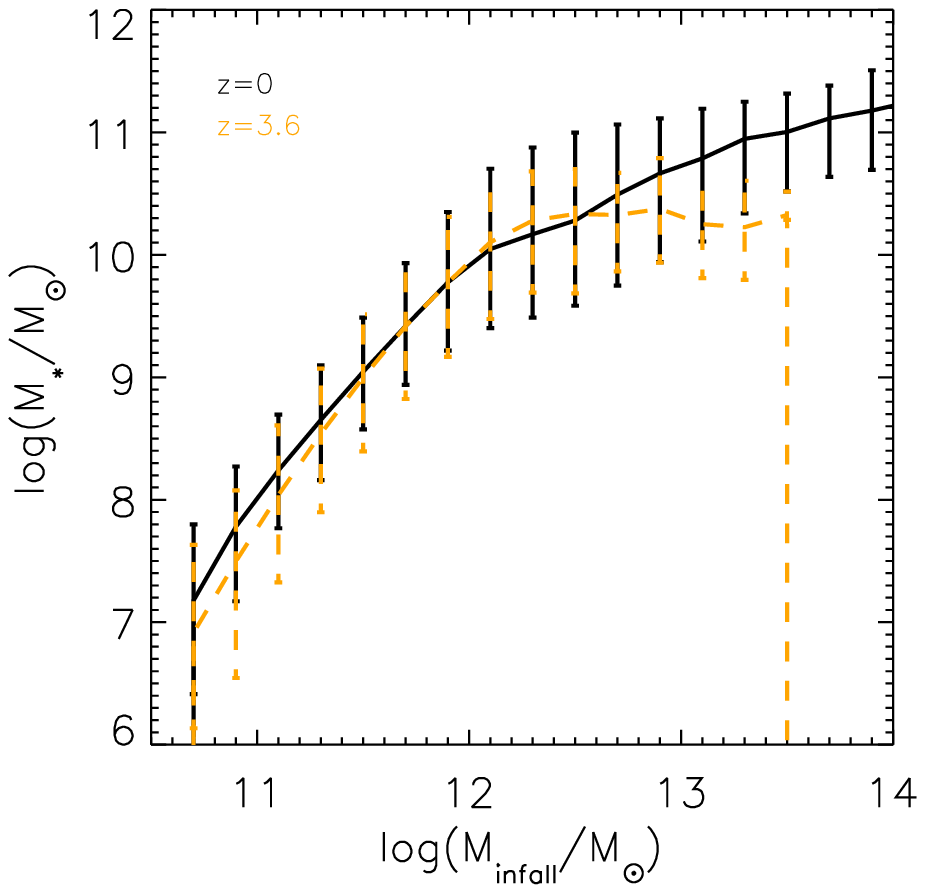}}
\caption{The median predicted stellar mass at $z=$0 (solid line) and $z=$3.6
  (dashed line), as a function of the mass of the host halo of the
  galaxy at its infall time. The
error bars show the 10 to 90 percentile range. 
}
\label{fig:ms_minfall}
\end{figure}
Fig. \ref{fig:ms_minfall} shows the predicted median stellar mass as a function
of the mass of the halo hosting the corresponding
  galaxy at its infall time. This figure shows that the predicted median stellar
mass increases with the infall host halo mass and that there is a
change of slope at around $10^{12}M_{\odot}$. This change of slope is likely related
to the AGN feedback \citep{bower12}. At $z=$0, the curve is quite close to that calculated
by \citet{qi11} using a different semi-analytical model based on the
Millennium N-body simulation. The
corresponding trends for galaxies splitted into central and satellites
are very similar to the overall trend shown in
Fig. \ref{fig:ms_minfall}. However, we find that, as it was also shown
in \citeauthor{qi11}, for infall host haloes with $M<10^{12}M_{\odot}$ the median stellar mass of central galaxies is
systematicaly below that for satellites, by a factor of $\sim 1.6$.  As expected in a hierarchical scenario
of galaxy formation, at high redshifts, such as $z=3.6$, there are
fewer massive galaxies and also the minimum
halo mass required to host a galaxy of a given stellar mass is higher
than at lower redshifts.

\begin{figure}
{\epsfxsize=8.5truecm
\epsfbox[46 42 501 489]{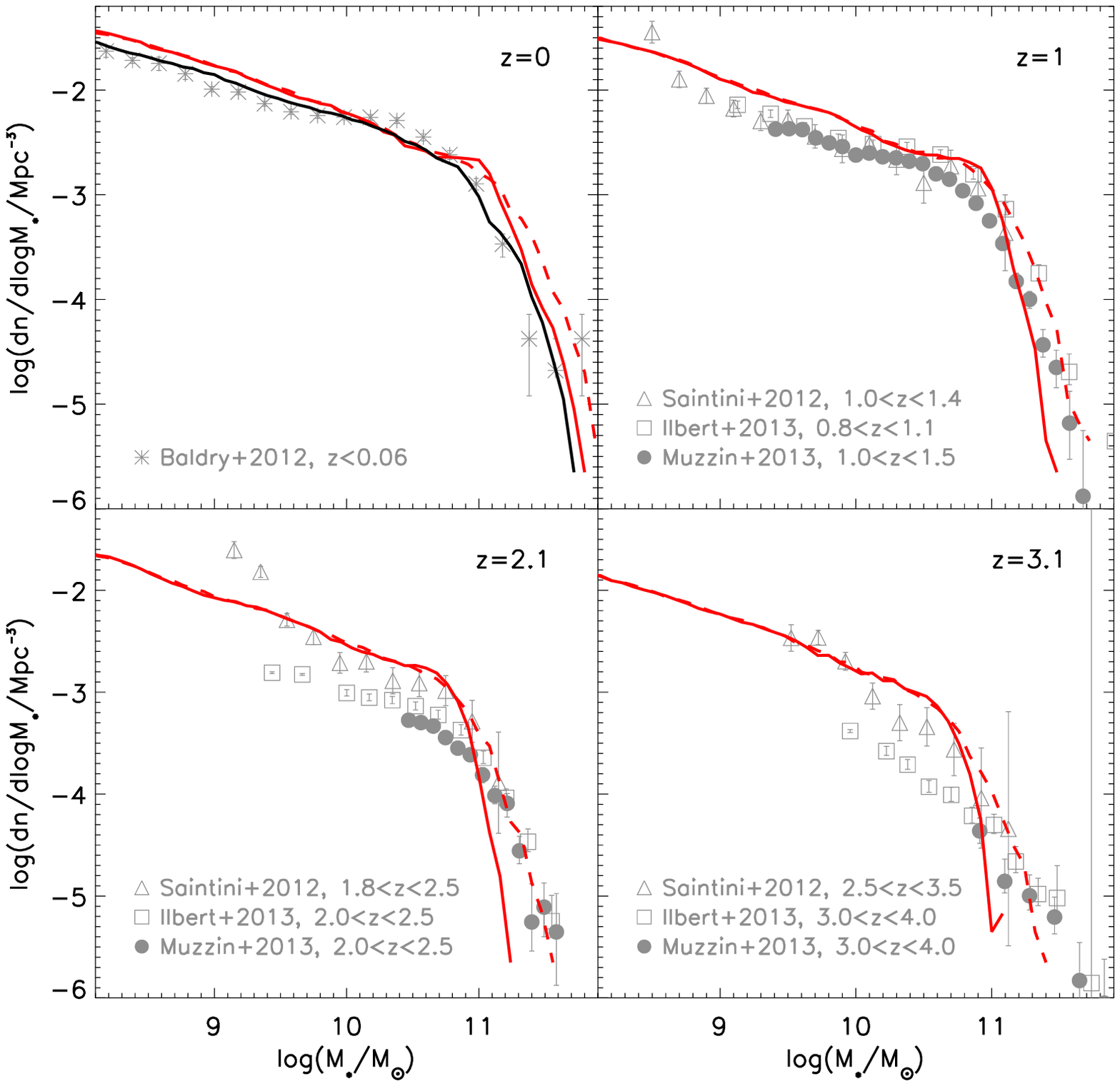}}
\caption{The solid red lines show the predicted stellar mass functions at $z=$0, 1.1, 2.1 and 3.1, as
  indicated in the legend of each panel. The dashed red lines show the
  corresponding mass functions convolved with a typical observational
  error in the stellar mass estimation of 0.2 dex. The solid black
  line shows the stellar mass function at $z=0$ obtained from the model broad-band
  photometry by SED-fitting using a similar estimator as that in \citet{baldry12}. Symbols show the
  observational data from \citet{baldry12} (asterisks),
  \citet{santini12} (triangles), \citet{ilbert13} (squares) and
  \citet{muzzin13} (filled circles). The observational data have been
converted to be compared with the predictions done assuming a
Kennicutt IMF \citep{ilbert10,gilbank11}.
}
\label{fig:massf}
\end{figure}
Fig. \ref{fig:massf} shows the stellar mass function at different
redshifts compared with observations. At $z=0$, the mass function 
computed using the stellar masses output directly from the model, 
which we refer to as the ``true'' stellar mass function (shown by 
the solid red line in Fig.~\ref{fig:massf}), appears to be slightly at odds 
with the mass function inferred from the observations, predicting too 
many galaxies both at low and high stellar masses. \citet{mitchell13} have shown that such a comparison is flawed, and that the 
appropriate way to test the model is using the stellar mass assigned to each 
model galaxy by fitting to its photometry in different bands, mimicking 
the process applied to the observed galaxies. The common perception of this 
procedure is that the true mass function should be convolved with a 
Gaussian, reflecting the error in the fitted stellar mass. The outcome 
of this simple exercise of convolving with a Gaussian is shown by the
red dashed line in Fig.~\ref{fig:massf}. The result of applying the full fitting machinery to the model galaxies is more complicated than this and is shown by the black line in Fig.~\ref{fig:massf}. The  stellar mass function inferred in this way from the model photometry 
is in excellent agreement with the observationally inferred one. We note 
in passing that a model in which the parameters have been set by comparing 
the ``true'' mass function to one inferred from observations is unlikey to 
still match the observations if the mass function is deduced by fitting to 
the predicted photometry. At higher redshifts the predicted mass function is in
reasonably good agreement with observations of M$_*$ and more massive
galaxies, once the contribution of stellar mass measurements errors is 
taken into account, which preferentially scatter galaxies to higher masses.  

\begin{figure}
{\epsfxsize=8.5truecm
\epsfbox[26 4 324 294]{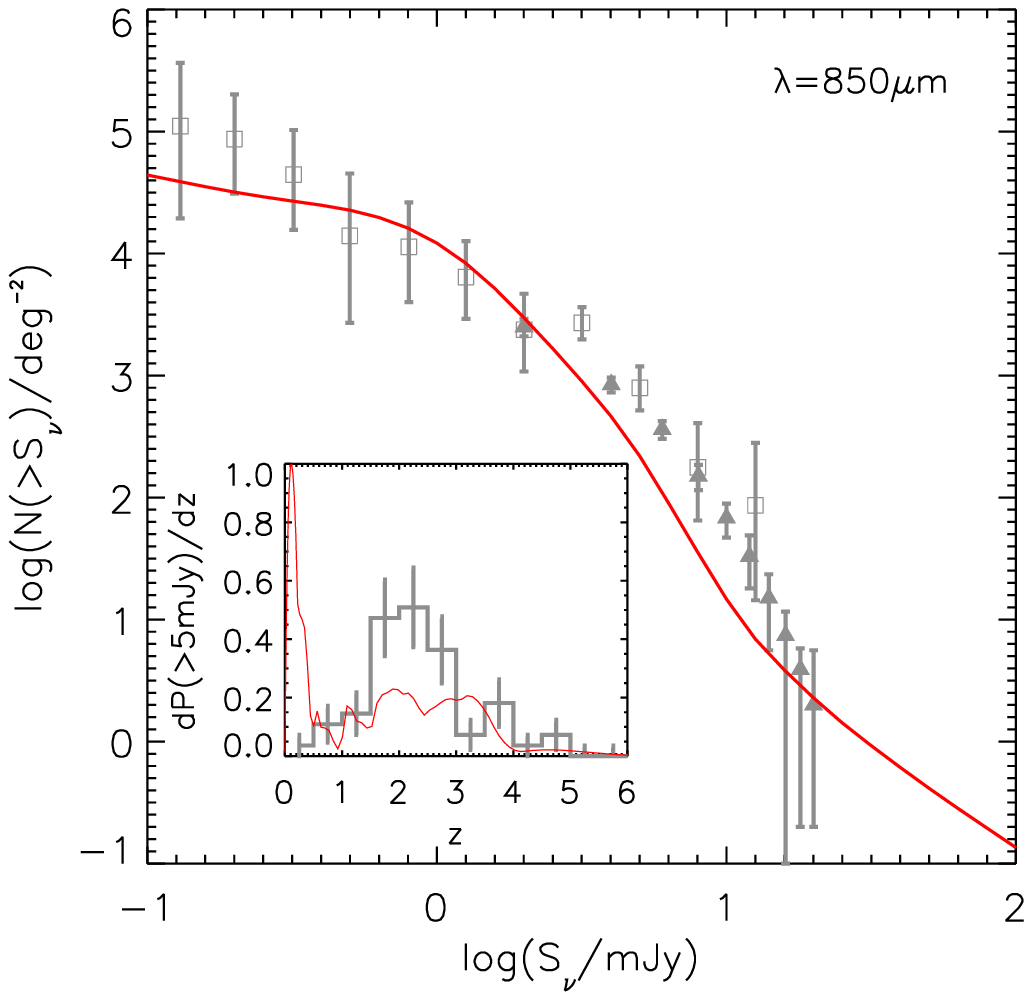}}
\caption{The predicted cumulative number counts at 850$\mu$m (red
  solid lines) and the observed ones by
  \citet{coppin06} (triangles), \citet{knudsen08} (squares). In the
  inset, they are shown the predicted redshift
  distributions at 850$\mu$m (red solid lines) together with the observational
  data from \citet{wardlow11}. The histograms are normalized to a unit area.
}
\label{fig:submm}
\end{figure}
Fig. \ref{fig:submm} shows the predicted cumulative number counts at
850$\mu$m. We find that both the number counts and redshift
distribution of model 870$\mu$m sources are virtually
indistinguishible from those at 850$\mu$m. Observations from ALMA by
\citet{karim13} at 870$\mu$m have shown that the number counts of
bright sources are affected by the confusion of sources, and thus are
lower than previously seen. Nevertheless, these
observations were taken in underdense fields
\citep{weiss09}, correcting for which will imply boosting
the number counts. Thus, even taking into account more recent data
than that shown in Fig. \ref{fig:submm}, the
model underpredicts by a factor of $\sim$ 3 the number counts for sources brighter than 2.5
mJy. Bright submillimeter galaxies  are also predicted to appear in large
  numbers at $z=0$, having a median redshift of $z=1.7$. The
  observations suggest that submilimiter galaxies are in place at
  higher median redshifts, in particular, the
  observational data shown in Fig. \ref{fig:submm} has a median redshift $z=2$. The inability to
  simultaneously match the number counts and redshift distribution of
  submillimiter galaxies is one of the reasons that motivated the
  inclusion of a top-heavy IMF in bursts of star formation in the
  \citet{baugh05} model. The
  Lacey et al. (in preparation) model is based on a WMAP7 cosmology
  (as it is the model presented in this paper) and it does not assume a universal IMF. In
  doing so, the Lacey et al. model
  matches very well the observed number counts of submillimiter
  galaxies and predicts a redshift distribution closer to the observed
  one, lacking the low
  redshift spike seen in Fig. \ref{fig:submm}, with a broad peak in the
  same range as observations and a median redshift of $z=2.35$.

\begin{figure}
{\epsfxsize=8.5truecm
\epsfbox[10 0 328 246]{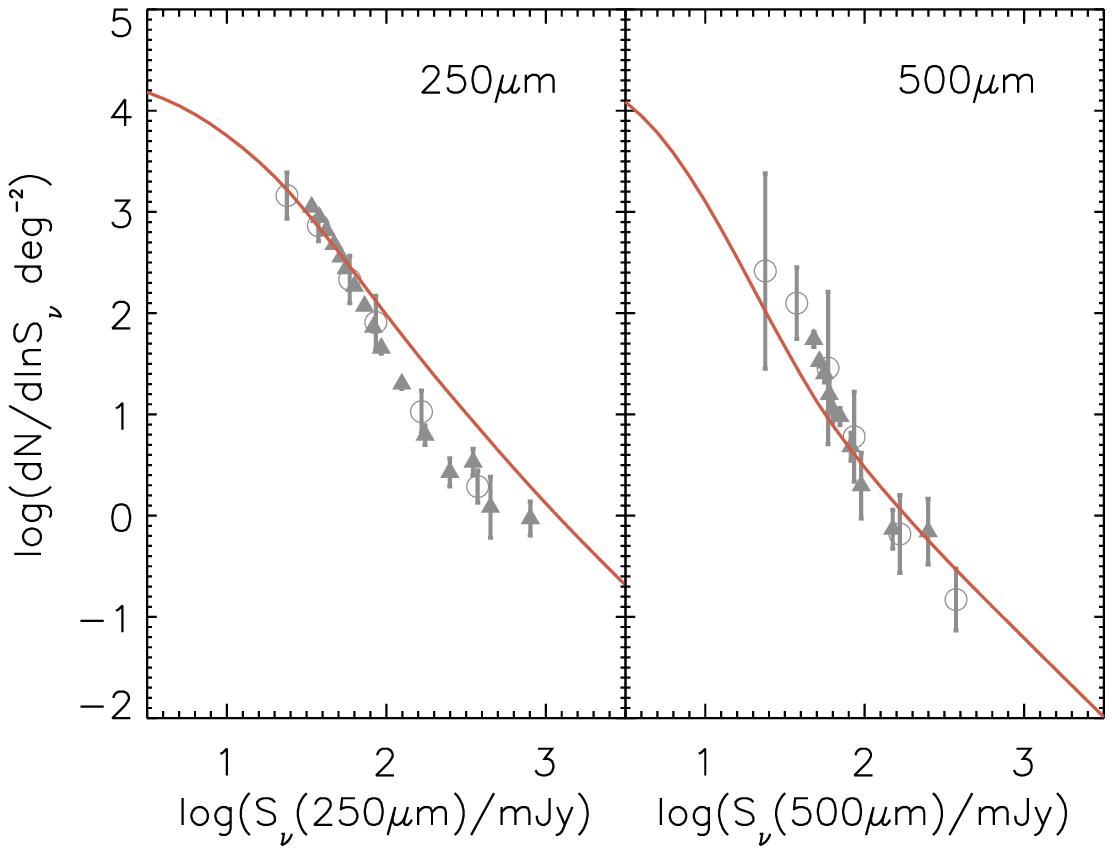}}
\caption{The solid line shows the predicted differential number counts
  counts at 250$\mu$m (left) and 500$\mu$m (right). The triangles correspond to observations from \citet{hatlas}
  and the circles to those from \citet{hermes}.
}
\label{fig:herschel}
\end{figure}
Fig. \ref{fig:herschel} shows the differential number counts at
250$\mu$m and 500$\mu$m. These two wavelengths correspond to the peak
of emission due to dust for galaxies at $z\sim 2$ and $z\sim 4$,
respectively. At 250$\mu$m the model overpredicts the number of the
observed bright galaxies. This trend
was also found for previous releases of the {\sc galform} model \citep{hatlas,kim12}. Nevertheless, at
500$\mu$m, the model reproduces remarkably well the
observations.

\end{document}